\documentclass[11pt,letter]{article}
\usepackage{tikz}
\usetikzlibrary{positioning}

\usepackage[utf8]{inputenc}
\DeclareUnicodeCharacter{3000}{ }

\usepackage{geometry, diagbox,tabularx}

\usepackage[hyphens,spaces]{url}

\usepackage{latexsym, amsmath, amstext, amssymb, amsfonts,natbib, amsthm,lineno,pgfkeys, tikz, comment, lineno, makecell, pgfplots}
\usepackage{lineno}

\usepackage{thmtools, stackrel, caption,,graphicx,lscape, makecell, tabularx}

\usetikzlibrary{shapes.geometric}

\usepackage{tikz,verbatim,kbordermatrix}

\usepackage[flushleft]{threeparttable}

\usepackage{subfig}
\usepackage{hyperref}
\usepackage{multirow}
\usepackage{caption}

\usepackage{lscape} %Needed to use the landscape command.
\usepackage{threeparttable} %To put notes right below the table

\usepackage[english]{babel}
\usepackage[]{fontenc}
\usepackage{xr}
\usepackage{euscript}
\usepackage{bm}

\usetikzlibrary{arrows, decorations.markings}

\def\addlegendimage{\csname pgfplots@addlegendimage\endcsname}

\newtheorem{theorem}{Theorem}
\newtheorem*{theorem*}{Theorem}

\newtheorem{corollary}{Corollary}
\newtheorem{definition}{Definition}
\newtheorem{lemma}{Lemma}
\newtheorem{proposition}{Proposition}

\newtheoremstyle{boldremark}
    {\dimexpr\topsep/2\relax} % space above
    {\dimexpr\topsep/2\relax} % space below
    {}          % body font
    {}          % indent amount
    {\bfseries} % theorem head font
    {.}         % punctuation after theorem head
    {.5em}      % space after theorem head
    {}          % theorem hed spec. (empty = "normal")
\theoremstyle{boldremark}
\newtheorem{remark}{Remark}

\declaretheoremstyle[spaceabove=1pt, spacebelow=6pt, headfont=\bfseries, notefont=\bfseries, notebraces={\UTF{0081}i}{\UTF{0081}j}, postheadspace=1em, numbered=no,qed=$\blacksquare$]{myproof}\declaretheorem[title=Proof, style=myproof]{myproof} 

\declaretheoremstyle[spaceabove=1pt, spacebelow=6pt, headfont=\itshape, notefont=\bfseries, notebraces={\UTF{0081}i}{\UTF{0081}j}, postheadspace=1em, numbered=no,qed=$\square$]{mysubproof} \declaretheorem[title=Proof, style=mysubproof]{mysubproof}\renewenvironment{proof}{\begin{mysubproof}}{\end{mysubproof}}

\def\ep{\epsilon}
\def\al{\alpha}

\def\Re{\mathbf{R}}\def\R{\mathbf{R}}\def\N{\mathbf{N}}\def\F{\mathcal{F}}\def\P{\mathcal{P}}\def\D{\mathcal{D}}\def\Ql{\mathcal{Q}}\newcommand{\M}{\mathcal{M}}

\newcommand{\te}{\text}

\def\epsilon{\varepsilon}

\def\la{\lambda}

\newcommand{\df}[1]{{\em #1}}

\newcommand{\beq}{\begin{eqnarray*}}
\newcommand{\eeq}{\end{eqnarray*}}
\newcommand{\beqq}{\begin{eqnarray}}
\newcommand{\eeqq}{\end{eqnarray}}

\newcommand{\bequa}{\begin{equation}}
\newcommand{\eequa}{\end{equation}}

\newcommand{\bit}{\begin{itemize}}
\newcommand{\eit}{\end{itemize}}
\newcommand{\bc}{\begin{center}}
\newcommand{\ec}{\end{center}}

\newcommand{\tb}{\textbf}

\newcommand{\step}{\textit{Step }}
\newcommand{\case}{\tb{Case }}

\newcommand{\cl}{\text{cl\hspace{2pt}}}
\newcommand{\conv}{\textrm{co\hspace{2pt}}}
\newcommand{\co}{\textrm{co\hspace{2pt}}}
\newcommand{\rint}{\text{rint\hspace{2pt}}}
\newcommand{\supp}{\text{supp\hspace{2pt}}}
\newcommand{\aff}{\text{aff\hspace{2pt}}}

\newcommand{\lra}{\Longleftrightarrow}

\newcommand{\etadj}{\eta_j}
\newcommand{\etadl}{\eta_l}

\usepackage{setspace}
\title{Approximating Choice Data by Discrete Choice Models} 
\author{Haoge Chang, Yusuke Narita, and Kota Saito\footnote{Chang: Columbia University:  hc3516@columbia.edu; Narita: Yale University: yusuke.narita@yale.edu; Saito: Caltech: saito@caltech.edu This paper subsumes parts of ``Axiomatizations of the Mixed Logit Model'' by Saito (The paper is available at  \protect\url{http://www.hss.caltech.edu/content/axiomatizations-mixed-logit-model}).}}
\begin{document}
%\linenumbers

\maketitle
\abstract{
Random-coefficient discrete choice models, including mixed-logit, are commonly used in empirical choice analysis for their ability to capture rich substitution patterns, with applications in demand prediction, revenue management, and merger analysis. We ask when these models are sufficiently flexible to approximate any nonparametric random utility model arbitrarily well across choice sets. We show that the necessary and sufficient condition is the affine independence of the characteristic vectors. When the condition fails, we identify strict preference rankings that are difficult to approximate and propose algorithms to quantify the resulting approximation error. Applying the framework to a dataset on DVD sales, we show that widely used random-coefficient models substantially restrict substitution patterns and may lead to suboptimal assortment decisions. We offer practical guidance for diagnosing and mitigating these problems.

}

\section{Introduction}\label{sec:introduction}

\vspace{-0.2cm}

Random-coefficient discrete choice models are widely used in economics, marketing, and operations research to analyze individual and aggregate choice behavior. They are commonly used to approximate rich preferences and to capture flexible substitution patterns. The models are key inputs for counterfactual analyses in applications such as demand prediction, revenue management, and merger analysis. However, the exact degree of flexibility and the limitations of random-coefficient models have not been fully understood.\footnote{An important early work in this direction is \cite{mcfadden2000mixed}. See the section on related literature for details.} When a random-coefficient model fails to adequately capture the underlying choice behavior, the resulting misspecification can generate prediction errors that lead to misleading policy conclusions and suboptimal managerial decisions.

%\textcolor{red}{Old paragraph } Random-coefficient discrete choice models are widely used across economics, marketing, and operations research to analyze individual and aggregate choice behavior. These models are commonly used to approximate various preferences and to capture rich substitution patterns. However, the exact degree of flexibility and the limitations of the random-coefficient models have not been fully understood.\footnote{An important early work in this direction is \cite{mcfadden2000mixed}. See the section on related literature for details.} When a model cannot represent the underlying choice behavior well, misspecification and resulting prediction errors can translate into misleading policy conclusions and suboptimal managerial decisions. 

In this paper, we provide a sharp characterization of the approximation power of random-coefficient additive random utility models (ARUMs). We obtain a necessary and sufficient condition under which random-coefficient models can approximate the choice behavior generated by any nonparametric random utility model arbitrarily well. Some widely used specifications—for example, linear mixed-logit models with a limited set of observed characteristics—violate this condition and therefore cannot flexibly capture key features of choice behavior, including substitution patterns. We show how to identify strict  preference rankings that may be difficult for a given model to approximate closely. We then propose algorithms to quantify the magnitude of the resulting approximation errors. Together, these results offer researchers practical tools to assess whether their chosen discrete choice specification is rich enough for the application at hand.

We consider the standard discrete-choice setup (see, e.g., \cite{train2009discrete}). Let $J$ denote the set of all alternatives. For each alternative $j\in J$,  we use $x_j\in \Re^K$ to denote the vector of characteristics of alternative $j$, where $K$ is the number of characteristics. Let $\rho(D,j)$ denote the probability of choosing $j$ from a choice set $D\subset J$. In this paper, we focus on the class of  {\it random-coefficient additive random utility models}. Such models—including 
mixed-logit specifications—have become workhorse tools for empirical work in 
industrial organization, marketing, transportation, and operations. They are routinely  used to capture rich substitution patterns and to accommodate both observed and unobserved heterogeneity in preferences.

To introduce these models, we first define an  {\it additive random utility model (ARUM)} as one in which the choice probabilities take the form $
\rho\left(D,j\right)= \mu\bigl(\{\ep| \beta\cdot x_j+\etadj+\ep_j > \beta\cdot x_l+\etadl+\ep_l, $ $\forall l \in D\setminus \left\{j\right\}\}\bigr)$, where $\beta\in\Re^K$ is a coefficient vector capturing an agent's preferences,  $\eta=(\etadj)_{j\in J}$ is a vector of {\it fixed effects} capturing unobserved characteristics of alternatives, and $\ep=(\epsilon_j)_{j\in J}$ is a vector of random utility shocks with a probability measure $\mu$. The class of  ARUMs is general and includes probit and logit models as special cases. A {\it random-coefficient} ARUM extends this framework by allowing taste coefficients to vary across agents. Formally, the choice probabilities are obtained by integrating ARUM choice probabilities against a distribution over $\beta$. In one standard interpretation, the distribution $m$ captures the heterogeneity of preferences among the population of agents. When $\mu$ is an iid type-I extreme-value distribution, $\rho$ reduces to a  {\it mixed-logit model}.

Given the popularity of the random-coefficient ARUMs, it is important to understand their flexibility and limitations.  For this purpose, we obtain a necessary and sufficient condition under which the  random-coefficient ARUMs are rich enough to approximate any choice probabilities generated by nonparametric random utility models arbitrarily well across choice sets. For the approximation target, we choose the random utility models, which are defined as probability measures over strict preference rankings over alternatives.  We choose this class of models as the approximation target because it is the most general and agnostic framework that only assumes individual rationality. 
%We study approximation {\it across choice sets} because many questions of interest, such as substitution patterns, rely on analyzing behavior across choice sets. Choice data across multiple choice sets are commonly encountered in applications such as conjoint analysis, revenue management, and merger analysis.
We study approximation {\it across choice sets} because many questions of interest, such as substitution patterns, rely on analyzing behavior across choice sets. Choice data across multiple choice sets are commonly encountered in applications such as conjoint analysis, revenue management, and merger analysis. 
Counterfactual analyses also require predicting behavior in new or modified choice sets. In such cases, the random-coefficient structure imposes restrictions across choice sets.\footnote{In applications with fixed effects, these fixed effects are often estimated from observed choice sets and then held fixed or otherwise specified for counterfactual choice sets.}
%and our results characterize when these restrictions allow arbitrary random utility behavior to be approximated.

We briefly summarize our main results. Our main theorem (Theorem \ref{theo:1}) states that the necessary and sufficient condition for approximating arbitrarily well is the \textit{affine-independence} of the set $\{x_j \in \Re^K| j \in J\}$.\footnote{A set $Y\equiv \{y_1,\dots, y_n\}$ is \textit{affinely independent} if for any $y_i \in Y$, there exist no real numbers $\{\mu_j\}_{j \neq i}$ such that $y_i =\sum_{j \neq i} \mu_j y_j$ and $\sum_{j \neq i} \mu_j=1$. A set $Y\equiv \{y_1,\dots, y_n\}$ is affinely independent if and only if the set $\{y_2-y_1,\dots, y_n-y_1\}$ is linearly independent.}  To interpret the condition, consider a typical setup, where a researcher first fixes one probability measure $\mu$ over the shock $\ep$. Then the researcher estimates the distribution  $m$ over coefficients $\beta$ and the fixed effects $\eta$ after observing a dataset. If the affine-independence condition is satisfied, then the researcher should be able to approximate any given dataset by using some random-coefficient ARUMs arbitrarily well across choice sets. On the other hand, if the affine-independence condition is violated, there exists a dataset generated by a random utility model that cannot be approximated arbitrarily well by any sequence of random-coefficient ARUMs,  no matter which random-coefficient distribution $m$ and fixed effects $\eta$  are used. The affine-independence condition is straightforward  to test: the condition is generically equivalent to the simpler condition: $K \ge |J|-1$, where $|J|$ is the number of alternatives and  $K$ is the number of characteristics observed for each alternative.

Many empirical papers use the mixed-logit models that are linear in the original characteristics and do not contain additional terms such as polynomials. We call such models {\it linear mixed-logit models}. In these papers,  we often observe a deviation from the condition $K\ge |J|-1$, resulting in the violation of the affine-independence condition. This means that the linear mixed-logit models may not be rich enough to approximate the true substitution pattern arbitrarily well across subsets of $J$, no matter how one chooses the distribution $m$ and the fixed effects $\eta$.

We show that the source of this approximation failure can be traced to strict preference rankings that are difficult for the model to approximate. These rankings can be identified efficiently using linear programming methods.

%To further quantify the flexibility of the random-coefficient ARUMs under researchers' consideration, we use two algorithms, the EM  (Expectation-Maximization)  algorithm drawn from  \cite{dempster1977maximum}, and an adaption of the greedy algorithm proposed in \cite{barron2008approximation} to calculate approximation error to stochastic choice functions. For example, one can calculate the approximation error to deterministic choice functions corresponding to strict preference rankings that are challenging to approximate precisely. With a slight modification, one can use the same greedy algorithm to calculate the maximal substitution patterns allowed in the class of random-coefficient ARUMs. 

To further quantify the flexibility of any given random-coefficient ARUM specification, we develop two algorithms to compute the approximation error to strict preference rankings that are particularly difficult to approximate. First, we implement an EM (Expectation–Maximization) algorithm, following \cite{dempster1977maximum}. Second, we adapt the greedy algorithm proposed in \cite{barron2008approximation}. With a slight modification,  we also compute the maximal substitution patterns that are attainable within a given class of random-coefficient ARUMs.

We apply our theorem and algorithms to a DVD sales dataset from \cite{rusmevichientong2010dynamic}. We focus on a subsample with four alternatives (products 1--4 hereafter) and two characteristics. In this setting, the affine-independence condition fails because \(K=2<3=|J|-1\). We show that  half of all strict preference rankings cannot be well approximated by linear mixed-logit specifications, and the resulting approximation errors can be substantial. Moreover, as shown in Section~\ref{sec:substitution}, the linear mixed-logit model imposes strong restrictions on substitution patterns. In our setting, \emph{no} linear mixed-logit specification can generate substantial substitution between products 1 and~3 or between products 2 and~4.

In this application, we show that the above limitation of the linear mixed-logit model can distort assortment decisions. Consider a market with three consumer segments: $50\%$ rank $3 \succ 4 \succ 1 \succ 2$, $25\%$ rank $1 \succ 3 \succ 2 \succ 4$, and $25\%$ rank $2 \succ 4 \succ 3 \succ 1$, where product~4 is the outside option. Suppose product~3 has the highest margin, followed by products~1 and~2, and shelf-space constraints require the retailer to remove either product~1 or product~2. A linear mixed-logit model estimated from historical sales cannot predict sizable reallocation from product~1 to product~3 when product~1 is removed, but may predict that removing product~2 redirects demand toward products~1 and~3 with little market exit. The retailer may therefore drop product~2. In reality, however, the $25\%$ segment with preferences $2 \succ 4 \succ 3 \succ 1$ exits the market when product~2 is removed, causing an avoidable revenue loss. Thus, restrictive substitution patterns in linear mixed-logit models can lead firms to mis-rank assortments and make suboptimal decisions.

%This limitation of the linear mixed-logit model can lead to  a suboptimal decision in  an assortment-optimization problem. For example, consider a market with three consumer segments: 50\% rank $3 \succ 4 \succ 1 \succ 2$,  25\% rank $1 \succ 3 \succ 2 \succ 4$, and 25\% rank $2 \succ 4 \succ 3 \succ 1$, where product~4 is treated as the outside (no-purchase) option. Suppose product~3 has the highest margin, followed by products 1 and 2. Because of shelf-space constraints, the retailer must remove either product~1 or product~2. The retailer estimates a linear mixed-logit model using historical sales data and uses it to predict the effect of dropping a product. However, no linear mixed-logit model can predict sizable demand reallocation from product~1 to product~3 when product~1 is removed. By contrast, removing product~2 is predicted to redirect demand toward the high-margin products 1 and~3 with relatively little market exit (i.e., choosing product $4$). Based on these predictions, the retailer drops product~2. In reality, however, the entire 25\% segment with preferences $2 \succ 4 \succ 3 \succ 1$ exits  the market (i.e., chooses product $4$) when product~2 is removed---a substitution pattern that no linear mixed-logit model can represent. As a result, the retailer incurs a substantial and avoidable revenue loss. More broadly, this example illustrates how restrictive substitution patterns in linear mixed-logit models can lead firms to mis-rank assortments and make systematically suboptimal decisions.

The structure of the paper is as follows. Section~\ref{sec:math} introduces the discrete-choice setup and the definitions of various models used in our analysis. Section~\ref{sec:main} presents our main theorem and related results and discusses their implications for representability and approximation across choice sets. Section~\ref{sec:measuring} develops our distance measure and the EM and greedy algorithms for quantifying approximation
errors. Section~\ref{sec:data} applies our framework to a DVD sales dataset and documents the resulting approximation errors and implications for substitution patterns and assortment planning. All proofs are collected in the appendix.

\vspace{-0.2cm}

\subsection*{Related Literature}

\vspace{-0.2cm}

The work most closely related to our paper is \cite{dagsvik1994discrete} and, especially, \cite{mcfadden2000mixed}, who show that any given nonparametric continuous random utility model can be approximated arbitrarily well by a mixed-logit model.\footnote{Our result is consistent with theirs: heuristically, the result of \cite{mcfadden2000mixed} corresponds to the case in which the researcher allows arbitrarily high-order polynomials (i.e., $K \to \infty$), which satisfies our condition $K \ge |J|-1$.} The key differences concern both the setup and the sharpness of the results. First, we provide a \emph{necessary and sufficient} characterization of when random-coefficient ARUMs can approximate arbitrary random-utility behavior, whereas \citet{mcfadden2000mixed} only establish a sufficient condition for approximation. 
    We obtain this sharper characterization by focusing on approximation for a fixed, finite set of alternatives. Second, we analyze approximation of choice probabilities \textit{across choice sets}. This cross-menu perspective with fixed alternatives is central for applications in assortment optimization and revenue management \citep{farias2009non,jagabathula2022nonparametric}. Third, our affine-independence condition is necessary and sufficient, applies beyond mixed-logit, and is tailored to environments with fixed finite alternatives. By contrast, \citet{mcfadden2000mixed} provide asymptotic approximation guarantees, focusing primarily on settings with continuous covariates, and they offer no comparable guidance on how to construct random-coefficient ARUMs to achieve approximation in the finite-alternatives, cross-menu setting.

Another paper closely related to ours is \citet{norets2013surjectivity}. They study whether
ARUMs can represent any stochastic choice on a single multinomial
choice problem. In particular, under a mild regularity condition, they show that for a given menu $D$, by
appropriately choosing the vector of mean utilities (the deterministic utility indices of the
alternatives), an ARUM without random coefficients can generate any desired choice
probability vector $\{\rho(D,j)\}_{j\in D}$ on the interior of $|D|-1$-simplex. Our scope is different in two important
respects. First, we study approximation across multiple menus for a
fixed set of alternatives. This cross-menu perspective requires the model to approximate the data well simultaneously across different choice sets. This issue does not arise in the single-menu setting. Second, we allow for random coefficients and ask when a given class of
random-coefficient ARUMs can approximate any random utility model across menus. As
we show (see Lemma~\ref{lem:degenerate}), menu-by-menu constructions à la
Norets–Takahashi cannot in general be made mutually consistent across menus, so their
single-menu surjectivity result does not imply our cross-menu approximation result, and
vice versa. Other related papers also focus on a fixed  choice set. \cite{berry1994estimating} provides an earlier and classical inversion result useful for representing any stochastic choice on a given choice set. \cite{athey2007discrete} investigate how a rich specification of the unobserved components is needed to represent any stochastic choice function.

%Another paper closely related to ours is \citet{norets2013surjectivity}. They study whether ARUMs can represent any stochastic choice (i.e., market shares). However, their analysis focuses on a fixed choice set while our paper studies approximation across various choice sets. 

A recent paper by \cite{lu2021pure} is also related to our work, but it is conceptually closer to \cite{mcfadden2000mixed} than to our paper. \cite{lu2021pure} differ from our analysis in three key respects. First, they work in a continuous characteristics space, as in \cite{mcfadden2000mixed}, rather than on a finite set of alternatives with fixed characteristics. Second, they restrict the approximation target to pure-characteristics models (continuous random utility), whereas we allow for arbitrary finite-domain random utility models (probability measures over strict rankings). Third, they focus on parametric mixed-logit specifications whose systematic utility is a polynomial of degree at most a given bound, while we consider a general class of random-coefficient ARUMs. They provide a necessary and sufficient condition under which a given pure-characteristics model  can be approximated by a mixed-logit model  with a specified polynomial degree across choice sets. Thus, although both papers address  related approximation questions, they operate in 
different domains and yield conceptually distinct characterizations. A more detailed comparison 
between the two frameworks is provided in Section \ref{sec:related} of the online appendix.\footnote{In the section, we discuss \cite{mcfadden2000mixed}, \citet{norets2013surjectivity}, and \cite{lu2021pure} in more detail.}

%At a conceptual level, this paper is related to nonparametric discrete choice models  \cite{compiani2019market, tebaldi2019nonparametric}, which attempt to minimize specification and approximation errors. In contrast to the nonparametric models, our model is parametric yet flexible enough to represent the nonparametric data-generating process. This feature shares some of its spirit with \cite{athey2007discrete}, who investigate what rich parametric specification of the unobserved utility components is needed to rationalize arbitrary choice patterns.  

Our paper builds on the decision theory literature, where logit models and random utility models have been analyzed extensively since \cite{Luce59} and \cite{block1960random}. For more recent generalizations of logit models, see \cite{cerreia2023multinomial}, \cite{cerreia2021canon}, and \cite{horan2021stochastic}. Recent studies by \cite{ApesteguiaBallester18} and \cite{FrickIijimaStrzalecki19} further highlight the distinctions in choice behavior between random utility models and logit models. Furthermore, several recent studies examine the substitution property in discrete choice analysis. In \cite{horan_substitution}, the substitution patterns captured by random utility models are discussed, while \cite{allen2020hicksian} analyze aggregate complementarity in latent utility models applied to discrete choice. Additionally, \cite{chambers2021weighted} introduce a new model of discrete choice, demonstrating that their model allows for flexible substitution patterns and variations in market shares across different choice sets.

%Our paper is related with papers in the decision theory literature, where the logit models and the random utility models have been analyzed extensively ever since \cite{Luce59} and \cite{block1960random}. See \cite{cerreia2023multinomial}, \cite{cerreia2021canon}, and \cite{horan2021stochastic} for more recent generalization of logit models. Recent studies, such as those by \cite{ApesteguiaBallester18} and   \cite{FrickIijimaStrzalecki19}, highlight the distinctions in choice behavior between random utility models and logit models.

%Furthermore, a few recent studies examine the substitution property in discrete choice analysis. \cite{horan_substitution} discusses the substitution patterns captured by random utility models. \cite{allen2020hicksian} analyze aggregate complementarity in latent utility models used in discrete choice.  \cite{chambers2021weighted} introduced a new mode of discrete choice. They show that their model  allows for flexible substitution patterns and changes in market shares across choice sets.
 
Our analysis also shares some of its spirit with the growing literature that identifies and estimates flexible discrete choice/demand models under minimal assumptions. See, for example, \cite{berry2014identification}, \cite{compiani2022market}, and  \cite{tebaldi2019nonparametric}. Despite the similarity in the spirit, our problem is different from the standard  econometric problems. We are not concerned with statistical estimation or identification problems, i.e., recovering model parameters in either a sample or a population setting. In contrast, our primary goal is to explore the limitations, in terms of the approximation power, of common modeling strategies within the discrete choice literature. Our work focuses on a specification or approximation question rather than identification, estimation, or inference. 

Applications of discrete choice models span a wide range of fields, including economics, transportation, management, and marketing. \cite{berbeglia2022comparative} and \cite{feng2022consumer} review applications in the context of management science and operations research.\footnote{Applications in management science include but are not limited to \cite{kamakura1996modeling,chintagunta2005beyond,fiebig2010generalized,musalem2010structural,farias2013nonparametric,alptekinouglu2016exponomial,jagabathula2022nonparametric,akchen2025consider}.}

\vspace{-0.2cm}

\section{Model}\label{sec:math}

\vspace{-0.2cm}

We first introduce the setup. 

\noindent \textbf{Set of alternatives:} The set of all alternatives is denoted by $J$.  $J$ is assumed to be finite and $|J|> 2$.

\noindent\textbf{Choice sets:} Let $\D\subset  2^J\setminus \{\emptyset\}$ be the set of choice sets. $\D$ can be a proper subset of $2^J\setminus \{\emptyset\}$. Unless otherwise noted, we assume throughout the paper that $\D$ contains all binary and ternary choice sets: $\{j, l\} \in \D$ and $\{j,l,r\} \in \D$ for any distinct $j, l ,r\in J$. 
The set $\D$ may contain both observed choice sets as well as hypothetical choice sets the researcher is interested in. For example, even when the researcher observes consumers' choices only over $\{\text{train},\text{bus},\text{car}\}$, he may also be interested in choices over $\{\text{train},\text{bus}\}$, $\{\text{train},\text{car}\}$, and $\{\text{bus},\text{car}\}$ to learn about the consumers' substitution pattern.

%In a part of the paper (i.e., Section \ref{sec:prop}), however, we drop this assumption and assume that $\D=\{J\}$ when we consider the case in which the researcher's purpose is fitting a model to  the observed choice probabilities from the single choice set.\footnote{It is possible to generalize our analysis to an intermediate case where ${\cal D}$ neither equals $\{J\}$ nor includes all binary and ternary choice sets. For simplicity and due to space constraints, this paper focuses on the two extreme cases.}

\noindent\textbf{Explanatory variables:}  An alternative $j \in J$ is described by a real vector $x_j \in \Re^K$ of explanatory variables, where $K$ is the number of  explanatory variables. For instance, if an alternative $j$ is a consumption good, the alternative may be described by its price $p_j$ and its quality index $q_j$; in that case $x_j=(p_j, q_j)$.   Moreover, the researcher can include functions of original characteristics in $x_j$ such as higher order polynomials as well as splines or wavelets \citep{chen2007large}.  For example, with the original characteristics $(p_j,q_j)$ of alternative $j$, the researcher may include higher order polynomials such as  $p_j^2,q_j^2, p_jq_j$ in the characteristic vector $x_j$ and can make the number $K$ of characteristics larger. If the researcher includes all second-order terms, then $x_j=(p_j,q_j,p_j^2,q_j^2,p_jq_j)$ and $K=5$. 

%For each $j \in J$ and $k \in \{1,\dots, K\}$, we write $x_j(k)$ to denote the $k$-th element of $x_j$. 

\noindent\textbf{Stochastic choice function:}  A function $\rho: \D \times J \to  [0,1]$ is called a \df{stochastic choice function} if $\sum_{j \in D}\rho(D,j)=1$ and $\rho(D,j)=0$ for any $D$ and $j \not \in D$. The set of stochastic choice functions is denoted by $\P$. For each $(D,j) \in \D \times J$, the number $\rho(D,j)$ is interpreted as the probability that an alternative $j$ is chosen from a choice set $D$. In market-level analysis, $\rho(D,j)$ can also be interpreted as the market share of product $j$ in a market in which the set of available products is $D$. %In such cases, we interpret the stochastic choice function $\rho$ as aggregate choice probabilities across individuals. 

\noindent\textbf{Rankings:} Let $\Pi$ be the set of bijections between $J$ and $\{1,\dots, |J|\}$, where $|J|$ is the number of elements of $J$. For any element $\pi \in \Pi$, if $\pi(j)=i$, then we interpret alternative $j$ to be the $|J|+1-i$-th best element of $J$ with respect to $\pi$. If $\pi(j)>\pi(l)$, then $j$ is better than $l$ with respect to $\pi$. An element $\pi$ of $\Pi$ is called a {\it ranking} over $J$. A ranking describes an agent's strict preference relation.  There are $|J|!$ elements in $\Pi$. 

In the main body of the paper, we consider all possible rankings for simplicity. In Section \ref{sec:res_ranking} of the online appendix, we show how our analysis can be extended to accommodate cases where only a subset of rankings is considered, particularly when certain rankings are deemed unreasonable.

\noindent\textbf{Fixed effects:} We define fixed effects as a function $\eta: \D \times J\to \R$. We interpret $\eta_{(D,j)}$ as the value of the fixed effect of alternative $j$ in set $D$. Thus,  $\eta_{(D,j)}=0$ if $j \not \in D$ for all $D \in \D$.  We denote the set of fixed effects by $\F$.  We say that a fixed effect $\eta$ is {\it menu-independent}, if $\eta_{(D,j)}= \eta_{(E, j)}$ for all $D, E \in \D$ and $j \in D \cap E$. For simplicity, we assume fixed effects are \textit{menu-independent} for the main results. In that case, we simply write $\eta_{(D,j)}$ as $\eta_j$ and the set $\F$ of fixed effects can be identified as $\Re^{|J|}$.  With \textit{menu-dependent} fixed effects, our main results continue to hold under an additional consistency condition. See Section \ref{sec:menu-dependent} for details.

%\vspace{-0.2cm}
%\subsection{Models}\label{subsec:models}
%\vspace{-0.2cm}

We now introduce the definition of random utility models.
 We denote the set of probability measures over $\Pi$ by $\Delta(\Pi)$. Since $\Pi$ is finite, $\Delta(\Pi) = \big\{(\nu_1,\dots, \nu_{|\Pi|}) \in \Re_+^{|\Pi|}\big| \sum_{i=1}^{|\Pi|} \nu_i =1\big\}$, where $\Re_+$ is the set of nonnegative real numbers. 

\begin{definition}\label{def:ruf}
\normalfont A stochastic choice function $\rho$ is called a \df{random utility model} if there exists a probability measure $\nu \in \Delta(\Pi)$ such that for all $(D,j)\in \D \times J$, if $j \in D$, then $\rho(D,j)= \nu\left(\{\pi \in \Pi |\pi(j)> \pi(l) \text{ for any } l \in D \setminus \{j\} \}\right)$. The set of random utility models is denoted by $\P_r$.\footnote{While the function above is often called a random ranking function, a random utility model is often defined differently by using the existence of a probability measure $\mu$ over utilities such that for all $(D,j) \in \D \times J$, if $j \in D$, then $\rho(D,j)=\mu(\{u\in \Re^{J}| u(j)\ge u(i) \text{ for all } i \in D \setminus \{j\}\})$. \cite{block1960random}'s Theorem 3.1 proves that the two definitions are equivalent.} 
\end{definition}

%Notice that when $\D=\{J\}$,  the restriction of random utility is vacuous: any stochastic choice function is a random utility model (i.e., $\P_r=\P$).\footnote{To see this, observe that $\P_r \subset \P$ by definition. We show the converse.  For any $j \in J$, let $\pi_j \in \Pi$ such that $\pi_j(j) > \pi_j(l)$ for all $l \in D\setminus \{j\}$. Then $\rho^{\pi_j}(l)=1_j(l)$ for any $l \in J$, where $1_j(l)=1$ if $l=j$ and $1_j(l)=0$ if $l\neq j$. (For the definition of $\rho^{\pi}$, see definition (\ref{df:rho_u}) in Section \ref{sec:ident}.) For any $\rho \in \P$, define $\rho'=\sum_{j \in J}\rho(j)\rho^{\pi_j}$. Then, $\rho' \in \P_r$ and $\rho'(j)=\rho(j)$ for any $j \in J$, as desired. Hence, $\P\subset \P_r$. In general, we have $\P_r\subsetneq \P$ and the random utility models have some testable implication. For example, when $\D=2^J\setminus \emptyset$, random utility models are characterized by the non-negativity of the Block-Marschak polynomials.}
%In certain scenarios, researchers might want to exclude rankings deemed unreasonable and restrict the set of rankings. We consider such a case in Section \ref{sec:res_ranking} in the appendix.  

In both theoretical and empirical literature, researchers approximate the random utility models by using random-coefficient models. First, we introduce one definition.

\begin{definition}\normalfont A Borel probability measure $\mu$ on the Borel $\sigma$-algebra of $\mathbf{R}^{|J|}$  is said to be  a \df{standard probability measure} if $\mu$ is absolutely continuous with respect to the Lebesgue measure and the support is convex.\footnote{Remember that the support $\supp \mu$ is defined as $\{\ep \in \Re^{|J|}|\mu(N_{\ep})>0  \text{ for any open neighborhood } N_{\ep}\text{ of }\ep \}$.} Let $\M$ be the set of all standard probability  measures.
\end{definition}

Although this definition is specific to our paper, it is general enough to nest many additive utility–shock distributions common in the literature, such as normal, logistic, and exponential shock distributions.

In the following, we denote the inner product of two vectors $x$ and $y$ by $x \cdot y$.

%\textcolor{red}{HG8: I like the notation $\mathbb{R}^J$ here.}
\begin{definition}\label{def:mll2}
\normalfont Let $\eta \in\F$ be a vector of fixed effects.  A stochastic choice function $\rho$ is called a \df{ random-coefficient additive random  utility model (random-coefficient  ARUM) with fixed effects $\eta$} if there exist a standard probability measure $\mu$ and a Borel probability measure $m$ such that for all $(D,j) \in \D \times J$, if $j \in D$, then 
\begin{equation}\label{eq:mll2}
\rho(D,j)= \int\mu\Bigl(\left\{\ep| \beta\cdot x_j+\eta_{j}+\ep_j > \beta\cdot x_l+\eta_{l}+\ep_l, \forall l \in D\setminus \left\{j\right\}\right\}\Bigr)d m(\beta).
\end{equation}
The random vector $(\ep_j )_{j\in J}\in \Re^{|J|}$ follows the distribution $\mu$. 
When the support of $m$ has only one point, the stochastic choice function $\rho$ is called an \df{additive random  utility model (ARUM) with fixed effects $\eta$}: for all $(D,j) \in \D \times J$, if $j \in D$, then 
\begin{equation}\label{eq:mll0}
\rho\left(D,j\right)= \mu\Bigl(\left\{\ep| \beta\cdot x_j+\eta_{j}+\ep_j > \beta\cdot x_l+\eta_{l}+\ep_l, \forall l \in D\setminus \left\{j\right\}\right\}\Bigr).
\end{equation}
\end{definition}

The set of random-coefficient ARUMs is denoted by $\P_{ra}(\eta|\mu)$ and the set of ARUMs by $\P_{a}(\eta|\mu)$.  When the context makes clear which standard probability measure $\mu$ we consider,  we do not specify the standard probability measure $\mu$.

The term $\beta\cdot x_j$ is the {\it systematic (deterministic)} part of the utility of alternative $j$ captured by the observed characteristics $x_j$. The vector $\beta$ captures preferences of an agent and the distribution $m$ over coefficients $\beta$ describes the heterogeneity of preferences among the population of  agents. The constant $\eta_{j}$ is called a {\it fixed effect} that  captures the utility of alternative $j$  from the unobserved characteristics;  $\ep_j$ is the preference  shock to the utility of alternative $j$. 

%In practice, a researcher first fixes one probability measure $\mu$ over the shock $\ep$; then the researcher estimates the distribution $m$ and the fixed effects $\eta$ after observing a dataset.

Almost all probability measures used in practice are standard. For a mixed-logit model, $\mu$ is an iid extreme-value type-I distribution; for a probit model, $\mu$ is the multivariate standard normal distribution; for an exponomial model, $\mu$ is an iid multivariate exponential distribution with scale parameter 1. In most empirical applications of these models, the mixing distribution $m$ is a parametric distribution like a multivariate normal distribution. In our case, the mixing distributions of the random coefficients do not come from a particular parametric family.  

For later exposition, we define the {\it mixed-logit models}  formally as follows:

\begin{definition}\label{def:mll} \normalfont Let $\eta \in\F$ be a vector of fixed effects. A stochastic choice function $\rho$ is called a \df{ mixed-logit model  with  fixed effects $\eta$} if there exists a Borel probability measure $m$ such that for all $(D,j) \in \D \times J$, if $j \in D$, then 
\begin{equation}\label{eq:mll}
\rho(D,j)= \int \dfrac{\exp({\beta \cdot x_j+\eta_{j}})}{\sum_{l \in D}\exp({\beta \cdot x_l+\eta_{l}}) } dm(\beta).
\end{equation} The set of  mixed-logit models with fixed effects $\eta$ is denoted by $\P_{ml}(\eta)$. When $m$ puts the unit mass on a particular $\beta$ in (\ref{eq:mll}), then $\rho$ is called a \df{logit model.} 
\end{definition}

We give two remarks on the models. First, in the definitions above, we write fixed effects as $\eta_j$ assuming menu-independence for simplicity (i.e., $\F=\R^{|J|}$). However, our results hold more generally with menu-dependent fixed effects. See Section \ref{sec:menu-dependent} for details. 
% I think it is best for us to keep the content as short as possible here: the later section provides details. 
Second, in the models above, following \cite{mcfadden2000mixed} as well as many other papers in discrete choice analysis, we write the systematic part of utility by $\beta\cdot x_j$. Although we have a linear structure in the explanatory variables $x_j$, it is important to remember that $x_j$ can be a function of original characteristics such as higher order polynomials of original characteristics as used in the proof of \cite{mcfadden2000mixed}. Since any continuous function can be approximated by  polynomials, our models are general enough to allow for flexible systematic parts of utility functions.

%The mixed-logit model has been known for a long time, but has become popular relatively recently since the development of simulation method. This is because calculating the integration used to be difficult. Corollary \ref{rem:MT} and Theorem \ref{theo:1} state that focusing on a finite mixture of logit models entails no loss of generality. Hence, the calculation of the integration is not necessary.\footnote{The nested logit model also can be seen as a finite mixture of the logit model when the nests do not overlap. \cite{gul2014random} axiomatize a model called {\it the complete attribute rule}, which is similar to the nested logit model. Neither the complete attribute rule nor the mixed-logit model is more general than the other.  The intersection between the two models is the (degenerate) logit model. See appendix \ref{sec:gul2014} for details.}

Finally, we review essential mathematical concepts. A {\it polytope} is a convex hull of finitely many points. The closure of a set $C$ is denoted by $\cl C$ with respect to the standard finite dimensional Euclidean topology. The {\it affine hull} of a set $C$ is the smallest affine set that contains $C$, and it is denoted by $\aff C$. The convex hull of a set $C$ is denoted by $\co C$. The {\it relative interior} of a convex set $C$ is the interior of $C$ in the relative topology with respect to $\aff C$. The relative interior of $C$ is denoted by $\rint C$. We recall a basic concept in geometry: a set $\{x_j \in \mathbf{R}^K| j\in J\}$ is {\it affinely independent} if no $x_j$ can be written as an affine combination of the other elements $\{x_l\}_{l \neq j}$. Formally, for any $j \in J$, there exist no real numbers $\{\al_l\}_{l \in J\setminus \{j\}}$ such that $x_j =\sum_{l \in J \setminus \{j\}} \al_l x_l$ and $\sum_{l  \in J\setminus \{j\}} \al_l=1$.

%If $C$ is not empty, then (i) $\rint C$ is not empty,  and (ii) $\rint C= \{x \in C| \text{for all }y \in C\text{ there exists }\al\in \Re \text{ such that }\al >1 \text{ and }\al x+(1-\al)y \in C\}$. (See Theorem 6.4 in \cite{rockafellar2015} for the proof.) 

\vspace{-0.2cm}

\section{Main Result}\label{sec:main}

\vspace{-0.2cm}

%\textcolor{red}{We state our main results in this section. Section \ref{sec:ident} discuses a key ingredient of the proofs: the identification of deterministic preferences and their mixtures that cannot be approximated well. We present some generalizations of our results in Section \ref{sec:prop} and Section \ref{section:menu-independent_fixed}.}

%The main result of the paper uses a basic concept in geometry: a set $\{x_j \in \mathbf{R}^K| j\in J\}$ is {\it affinely independent} if no $x_j$ can be written as an affine combination of the other elements $\{x_l\}_{l \neq j}$. Formally, for any $j \in J$, there exists no real numbers $\{\al_l\}_{l \in J\setminus \{j\}}$ such that $x_j =\sum_{l \in J \setminus \{j\}} \al_l x_l$ and $\sum_{l  \in J\setminus \{j\}} \al_l=1$.
%\footnote{\textcolor{red}{HG8 Remove per AE's request} It is easy to see that a set $\{x_j \in \mathbf{R}^K| j\in J\}$ is affinely independent if and only if $\{x_l-x_j\}_{l \in J \setminus \{j\}}$ is linearly independent for any $j$. }

%\footnote{This definition is similar to the concept of {\it shattering} in machine learning. Notice, however, that the above definition requires that all points be ordered according to a given ranking $\pi$, while in the standard setup, {\it shattering} only requires that points be separated into two groups. We are grateful to Prof. Brendan Beare, who informed us about the concept of shattering. Hence, the definition above is stronger than the concept of {\it shattering} when $J$ contains more than two elements.}  

\begin{theorem}\label{theo:1}
\bit
\item[(i)]  Let $\mu \in \M$ be any standard probability measure.  If the set  $\{x_j \in \Re^K| j\in J\}$ is affinely independent, then any random utility model can be approximated arbitrarily well by a sequence of random-coefficient ARUMs with the given $\mu$; moreover, the approximation can be done  without fixed effects (i.e., $\eta=0$). Formally, 
\begin{equation*}\label{eq:theo11}
 \forall \mu \in \M, \forall \rho \in \P_r, \exists \{\rho_n\}_{n=1}^{\infty} \subset \P_{ra}(0|\mu), \forall D \in \D, \forall j\in D,   \lim_{n\to \infty}\rho_n(D,j)= \rho(D,j).
\end{equation*}
\item[(ii)] If the set  $\{x_j \in \Re^K| j\in J\}$ is not  affinely independent, then there exists a random utility model $\rho$ such that for every standard probability measure $\mu$, $\rho$ cannot be approximated arbitrarily well by any sequence of random-coefficient ARUMs with fixed effects and with the distribution $\mu$.  Formally, 
\[
\exists \rho \in \P_r,\,  \forall \mu \in \M, \rho \not \in \cl \bigcup_{\eta \in \F} \P_{ra}(\eta|\mu).
\]
\eit
\end{theorem}

We provide the proof in the appendix. %\textcolor{red}{The proofs of the if direction (Theorem 1(i)) are intuitive. The only-if direction (Theorem 1(ii)) requires more work. It involves identifying (i) deterministic preference rankings that cannot be well approximated without fixed effects, and (ii) random utility models that that cannot be well approximated even with fixed effects. For a detailed discussion, see Propositions \ref{prop:unrep} and \ref{prop:unrep2} in Section \ref{sec:ident}. }
To interpret the main theorem, consider a typical practice where the shock distribution $\mu$ over $\ep$ is fixed a priori by a researcher; then he or she estimates the distribution $m$ over coefficients $\beta$  as well as the fixed effects $\eta$ after seeing a dataset $\rho$ generated from a random utility model.

Statement (i) shows that if the set  $\{x_j \in \Re^K| j\in J\}$ is affinely independent, then the researcher should be able to approximate the given dataset $\rho$ arbitrarily well across choice sets $D \in \D$ by choosing an appropriate  distribution $m$ over coefficients $\beta$. 
This direction builds upon the classical result of \cite{mcfadden2000mixed} about mixed-logit. Note that the result allows a general class of utility-shock distributions, not confined to the type-I extreme-value distribution or a particular choice of mixing distributions.

%\textcolor{orange}{This direction can be seen as an extension of  straightforward given the geometric intuition we shall discuss in Section \ref{sec:proof_sketch} .} 

Statement (ii) shows that if the affine-independence condition fails, then there exists a random utility model that cannot be approximated arbitrarily well by any sequence of  random-coefficient ARUMs, no matter how the researcher changes the distribution $m$ over  coefficients $\beta$ as well as the fixed effects $\eta$, given any arbitrarily chosen standard probability measure $\mu$ over $\ep$.   This direction is novel and has not been studied in \cite{mcfadden2000mixed}. Again, this result holds for the broad class of random-coefficient models. For example, the approximation is impossible using any  mixed-logit model or any  random-coefficient probit model. 
In Propositions \ref{prop:unrep} and \ref{prop:unrep2} in Section \ref{sec:ident}, we  give examples of the random utility models that cannot be approximated arbitrarily well. 

The affine-independence condition can be simplified further to a generically equivalent condition. To see this, remember the following basic facts: (i) if $|J|>K+1$, then $\{x_j \in \Re^K| j \in J\}$ is not affinely independent; (ii) if $|J|\le K+1$, then the set is generically affinely independent.\footnote{This is a standard concept of genericity in the literature of discrete geometry.  Even if the set is not affinely independent, as long as $|J|\le K+1$, for any $\varepsilon>0$,  there exists an affinely independent  set $X'$, obtained from $X$ by moving each point by a distance of at most $\varepsilon$ (see Section 3 of \cite{matousek2013lectures}).} Given these observations, Theorem \ref{theo:1} implies the following corollary.\footnote{One caveat of the result is that even though the generic condition holds, the original condition of the affine independence may not hold when explanatory variables  include zeros  and ones. In that case, one should check the affine-independence  of $\{x_j \in \Re^K|j \in J\}$, rather than the generic condition.}

\begin{corollary}\label{rem:MT} Let $K$ be the number of explanatory variables and $|J|$ be the number of alternatives.
\bit
\item[(i)] If $K \ge |J|-1$, then the statements in Theorem \ref{theo:1} (i) hold generically.
\item[(ii)] If $K <|J|-1$, then the statements in Theorem \ref{theo:1} (ii) hold.
\eit
\end{corollary}

We now mention a few remarks on the results. First, as mentioned earlier, to increase the number $K$ of explanatory variables, researchers may include additional terms such as higher order polynomials (\cite{mcfadden2000mixed}) as well as splines or wavelets \citep{chen2007large}.  In their proof, \cite{mcfadden2000mixed} use higher order polynomials of arbitrarily high degrees to approximate continuous random utility models. In particular, in their construction, $x_j$ is a vector of monomials of any degree of original characteristics $(y_{j1},\cdots, y_{jn})$:
\[
x_j=(\underbrace{y_{j1},\cdots, y_{jn}}_{\text{1st order terms}}, \underbrace{y_{j1}^2,\cdots, y_{jn}^2}_{\text{2nd order terms including interactions}},y_{j1}^3,\cdots, y_{jn}^3,.... )\in \Re^{K},
\]
where $K\to \infty$. Their result is thus consistent with the sufficiency part of our result: we proved that $K \ge |J|-1$ is sufficient in our setup.\footnote{It should be noted that our results are not suggesting the imprudent use of a large number of characteristics in practice. In empirical applications for predictions, one must account for the statistical trade-off between bias and variance when comparing models with different numbers of characteristics. Some model selection procedures (e.g., AIC, BIC, or cross-validation) or penalization methods may be required.}

Second, in the theorem, we consider all possible random utility models (i.e., probability distributions over all rankings) as the prediction target mainly for simplicity. As mentioned after Definition 1, in some cases, the researchers may want to restrict the set of rankings  by excluding those deemed unreasonable.  In Section \ref{sec:res_ranking} of the online appendix, we provide a necessary and sufficient condition for the approximation of the restricted random utility model. 

Third, many empirical papers use the mixed-logit models that are linear in original characteristics and do not contain additional terms such as polynomials. The condition that $K \ge |J|-1$ is frequently violated in various contexts, which in turn results in a violation of the affine-independence condition. Remember that $|J|$ is the number of  alternatives and $K$ is the number of characteristics. There are many choice situations in which $|J|$ is very large such as choices of groceries, hospitals, cars, schools, and restaurants.  In such a dataset, the condition is likely to be violated. This means that the class of the models may  not be rich enough to approximate the true substitution pattern across choice sets,  no matter how one  chooses parameters and fixed effects.\footnote{By substitution patterns, in general, we mean how choice probabilities change in different choice sets. In Section \ref{sec:substitution}, we provide a more specific definition of substitution patterns.}  Theorem \ref{theo:1} considers the case of approximation on a rich collection of choice sets. Less stringent conditions are required for approximation if one considers the case of a single choice set. See Section \ref{sec:prop} below.

\vspace{-0.2cm}

\subsection{Unrepresentable Rankings and Approximation Failure}\label{sec:ident}

\vspace{-0.2cm}

Theorem 1 shows that when the affine-independence condition fails, there exist random utility models that cannot be approximated arbitrarily well by random-coefficient ARUMs. A key step toward this result is to identify deterministic rankings that ARUMs without fixed effects may fail to approximate well.
 In what follows, we provide a method for identifying such rankings. The following definition is central.

\begin{definition}\label{def:rep} 
\normalfont A ranking $\pi \in \Pi$ is {\it representable in choice sets $\D$} if there exists a real vector $\beta$ such that, for all $D \in {\cal D}$ and $j \in D$, 
\begin{equation}\label{eq:rep}
\pi (j) > \pi (l), \forall l \in D\setminus \{j\} \text{ if and only if  } \beta \cdot x_j > \beta \cdot x_l, \forall l \in D\setminus \{j\}.
\end{equation}
If $\pi$ is not representable in $\D$, we say that $\pi$ is {\it unrepresentable in $\D$}.
\end{definition}

To determine the representability of a specific ranking $\pi$, one can employ linear programming techniques.\footnote{The ranking $\pi(1)>\pi(2)>...>\pi(J)$ is representable if and only if the linear programming problem $\max_{\beta\in \mathbf{R}^K,t\in \mathbf{R} } t$ subject to $\beta\cdot (x_j-x_{j+1})\geq t$ for each $j=1,...,|J|-1$ is unbounded. If the ranking is unrepresentable, the problem has the optimal value 0.} In Section \ref{sec:data}, utilizing an empirical  dataset, we identify such unrepresentable rankings, as detailed in Table 1 of that section. The identification of these rankings enables researchers to  evaluate which substitution patterns are challenging to capture within their models.

 Notice that the requirement (\ref{eq:rep}) depends both on the specification of the characteristic vectors $\{x_j\}_{j \in J}$ as well as the set $\D$ of choice sets.
 %The requirement (\ref{eq:rep}) becomes less restrictive as the characteristic vector $x_j$ becomes longer because it becomes easier to find the desired $\beta$ with additional characteristic variables; the requirement (\ref{eq:rep}) becomes more restrictive as the set $\D$ of choice sets becomes richer, simply because the number of inequalities to be satisfied becomes larger.
 As mentioned earlier, we usually consider the general choice sets $\D$ with binary and ternary choice sets, while in some places, we assume a simpler case in which $\D =\{J\}$. In the rest of this section, we consider the general $\D$.  When there is no risk of confusion, we will simply say that a ranking $\pi \in \Pi$ is representable without specifying choice sets $\mathcal{D}$.

The {\it reverse} of a ranking $\pi$ plays an important role:

\begin{definition}\label{def:rev} 
\normalfont For any ranking $\pi \in \Pi$, define $\pi^- \in  \Pi$ such that $\pi(j)>\pi(l)$ if and only if  $\pi^-(l)>\pi^-(j)$ for any $j, l\in J$.  The ranking $\pi^-$  is called the {\it reverse} ranking of $\pi$. 
\end{definition}

Note that if a ranking $\pi$ is representable, then $\pi^{-}$ is also representable.

The proposition below demonstrates that the failure of the affine-independence condition leads to the existence of unrepresentable rankings. To show the proposition, for each ranking $\pi \in \Pi$, define   
\begin{eqnarray}\label{df:rho_u}
\rho^{\pi}(D,j)=1\{\pi (j)> \pi(l) \text{ for all } l \in D \setminus \{j\}\}.
\end{eqnarray}
 The function $\rho^{\pi}$ gives probability one to the best alternative $j$ in a choice set $D$ according to the strict ranking $\pi$. 
 
\begin{proposition}\label{prop:unrep}   
The set $\{x_j \in \Re^K|j \in J\}$ is not affinely independent if and only if there exists a ranking $\pi \in \Pi$ that is not representable in $\D$.  For every unrepresentable ranking $\pi$ and any standard probability measure $\mu$, there exists $\sigma \in \{\pi, \pi^-\}$ and  a neighborhood $U$ of  $\rho^{\sigma}$ such that any random utility model in $U$ cannot be approximated arbitrarily well by any sequence of  random-coefficient ARUMs without fixed effects and with the same shock distribution $\mu$.
\end{proposition}

\begin{remark}\label{rem:mu_approx}
{\it A careful reader may wonder why Proposition 1 considers both an unrepresentable ranking $\pi$ and its reverse ranking $\pi^-$. The reason is that, for a suitably chosen shock distribution, an ARUM without fixed effects can approximate $\rho^\pi$ arbitrarily well even when $\pi$ itself is not representable by the characteristic vectors. To see this, suppose that the shock distribution $\mu$ satisfies, for every pair $j,l\in J$, $\epsilon_j>\epsilon_l$ a.s. if and only if $\pi(j)>\pi(l)$. Fix any coefficient vector $\beta\in \R^K$, and consider the sequence of ARUMs with coefficient vector $\beta/n$ and shock vector $(\epsilon_j)_{j\in J}$. As $n\to\infty$, the systematic utility term $(\beta/n)\cdot x_j$ vanishes for each alternative $j$. Hence the induced choice probabilities converge to those generated solely by the shock ranking. Therefore, $\rho_n \to \rho^\pi$ even if $\pi$ is not representable by the characteristic vectors. This observation explains why the conclusion must be formulated as a joint statement involving $\rho^\pi$ and $\rho^{\pi^-}$, rather than as a statement about $\rho^\pi$ alone. Such a construction is ruled out under either of the following additional assumptions: (a) the one-dimensional marginal distributions of the shocks are identical across alternatives; or (b) the zero vector belongs to the interior of $\operatorname{supp}\mu$, as in Proposition \ref{theo:2}(ii)(a). Under either assumption, if $\pi$ is not representable, then $\rho^{\pi}$ cannot be approximated arbitrarily well.}
\end{remark}

Remember that we have assumed no fixed effects (i.e., $\eta=0$) in this proposition. In the next proposition, we consider the case with fixed effects. The next proposition shows that when the affine-independence condition fails, approximating a mixture of $\rho^{\pi}$ and $\rho^{\pi^-}$ is impossible even using fixed effects when $\pi$ is not representable.

\begin{proposition}\label{prop:unrep2}  Suppose that $\{x_j \in \Re^K|j \in J\}$ is not affinely independent. Then there exists an unrepresentable ranking $\pi$.  For any unrepresentable ranking $\pi \in \Pi$,  any standard probability measure $\mu$,  and  any $\al \in (0,1)$, there exists a neighborhood $U$ of $\al\rho^{\pi}+(1-\al)\rho^{\pi^-}$ such that any random utility model that belongs to $U$ cannot be approximated arbitrarily well by any sequence of random-coefficient ARUMs with fixed effects and  with the same shock distribution $\mu$.
\end{proposition}

\vspace{-0.2cm}

\subsection{Menu-dependent Fixed Effects}\label{sec:menu-dependent}

\vspace{-0.2cm}

In the sections above, we assume menu-independent fixed effects for simplicity.  All results  hold with menu-dependent fixed effects with the following condition. For all $(D,j) \in \D \times J$ such that $j \in D$, remember that $\eta_{(D,j)} \in \Re$ denotes a fixed effect of alternative $j$ in a choice set $D$.  We allow $\eta_{(D,j)}$ to be different across $D$ but  impose a local restriction:
there are three alternatives $j,l,r \in J$ such that for all $\eta \in {\cal F}$,  $\eta_{(D,j)}$, $\eta_{(D,l)}$, and $\eta_{(D,r)}$ do not depend on $D$ if $D \subseteq \{j,l,r\}$. Formally, there exist $j,l,r \in J$ such that for all $\eta \in {\cal F}$,
\begin{equation}\label{def:Fixed}
 \eta_{(D,s)}=\eta_{(E,s)} \text{ for all } D,E \subseteq \{j,l,r\} \text{ and all }s \in D \cap E.
\end{equation}
Note that the restriction only requires that there exist three alternatives for which \eqref{def:Fixed} holds; it does \emph{not} require \eqref{def:Fixed} to hold for all triples of alternatives. This restriction is a minimal restriction that allows us to study substitution patterns within the triple \(\{j,l,r\}\)  separately from the menu effects by comparing \(\{j,l,r\}\) to the binary sets \(\{j,l\},\{j,r\}\), and \(\{l,r\}\). 
Theorem 1, Proposition 1,  and Proposition 2 remain valid with menu-dependent fixed effects as long as restriction (\ref{def:Fixed}) holds, replacing $\eta_j$ with $\eta_{(D,j)}$ in the representations (\ref{eq:mll2}) and  (\ref{eq:mll0}) in Definition \ref{def:mll2}.\footnote{The proofs in the appendix only assume (\ref{def:Fixed}), not menu-independence.}

In Section~\ref{section:menu-independent_fixed} of the online appendix, we also analyze the case in which $\eta$ is fully menu-dependent without any restriction. Building on the surjectivity result of \cite{norets2013surjectivity}, we show that fully menu-dependent fixed effects are rich enough to approximate any choice behavior, including non-RUM behavior, even without random coefficients. Intuitively, the fixed effects can absorb all variation in choice probabilities. 

In practice, however, when the model is used for counterfactual prediction, the fixed effects for unobserved choice sets must be chosen a priori by the researcher---either by reusing fixed effects from some observed choice sets or by setting them to constants. In this case, our theorem implies that a random-coefficient ARUM with such a priori fixed effects cannot approximate arbitrary random utility models across choice sets unless our conditions are satisfied. For example, suppose we observe data only for the full menu $J$, but wish to make counterfactual predictions for smaller choice sets--such as all binary and ternary choice sets--to learn about substitution patterns. Our theorem implies that if the fixed effects for the counterfactual choice sets are chosen a priori---or set equal to the fixed effects from some observed choice set---then a random-coefficient ARUM cannot approximate arbitrary random utility models unless the affine-independence condition is satisfied. Thus, our main result remains directly relevant to counterfactual analysis.

\vspace{-0.2cm}

\subsection{Additional Result for Single Choice Set Case}\label{sec:prop}

\vspace{-0.2cm}

In the following, we provide a supplemental result for the case in which $\D=\{J\}$. Such a case corresponds to a situation in which the researcher  is interested only in the observed choice probabilities (i.e., market shares) on a single set $J$ (but not on its subsets).

\begin{proposition}\label{theo:2}  Assume that $\D=\{J\}$.  
\bit
\item[(i)]  Let $\mu \in \M$ be any standard probability measure. If the set $\{x_j \in \Re^K| j\in J\}$ is convex-independent (i.e., if $x_j \not \in \co\{x_l| l\in J\setminus \{j\}\}$ for any $j \in J$), then any random utility model can be approximated arbitrarily well by a sequence of random-coefficient ARUMs with the same shock distribution $\mu$; moreover, the approximation can be done  without fixed effects (i.e., $\eta=0$). That is, 
\begin{equation*}\label{prop11}
\forall \mu \in \M,\ \forall \rho \in \P_r,\ \exists \{\rho_n\}_{n=1}^{\infty} \subset  \P_{ra}(0|\mu),\   \forall j\in J, \lim_{n \to \infty}\rho_n(J,j)=\rho(J,j).
\end{equation*}
\item[(ii)] (a) If the set  $\{x_j\in \Re^K| j\in J\}$ is not convex-independent, then there exists a random utility model $\rho$ such that for every standard probability measure $\mu$ with $\textbf{0} \in\text{int} (\supp \mu)$, $\rho$ cannot be approximated arbitrarily well by any sequence of random-coefficient ARUMs without fixed effects (i.e., with $\eta=0$) and with the same shock distribution $\mu$. Formally, $\  \exists \rho \in \P_r,\, \forall \mu \in \M$ s.t. $\textbf{0} \in\text{int} (\supp \mu)$,\, $\rho \not \in \cl \P_{ra}(0|\mu)$. \\ 
(b) However, if fixed effects are used, any random utility model can be approximated arbitrarily well by a sequence of ARUMs  with any shock distribution $\mu \in \M$. Formally, $\forall \rho \in \P_r,\, \forall  \mu \in \M, \rho \in \cl \bigcup_{\eta \in \F} \P_{ra}(\eta|\mu)$.
\eit
\end{proposition}

Note that the  convex-independence condition is weaker than the affine-independence condition.  This makes sense  because the convex-independence condition guarantees the approximation only on the single choice set (i.e., $\{J\}$), while the affine-independence condition guarantees the approximation across all subsets $D \in \D$ of $J$ (including $J$ itself).

The implications of Theorem \ref{theo:1} and  Proposition \ref{theo:2} are similar. In both  cases, there exists a random utility model that cannot be approximated arbitrarily well {\it without} using fixed effects. However, there are two important differences. First, Proposition \ref{prop11}(ii)(a) restricts attention to standard probability measures with $\textbf{0} \in\text{int} (\supp \mu)$, whereas Theorem \ref{theo:1} allows arbitrary standard probability measures.\footnote{This assumption can be weakened to require only that $a\mathbf{1}\in \operatorname{int}(\operatorname{supp}\mu)$ for some $a\in\R$ because only the differences $\ep_j-\ep_l$ matter. For simplicity, we state the result under the normalization $a=0$. Another possible sufficient condition is that the marginal distribution of $\varepsilon_j$ is the same for every $j\in J$.} This restriction is necessary in the single-choice-set case because the shock distribution can otherwise be chosen to fit the target choice probabilities directly: for any given $\rho\in\P_r$, one can choose $\mu\in\M$ so that an ARUM without fixed effects matches $\rho$, by constructions of the kind described in Remark \ref{rem:mu_approx}.

Second, as stated in Proposition \ref{prop11} (ii)(b), if fixed effects are used, any random utility model can be approximated arbitrarily well on one particular choice set $J$. This result directly follows from \cite{norets2013surjectivity}. This is in contrast to Theorem \ref{theo:1} (ii), which claims that there exists a random utility model that cannot be approximated arbitrarily well even using fixed effects  across choice sets $\D$.\footnote{This difference originates from the fact that we  require approximation on $\D$ in Theorem \ref{theo:1} which contains binary and ternary choice sets, while in Proposition \ref{prop11}, we require approximation only on $J$.}

Unlike  affine-independence, the convex-independence does not restrict the number of elements in a convex-independent set.\footnote{\label{ft:convex_ind}For example, in three-dimensional space $(x,y,z)$, consider a circumference of radius one whose origin is $(0,0,1)$ on a hyperplane of $z=1$. The number of points on the circumference is  a continuum. However, the set of points on the circumference is convex-independent. 
%(If $z=0$, the set is not convex-independent. If the set is a square, the set is not convex-independent.)
} So there exists no counterpart of Corollary \ref{rem:MT}.

\vspace{-0.2cm}

\section{Measuring Approximation Errors}\label{sec:measuring}

\vspace{-0.2cm}

Propositions \ref{prop:unrep} and \ref{prop:unrep2} in Section \ref{sec:ident} show that the approximation errors to some random utility models may not be negligible when the affine-independence condition fails. In this section, we provide a way to quantify the approximation errors. We first define the distance function as follows: for any $\hat{\rho},\rho \in \P_r$, define $d(\hat{\rho},\rho) \equiv\sqrt{(\sum_{D\in \mathcal{D}} \sum_{j\in D} \left( \rho(D,j)-\hat{\rho}(D,j) \right)^2 )/|\mathcal{D}|}$.

 In our analysis, $\hat{\rho}$ is a given random utility model and  $\rho$  is a random-coefficient ARUM by which we approximate $\hat{\rho}$. We divide the norm by $\sqrt{|\D|}$ to make  the distance independent from the number of choice sets.\footnote{$d(\hat{\rho},\rho)$ can be written as $\|\rho-\hat{\rho}\|/\sqrt{|\mathcal{D}|}$, where $\|\cdot \|$ is the Euclidean norm (i.e., $l_2$ norm).  One can consider an alternative distance function based on $l_1$ norm as follows: $d_1(\hat{\rho},\rho)\equiv (\sum_{(j,D) \in J \times \D } |\hat{\rho}(D,j)-\rho(D,j)|)/|\D|$. Since $\sqrt{\sum_{(j,D)\in  J \times \D} (\hat{\rho}(D,j)-\rho(D,j))^2} \le \sum_{(j,D)\in  J \times \D} |\hat{\rho}(D,j)-\rho(D,j)|$, we can show that $d(\hat{\rho},\rho)/\sqrt{|\D|} \le  d_1(\hat{\rho},\rho)$ for any $\rho$ and  $\hat{\rho}$. So our approximation error divided by $\sqrt{|\D|}$ will provide a lower bound of an  alternative approximation error measured by $d_1$. We use our distance function $d$ rather than $d_1$ because the inner-product structure of $d$ is useful for constructing an algorithm.} Notice that the maximal distance is at most $\sqrt{2}$.

Given an approximation target $\hat{\rho} \in \P_r$ and a standard probability measure $\mu$, when researchers use  random-coefficient ARUMs with fixed effects $\eta$, the approximation error is defined as:
\begin{equation}\label{eq:dist.1}
\inf_{\rho \in \mathcal{P}_{ra}(\eta|\mu)} d(\hat{\rho},\rho).
\end{equation}
We call (\ref{eq:dist.1}) the {\it approximation error to $\hat{\rho}$} by  random-coefficient ARUMs with fixed effects $\eta$. 

Given $\hat{\rho}$, we propose two algorithms to solve (\ref{eq:dist.1}) and measure the approximation errors. The first is the standard EM (Expectation-Maximization) algorithm \citep{dempster1977maximum} to estimate the best possible finite-mixture logit model. However, the EM algorithm may not converge to the global optimum in our setting. To alleviate this concern, we propose a greedy algorithm inspired by \cite{barron2008approximation}. This algorithm solves a sequence of optimization problems to converge to the global optimal solution. We provide explanations of these algorithms in Section \ref{sec:alg} of the online appendix. In Section \ref{sec:data} below, we find that the approximation error can be large for some unrepresentable rankings in a real dataset. 

\vspace{-0.2cm}
\section{Application to Data}\label{sec:data}
\vspace{-0.2cm}
To quantify approximation errors with and without fixed effects, we use the DVD sales dataset introduced by \cite{rusmevichientong2010dynamic}.\footnote{In Section \ref{appendx:fish} of the online appendix, we measure approximation errors using an additional fishing-site choice dataset from \cite{thomson1991results}, illustrating that our methods apply more generally and are not specific to this particular dataset.} This dataset has been widely used in research on consumer choice modeling and assortment optimization by, for example, \cite{rusmevichientong2010dynamic,farias2013nonparametric} and \cite{abdallah2021demand}. We obtained the dataset from Table 2 of \cite{farias2009non} and replicate it in Appendix~\ref{appendix_dvd_dataset}. The dataset contains 15 DVD products. As noted by \cite{rusmevichientong2010dynamic}, the most predictive attributes for choices are the total number of helpful review votes and the price per disc. Motivated by this fact and to keep our computation manageable, we focus on these two characteristics and 4 DVDs to illustrate our theoretical results.\footnote{We select the four DVDs from the first four rows of Table 2 in \cite{farias2009non}, which correspond to the four most expensive alternatives.} Hence we have $|J|=4$ and $K=2$. In Appendix Section \ref{sec:J10K2}, we report the approximation errors without fixed effects for a larger number of alternatives with $|J|=10$ and $K=3$. The findings are similar.

Our empirical analysis concentrates primarily on mixed-logit models given their widespread use. We consider two specifications of mixed-logit models. The first is the linear mixed-logit models that are linear in the original characteristics. The second is the mixed-logit model defined with quadratic polynomials.\footnote{That is, in addition to the original characteristics, we include their squared terms and interactions.} We call the model {\it quadratic mixed-logit model.}  If we use the linear mixed-logit models, then $K=2$ and the condition in Corollary \ref{rem:MT} is violated (i.e., $K=2 \not \ge 3= |J|-1$) and the set of characteristics is not affinely independent. Thus, by Theorem \ref{theo:1}, the linear mixed-logit models with fixed effects are not flexible enough to approximate some random utility models. On the other hand, with quadratic polynomials, the generic condition for representability in Corollary \ref{rem:MT} is satisfied, since $K=5\ge 3=|J|-1$ and the characteristics are affinely independent. Thus, by Theorem \ref{theo:1}, the quadratic mixed-logit models are flexible enough to approximate any random utility model. These theoretical implications are numerically verified and quantified below, where we report approximation errors for the linear and quadratic logit models with and without fixed effects.

\vspace{-0.2cm}

\subsection{Approximation Errors with and without Fixed Effects}\label{sec:without_fixed}

\vspace{-0.2cm}

We assume that $\D=2^J \setminus \{\emptyset\}$. We say that a ranking $\pi$ is {\it linearly representable} if $\pi$ is representable in $\D$ with the original characteristics and $\pi$ is {\it linearly unrepresentable} if $\pi$ is not  linearly representable. With $|J|=4$, there are 24 possible rankings and we find that 12 rankings are not linearly representable. Table \ref{tbl:apx_error1} below reports the approximation errors for rankings that are not linearly representable, calculated using both the Greedy and EM algorithms. (In the table, we label the four DVDs by the indices 1, 2, 3, and 4.) Approximation errors for deterministic rankings are of particular interest, because the worst-case approximation error over all stochastic choice rules is attained at such a ranking.\footnote{A proof of this statement is provided in Section~\ref{section:worst_case} of the online appendix.}

\begin{table}[ht]
\begin{center}
  \caption{Approximation errors to  $\rho^{\pi}$}\label{tbl:apx_error1}
\scalebox{0.9}{
\begin{tabular}{|c|c|c|c|c|}
\hline
\multirow{2}{*}{Ranking $\pi$ } &\multicolumn{2}{c|}{Linear mixed-logit}&\multicolumn{2}{c|}{Quadratic mixed-logit} \\ \cline{2-5}
 & Greedy  & EM    & Greedy  & EM \\ 
   &  (1) &  (2) &(3) &(4) \\  
\hline
Linearly Unrepresentable Rankings &&&  &\\
$\pi(1)> \pi(3)> \pi(4)> \pi(2)$    & 0.649 & 0.649& 0.000 &0.000   \\
$\pi(1)>\pi(3)>\pi(2)>\pi(4)$        & 0.587 & 0.603& 0.000 &0.000\\
$\pi(1)> \pi(4) > \pi(3) > \pi(2)$   &0.437 & 0.445&0.000  &0.000\\
$\pi(2)> \pi(4)> \pi(3)> \pi(1)$   & 0.685 & 0.691& 0.000 &0.000\\
$\pi(2)>\pi(4)> \pi(1)> \pi(3)$   &0.674 & 0.685 & 0.000 &0.000\\
$\pi(2)> \pi(3) > \pi(4)> \pi(1)$  &0.436 & 0.445&0.000  &0.000\\
$\pi(2)>\pi(1)>\pi(4)>\pi(3)$ & 0.422 & 0.431 &0.000  &0.000\\
$\pi(3)>\pi(1) >\pi(4)> \pi(2)$  & 0.596 & 0.642& 0.000 &0.000\\
$\pi(3)> \pi(1)> \pi(2) >\pi(4)$ & 0.493 & 0.522&0.000  &0.000\\
$\pi(3)>\pi(4)> \pi(1)> \pi(2)$  &0.388 & 0.404&0.000  &0.000\\
$\pi(4)>\pi(2)> \pi(3)> \pi(1)$  & 0.511 & 0.566&0.000  &0.000\\
$\pi(4)> \pi(2)> \pi(1)> \pi(3)$  &0.455 & 0.490 &0.000  &0.000\\
\hline
Linearly Representable Rankings & 0.000 & 0.000 & 0.000& 0.000\\
\hline
\end{tabular}
}
\end{center}
\begin{scriptsize}
\textit{Note}: The numbers in the table show the approximation errors without fixed effects for each $\rho^{\pi}$, where each ranking $\pi$ is defined in the leftmost column. For each ranking, columns (1) and (2) show the approximation errors of the linear mixed-logit models computed by the greedy algorithm and the EM algorithm, respectively. We choose 18 mixtures for the EM algorithm, as implied by Propositions \ref{cor:red} and  \ref{lem:dim_P_r} in Section \ref{sec:simplify} because $(\dim \P_r)+1=1+\sum_{D \in \D}(|D|-1)=18$. Columns (3) and (4) show the approximation errors of the quadratic mixed-logit models calculated by each algorithm. All numbers are rounded to three decimal places. For the greedy algorithm we set the number of iterations to 1000. For the EM algorithm we set the number of random initial points to 10.
\end{scriptsize}
\end{table}

As implied by Theorem 1, quadratic mixed-logit models yield zero approximation errors for all rankings and linear mixed-logit models yield zero errors for linearly representable rankings. The approximation errors of linear mixed-logit models for unrepresentable rankings $\pi$ are almost always larger than $0.4$. Thus, even the best-fitting linear mixed-logit model remains far from the corresponding deterministic choice rule $\rho^\pi$ in the normalized Euclidean metric. Some errors are substantially larger. For example, the approximation error for the ranking $\pi(2)>\pi(4)>\pi(3)>\pi(1)$ is more than $0.685$.

Table \ref{tab:representability_macro} reports the approximation errors \textit{with fixed effects} to a half-half mixture, $\frac{1}{2}\rho^{\pi}+ \frac{1}{2} \rho^{\pi^-}$, for each unrepresentable ranking $\pi$.  The random utility model $\frac{1}{2}\rho^{\pi}+ \frac{1}{2} \rho^{\pi^-}$ can be interpreted as the aggregate behavior of a population in which half of consumers have ranking $\pi$ and  the other half have ranking  $\pi^{-}$.

\begin{table}[ht]
\caption{ Approximation errors to random utility models $\frac{1}{2}\rho^{\pi}+\frac{1}{2}\rho^{\pi^-}$}\label{tbl:apx_error2}
\label{tab:representability_macro}
\begin{center}
\scalebox{0.9}{
\begin{tabular}{|c|c|c|c|c|}
\hline
\multirow{2}{*}{Ranking $\pi$ } &\multicolumn{2}{c|}{Linear mixed-logit}&\multicolumn{2}{c|}{Quadratic mixed-logit} \\ \cline{2-5}
 & Greedy  & EM    & Greedy  & EM \\ 
   &  (1) &  (2) &(3) &(4) \\  
\hline
Linearly unrepresentable rankings &&&&\\
$\pi(1)> \pi(3)> \pi(4)> \pi(2)$     &0.265& 0.300& {0.000} & {0.000} \\
$\pi(1)>\pi(3)>\pi(2)>\pi(4)$        &0.225& 0.287& {0.000} & {0.000}  \\
$\pi(1)> \pi(4) > \pi(3) > \pi(2)$   & 0.191& 0.243& {0.000}& {0.000}  \\
$\pi(2)> \pi(4)> \pi(1)> \pi(3)$     &0.191& 0.243 &{0.000}& {0.000} \\
$\pi(2)>\pi(1)> \pi(4)> \pi(3)$      &0.191& 0.206 &{0.000}& {0.000}  \\
$\pi(3)> \pi(1) > \pi(2)> \pi(4)$    &0.191&  0.206 &{0.000}& {0.000}  \\
\hline
Linearly representable rankings & 0.000 & 0.000 & {0.000}& {0.000}\\
\hline
\end{tabular}
}
\end{center}
\begin{scriptsize}
\textit{Note:} The numbers in the table show the approximation errors to $\frac{1}{2}\rho^{\pi}+ \frac{1}{2}\rho^{\pi^-}$, where $\pi$ is defined in the leftmost column.  All numbers are rounded to three decimal places. For the greedy algorithm we set the number of iterations to 1000. For the EM algorithm we set the number of random initial points to 10. We optimize over fixed effects from -10 to 10 with a step size of 1. Therefore, the reported errors with fixed effects should be interpreted as grid-search approximation errors; the true infimum over unrestricted fixed effects can only be weakly smaller.
\end{scriptsize}
\end{table}

Proposition \ref{prop:unrep2} implies  that such random utility models cannot be approximated arbitrarily well by any sequence of linear mixed-logit models with fixed effects. In both algorithms, the approximation errors to $\frac12\rho^\pi+\frac12\rho^{\pi^-}$ are around $0.2$ or above whenever $\pi$ is not representable. Thus, even after allowing fixed effects, the best-fitting linear mixed-logit model remains bounded away from $\frac{1}{2}\rho^\pi+\frac{1}{2}\rho^{\pi^-}$ in the normalized Euclidean metric. By contrast, the approximation errors for $\frac{1}{2}\rho^\pi+\frac{1}{2}\rho^{\pi^-}$ are essentially zero when $\pi$ is representable, as predicted by Theorem 1.

\vspace{-0.2cm}

\subsection{Maximal Substitution}\label{sec:substitution}

\vspace{-0.2cm}

This section quantifies the flexibility of the linear mixed-logit models by measuring the \textit{maximal substitution patterns} that can be generated by these models. Specifically, for two alternatives $j$ and $l$, we define the maximal substitution pattern for mixed-logit models as
\begin{equation}\label{eq:dist.2}
\sup_{\rho \in \mathcal{P}_{ml}(0)}\Big(\rho(J\setminus\{j\},l)-\rho(J,l)\Big).
\end{equation} 
The quantity $\rho(J\setminus\{j\},l)-\rho(J,l)$ describes how many consumers would substitute to alternative $l$ if alternative $j$ becomes unavailable. The supremum of such quantities captures the maximal substitution pattern that can be generated by  mixed-logit models without fixed effects.\footnote{ \cite{conlon2021empirical} analyze similar concepts called {\it diversion ratios}, which measure the fraction of consumers who switch their choices from alternative $j$ to $l$ after the price of alternative $j$ increases marginally. One quantity highlighted by \cite{conlon2021empirical} (p. 701) is a functional of diversion ratios that characterizes substitution patterns when a product becomes unavailable. Our measure corresponds to the limit of the numerator of that quantity. Some recent papers study  substitution and complementarity properties in the discrete choice analysis. See \cite{horan_substitution} and  \cite{allen2020hicksian}.} Since $\rho$ is a random utility model, we always have $\rho(J\setminus\{j\},l) \ge \rho(J,l)$. Thus the quantity in (\ref{eq:dist.2}) takes values in $[0,1]$.  We use the greedy algorithm to solve (\ref{eq:dist.2}), as detailed in Section \ref{appendix:maximal_sub} of the online appendix.\footnote{ The quantity in (\ref{eq:dist.2}) is always 1 when we can choose fixed effects freely. This is because we can always choose fixed effects large enough to approximate degenerate preferences. We do not use the EM algorithm as it cannot be easily transformed to solve the problem in (\ref{eq:dist.2}). We can replace the $\sup$ in (\ref{eq:dist.2}) with $\inf$, for which we calculate the \textit{minimal substitution pattern}. The minimal substitution pattern is 0, since the dataset satisfies the convex independence. }

\begin{table}[ht]
\begin{center}
  \caption{Maximal substitution  of the linear mixed-logit models}\label{tbl:subs}
\scalebox{0.9}{
\begin{tabular}{|c|c|c|c|c|}
 \hline
\diagbox[width=10em, trim=l]{$j$ (drop)}{$l$ (choose)} & 1 & 2 & 3 & 4 \\
  \hline
1 & -  & 0.999 & 0.138 & 0.999 \\ 
\hline
  2 & 1.000 & - & 1.000 & 0.111 \\ 
\hline
  3 & 0.250 & 1.000 & - & 1.000 \\ 
\hline
  4 & 1.000 & 0.303 & 1.000 &  -\\ 
  \hline
\end{tabular}
}
\end{center}
\begin{scriptsize}
\textit{Note}: The numbers in the table show the value of (\ref{eq:dist.2}) for each $j, l \in \{1,2,3,4\}$ s.t. $j\neq l$. All numbers are rounded to three decimal places.   
\end{scriptsize}
\end{table}

Table \ref{tbl:subs} reports the maximal substitutions for each pair of alternatives. It is evident that the linear mixed-logit model has difficulty capturing the substitution behavior between products 1 and 3 and between products 2 and 4.  In particular, the maximal substitution from product 2 to product 4 is only 0.111. This finding aligns with the result presented in Table \ref{tbl:apx_error1} as we observe substantial approximation errors for the two specific rankings: $\pi(2)>\pi(4)>\pi(3)> \pi(1)$ and $\pi(2)>\pi(4)>\pi(1)> \pi(3)$. These are the only rankings in which 4 replaces 2 when 2 becomes unavailable.

The fact that linear mixed-logit choice models fail to approximate substitution patterns may have important implications for the  classical {\it assortment planning problem} in operations management. See  \cite{kok2008assortment} for a survey on the topic. For example, consider a simple example where there are three alternatives (labeled 1, 2 and 3) and a no-purchase option (labeled 4). Suppose the population consists of three groups; 50\% of customers have preferences  $\pi(3)>\pi(4)>\pi(1)> \pi(2)$, 25\%  have preferences  $\pi(1)>\pi(3)>\pi(2)> \pi(4)$; and the remaining 25\%  have preferences $\pi(2)>\pi(4)>\pi(3)> \pi(1)$. Assume that product 3 yields the highest profit, followed by products 1 and 2. The store initially offers products 1, 2, and 3, but a shelf-space constraint forces it to remove one product. Since product 3 is both the most popular and the most profitable, the choice is between removing product 1 or 2.

Suppose the store manager fits a linear mixed-logit model to historical sales data and uses this model to forecast the impact of dropping one product. If product 1 is removed, the model may predict that only a small fraction of customers (at most $13.8\%$, as in Table~\ref{tbl:subs}) switch from 1 to 3, with most reallocating from 1 to 2 or 4. If product 2 is removed, the model may predict that only a small fraction of customers (at most $11.1\%$, as in Table~\ref{tbl:subs}) switch from 2 to 4, with most reallocating from 2 to 1 or 3. On this basis, the manager might decide to remove product 2, hoping to shift demand toward the higher-margin products 1 and 3. In reality, however, one quarter of customers, those with preference 
$\pi(2)>\pi(4)>\pi(3)>\pi(1)$, will choose the no-purchase option 4. As a result, the store sacrifices revenue due to the poor substitution predictions of the linear mixed-logit model.

\vspace{-0.2cm}

\vspace{-0.2cm}

\appendix

\section*{Appendix: Proofs}

\vspace{-0.2cm}

In Proposition \ref{cor:red} of Section \ref{sec:simplify} of the online appendix, we prove that $\P_{ra}(\eta|\mu)=\co\P_a(\eta|\mu)$ for any fixed effects vector $\eta$ and standard probability measure $\mu$, where $\co$ denotes the convex hull.  We use this result throughout the proof. 

For all $(D,j) \in \D \times J$ such that $j \in D$, if $\pi(j)> \pi(l)$ for all $l \in D \setminus \{j\}$, then we  write $j=\max_{D} \pi$. We use this notation throughout the proof. 

\vspace{-0.2cm}

\section{ Proofs of Theorem \ref{theo:1} (i), Proposition 1 and 3}

\vspace{-0.2cm}

To prove Proposition 1, Proposition 3, and Theorem 1-(i), we will show several lemmas. We first consider models without fixed effects (i.e., $\eta=0$).  The following fact is elementary but fundamental:

\noindent \textbf{Observation:} {\it The set $\P_r$ of random utility models is a polytope, that is, $\P_r=\co \{\rho^{\pi}| \pi \in \Pi\}$. For each $\pi \in \Pi$, $\rho^{\pi}$ is a vertex of the polytope.}   

\medskip

The observation holds because for any random utility model $\rho \in \P_r$, we have $\rho = \sum_{\pi \in \Pi} \nu(\pi) \rho^{\pi}$, where $\nu$ is the probability measure on $\Pi$ rationalizing the random utility model. See section \ref{sec:geometric} in the online appendix for details.  The hexagons in Figure \ref{fig:P_r} in the section illustrate the polytope.

Lemma \ref{lem:appxQ} gives a necessary and sufficient condition under which any random utility models can be approximated arbitrarily well by a sequence of random-coefficient ARUMs without fixed effects. 

\begin{lemma}\label{lem:appxQ} Fix any $\mu \in \M$. $\P_r \subset \cl\P_{ra}(0| \mu)$ if and only if $\rho^{\pi} \in \cl \P_a(0| \mu)$ for any $\pi \in \Pi$.
\end{lemma}

The next lemma makes it easier for us to check the conditions of Lemma \ref{lem:appxQ}.

\begin{lemma}\label{lem:real_x3}
For any ranking $\pi \in \Pi$, the following statements hold: (1) If $\pi$ is representable in $\D$, then for any $\mu \in \M$, $\{\rho^{\pi},\rho^{\pi^-}\} \subset \cl \P_{a}(0|\mu)$; (2) If $\pi$ is not  representable in ${\cal D}$, then for any  $\mu \in \M$,  $\{\rho^{\pi},\rho^{\pi^-}\} \not \subset\cl \P_{ra}(0|\mu)$.
\end{lemma}

As mentioned earlier,  checking the representability of a particular ranking is easy. However, checking the representability of {\it all} rankings may be computationally demanding. This is because the number of rankings equals $|J|!$ and  can be large. To overcome this problem, we obtain a simpler necessary and sufficient condition for all  rankings to be representable:

\begin{lemma}\label{lem:real_x}(1) Every  ranking is representable in $\D$ if and only if the set $\{x_j \in \Re^K | j \in J\}$ is affinely independent; (2)  Every ranking is representable in $\{J\}$ if and only if the set $\{x_j \in \Re^K| j \in J\}$ is convex-independent.
\end{lemma}

%In Figure \ref{fig:aff_ind} (c),  we consider the model with the quadratic polynomials of original characteristic (i.e., $K=5$ and $x_j=(p_j, q_j, p_j^2, q_j^2, p_jq_j)$ for each $j \in J$).  

See Section \ref{sec:intuition} in the online appendix for the illustration of Lemma \ref{lem:real_x}. 

\vspace{-0.2cm}

\subsection{Proof of Theorem 1 (i)}

\vspace{-0.2cm}
By Lemma 3, the affine independence of $\{x_j \in \R^K| j \in J\}$ implies that all rankings are representable and thus can be approximated by a sequence of ARUMs by Lemma 2-(1). Applying Lemma 1 proves Theorem 1-(i).

\vspace{-0.2cm}

\subsection{Proof of Proposition \ref{prop:unrep}  and  \ref{theo:2}}

\vspace{-0.2cm}

We first prove Proposition \ref{prop:unrep}.
 By Lemma 3-(1), the non-affine independence of $\{x_j| j \in J\}$ implies that some rankings are not representable. By Lemma \ref{lem:real_x3}-(2),  for the given $\mu$ at least one element $\rho^\sigma$ with $\sigma\in\{\pi,\pi^-\}$ is not in $\cl\P_{ra}(0\mid\mu)$. Since $\cl\P_{ra}(0\mid\mu)$ is closed, there exists an open neighborhood $U$ of $\rho^\sigma$ such that $U\cap \cl\P_{ra}(0\mid\mu)=\emptyset$. Hence, Proposition 1 holds. 

Lemma \ref{lem:appxQ}, \ref{lem:real_x3}, and \ref{lem:real_x} (2) imply statement (i) of Proposition \ref{theo:2}. The proof of statement (ii) is in Section \ref{proof:proposition3} of the online appendix.

\vspace{-0.2cm}

\subsection{Proof of Lemma \ref{lem:appxQ}}

\vspace{-0.2cm}

We first prove the if direction. Suppose that $\rho^{\pi} \in \cl\P_a(0|\mu)$ for all $\pi \in \Pi$. Then $\co \{\rho^{\pi}| \pi \in  \Pi\} \subset \co\cl \P_a(0|\mu)$.  By the fact that $\P_a(0|\mu)$ is bounded, it follows from Theorem 17.2 of \cite{rockafellar2015} that $\cl \co \P_a(0|\mu)= \co \cl \P_a(0|\mu)$. Thus, $\P_r\subset \cl \co \P_a(0|\mu)\subset \cl \P_{ra}(0|\mu)$, where we used the fact that  $\P_{ra}(0|\mu)= \co\P_{a}(0|\mu)$ for any $\mu \in \M$.\footnote{See Proposition \ref{cor:red} of Section \ref{sec:simplify} of the online appendix.}

We now prove the only-if direction.  By assumption, $\P_r\subset \cl \conv \P_a(0|\mu)= \conv \cl \P_a(0|\mu)$, where the equality holds by the previous argument. Since $\P_r\subset \conv \cl \P_a(0|\mu)$, for any $\pi \in \Pi$, there exist positive numbers $\{\la_i\}_{i=1}^m$ and $\rho^i \in \cl \P_a(0| \mu)$ for all $i \in \{1,\dots,m\}$ such that $\sum_{i=1}^m \la_i=1$ and  $\sum_{i=1}^m \la_i \rho^i=\rho^{\pi}$.  Note that $\rho^i \in \cl \P_a(0| \mu) \subset \P_r$ for each $i$ because $\P_r$ is closed. Since $\rho^{\pi}$ is a vertex of $\P_r$,  thus $\rho^{\pi}$ is an exposed point.\footnote{A point of a convex set is an exposed point if there is a supporting hyperplane which contains no other points of the set (\cite{rockafellar2015}, Page 162)} Hence, $\sum_{i=1}^m \la_i \rho^i=\rho^{\pi}$ implies $\rho^i=\rho^{\pi}$  for all $i$. This means that $\rho^{\pi} \in \cl \P_a(0|\mu)$ for all $\pi \in \Pi$.

\vspace{-0.2cm}

\subsection{Proof of Lemma \ref{lem:real_x3}}

\vspace{-0.2cm}

We prove Lemma 2-(1) in the next step.

\step 1:  For any $\pi \in \Pi$ and any $\mu \in \M$, if a ranking $\pi$ is representable, then there exists a sequence $\{\rho_n\}$ of $\P_{a}(0|\mu)$ such that $\rho_n \to \rho^{\pi}$.

\begin{proof}Assume that a ranking $\pi$ is representable.  Fix $D \in \D$ and let $j=\max_D \pi$. Then $\beta \cdot x_j-\max_{l\in D \setminus \{j\}}  \beta \cdot x_l>0$. 
Let $\rho_{n}$ be the sequence of ARUMs with coefficient $n\beta$. For any positive integer $n$ and any $(D,j) \in \D \times J$ such that $j \in D$, note that $\rho_{n }(D,j)\geq \int 1\Big\{n \left(\beta \cdot x_j-\max_{l\in D \setminus  \{j\}}  \beta \cdot x_l\right) \geq \max_{l\in D \setminus  \{j\}} \ep_l  - \ep_j \Big\}d\mu$, where $1\{\cdot\}$ is the indicator function.\footnote{Note that $\rho_{n }(D,j)
\equiv  
\mu(\{\ep |n \beta \cdot x_j+\ep_j  \geq \max_{l\in D \setminus  \{j\}} \{ n \beta \cdot x_l +\ep_l \}\})
\geq \mu(\{\ep|n\beta \cdot x_j+\ep_j  \geq \max_{l\in D \setminus  \{j\}} n \beta \cdot x_l +\max_{l\in D \setminus  \{j\}}\ep_l \})
 = \mu(\{\ep| n (\beta \cdot x_j-\max_{l\in D \setminus  \{j\}}  \beta \cdot x_l) \geq \max_{l\in D \setminus  \{j\}}\ep_l  - \ep_j \})
=\int 1\{n (\beta \cdot x_j-\max_{l\in D \setminus  \{j\}}  \beta \cdot x_l) \geq \max_{l\in D \setminus  \{j\}} \ep_l  - \ep_j \}d\mu$.}  By the dominated convergence theorem, $\lim_{n\to \infty}\rho_{n}(D,j) \geq  \int \lim_n 1\Big\{n \left(\beta \cdot x_j-\max_{l\in D \setminus  \{j\}}  \beta \cdot x_l\right) \geq \max_{l\in D \setminus  \{j\}}\ep_l  - \ep_j \Big\}d \mu=1$. Since choice frequencies must be nonnegative and sum up to one,  $\lim_{n\to \infty}\rho_{n}(D,j)\ge 1$ implies $\rho_{n}\to \rho^\pi$.
\end{proof}

We prove the contrapositive of Lemma 2-(2) in Step 2 and 3. 

%We first show that the existence of converging sequences of ARUMs to $\rho^\pi$ and $\rho^{\pi^-}$ implies the representability of $\pi$. Step 3 shows that such a converging sequence of ARUMs exists if a sequence of random coefficient ARUMs converges to $\rho^\pi$.

\step 2: If there exists a $\mu \in \M$ and sequences $\{\rho_n\}$ and $\{\rho'_n\}$ of $\P_{a}(0|\mu)$ such that $\rho_n \to \rho^{\pi}$ and $\rho'_n \to \rho^{\pi^-}$, then $\pi$ and $\pi^-$ are representable.

\begin{proof}
Suppose $\rho_n\in \P_a(0\mid \mu)$ and $\rho'_n\in \P_a(0\mid \mu)$ satisfy $
    \rho_n\to \rho^\pi$ and $\rho'_n\to \rho^{\pi^-}$. Let $\beta_n$ and $\beta'_n$ be the corresponding coefficient vectors. Fix a binary menu $\{j,l\}$ with $\pi(j)>\pi(l)$. Define
$Z_{lj}=\ep_l-\ep_j$, $\gamma_n=\beta_n\cdot(x_j-x_l)$, and $\gamma'_n=\beta'_n\cdot(x_l-x_j)$. Since $\rho_n\to \rho^\pi$ and $\rho'_n\to \rho^{\pi^-}$, we have $
    \mu(\{\ep| Z_{lj}<\gamma_n\})\to 1$ and $\mu(\{\ep| Z_{lj}>-\gamma'_n\})\to 1$. If $\gamma_n+\gamma'_n\leq 0$ along an infinite subsequence, then along that subsequence $\{\ep| Z_{lj}<\gamma_n\}\cap\{\ep| Z_{lj}>-\gamma'_n\}=\emptyset$, because $\gamma_n\leq -\gamma'_n$. Therefore, $\mu(\{\ep| Z_{lj}<\gamma_n\})+\mu(\{\ep| Z_{lj}>-\gamma'_n\})\leq 1$,  contradicting the fact that both probabilities converge to one. Hence, for this pair $(j,l)$, $\gamma_n+\gamma'_n=(\beta_n-\beta'_n)\cdot(x_j-x_l)>0$ 
for all sufficiently large $n$. Since there are finitely many pairs with $\pi(j)>\pi(l)$, there is a single $N$ such that for all $n\geq N$, $
    (\beta_n-\beta'_n)\cdot(x_j-x_l)>0$
   for every pair $j,l$ with $\pi(j)>\pi(l)$. Thus, $\beta_n-\beta'_n$ represents $\pi$ and $\beta'_n-\beta_n$ represents $\pi^-$ for all sufficiently large $n$. 
\end{proof}

In the following, we again use the fact that  $\P_{ra}(0|\mu)= \co\P_{a}(0|\mu)$ for any $\mu \in \M$.

\step 3: If there exists $\mu \in \M$ and  a sequence $\{\rho_n\}$ of $\co \P_{a}(0|\mu)$ such that $\rho_n \to \rho^{\pi}$, then there exists a sequence $\{\rho'_n\}$ of $\P_{a}(0|\mu)$ such that $\rho'_n \to \rho^{\pi}$.

\begin{proof}
Fix $\pi \in \Pi$. Suppose that there exists a sequence $\rho_n$ of $\co \P_{a}(0|\mu)$ such that $\rho_n \to \rho^{\pi}$ as $n\to \infty$. Let $M=\dim \co  \P_{a}(0|\mu)$. Then for each $\rho_n$, by Caratheodory's theorem,  there exist $\{\rho^i_n\}_{i=1}^{M+1} \subset \P_{a}(0|\mu)$ and nonnegative numbers $\{\al^i_n\}_{i=1}^{M+1}$ such that $\rho_n= \sum_{i=1}^{M+1} \al^i_n \rho^i_n$ and $\sum_{i=1}^{M+1} \al^i_n=1$.  Denote $(\al^i_n)_{i=1}^{M+1}$ by $\al_n$. Then $\al_n$ belongs to a compact set (i.e., $M$-dimensional simplex). There exists a convergent subsequence $\{\al_{n'}\}$. Thus $\rho'_n \equiv \sum_{i=1}^{M+1} \al^i_{n'} \rho^i_{n'}$ is a subsequence of $\{\rho_n\}$.  For each $i$, let $\al^i_*$ be the limit of $\{\al^i_{n'}\}$. Since $\sum_{i=1}^{M+1}\al^i_{n'}=1$ for all $n'$, we have $\sum_{i=1}^{M+1} \al^{i}_*=1$, so that there must exist $i^*$ such that $\al^{i^*}_* \neq 0$. 

In the following, we will show that $\rho^{i^*}_{n'} \to \rho^{\pi}$ as $n' \to \infty$. To show the claim, we prove that if $\rho^{i^*}_{n'} \not \to \rho^{\pi}$, then $\al^{i^*}_{n'} \to 0$, which is a contradiction. Assume that $\rho^{i^*}_{n'} \not \to \rho^{\pi}$. Then there exist $D \in \D$, $j \in D$, and $\ep>0$ such that for any integer $N$ there exists $n'>N$ such that $|\rho^{i^*}_{n}(D,j)-\rho^{\pi}(D,j)|> \ep$. This implies that for any $N$ there exists $n'>N$ such that $\Big|\sum_{i=1}^{M+1} \al^i_{n'} \rho^i_{n'}(D,j) - \rho^{\pi}(D,j)\Big|=\sum_{i=1}^{M+1} \al^i_{n'} |\rho^i_{n'}(D,j) - \rho^{\pi}(D,j)|\ge  \al^{i^*}_{n'}\ep$, where the first equality holds because if $j=\max_{D} \pi$ then $\rho^i_{n'}(D,j) - \rho^{\pi}(D,j)\le 0$ for all $i$; if  $j\neq \max_{D} \pi$, then $\rho^i_{n'}(D,j) - \rho^{\pi}(D,j)\ge 0$ for all $i$. Since $\sum_{i=1}^{M+1} \al^i_{n'} \rho^i_{n'}(D,j) \to \rho^{\pi}(D,j)$, it must hold that $\al^{i^*}_{n'} \to 0$.
\end{proof}

Step 2 and 3 show that for any $\mu \in \M$, if there exist   sequences $\{\rho_n\}$ and $\{\rho'_n\}$ of $\co \P_{a}(0|\mu)$ such that $\rho_n \to \rho^{\pi}$ and $\rho'_n \to \rho^{\pi^-}$,  then $\pi$ is representable. The contraposition of this statement is Lemma 2-(2).

\vspace{-0.2cm}

\subsection{Proof of Lemma  \ref{lem:real_x}}

We first prove statement (1). For simplicity of notation, let $J=\{1,2, \dots, |J|\}$. For any ranking $\pi \in \Pi$, by relabeling $J$ if necessary, we assume  that $\pi(i)>\pi({i+1})$ for all $i \le |J|-1$ without loss of generality. We  label the following condition  as Condition ($*$):  if $\la_1 x_1+\sum_{i=2}^{|J|-1} (\la_i-\la_{i-1}) x_i -\la_{|J|-1}x_{|J|}=0$ and $\la_i \ge 0$ for all $i \in \{1,\dots, |J|-1\}$, then $\la_i=0$ for all $i \in \{1,\dots, |J|-1\}$. 

\step 1: For each $\pi \in \Pi$, Condition ($*$) holds if and only if $\pi$ is representable.

\begin{proof}
Since $\mathcal D$ contains all binary choice sets, $\pi$ is representable if and only if there exists $\beta\in\mathbb R^K$ such that $\beta\cdot x_1>\beta\cdot x_2>\cdots>\beta\cdot x_{|J|}$. Equivalently,  $\beta\cdot (x_i-x_{i+1})>0 \quad \text{for all } i=1,\ldots,|J|-1$.  By the theorem of alternatives, this system of strict inequalities has a solution if and only if there is no nonzero vector $(\lambda_i)_{i=1}^{|J|-1}$ with $\lambda_i\geq 0$ for all $i$ such that $\sum_{i=1}^{|J|-1}\lambda_i(x_i-x_{i+1})=0$. See Section \ref{proof:lemma3}  in the online appendix for the detail.
Expanding the last display gives exactly Condition $(*)$.
\end{proof}

\step 2: The set $\{x_j| j \in J\}$ is affinely independent if and only if Condition ($*$) holds for any $\pi \in \Pi$.

\begin{proof} 
First suppose that $\{x_j:j\in J\}$ is affinely independent. If the equality in Condition $(*)$ holds, define $\mu_1=\lambda_1,\qquad \mu_i=\lambda_i-\lambda_{i-1}\quad (i=2,\ldots,|J|-1),\qquad \mu_{|J|}=-\lambda_{|J|-1}$. Then $\sum_{i=1}^{|J|}\mu_i x_i=0$ and $\sum_{i=1}^{|J|}\mu_i=0$. Affine independence implies $\mu_i=0$ for all $i$, and hence $\lambda_i=0$ for all $i$. Thus Condition $(*)$ holds. Conversely, suppose Condition $(*)$ holds for every ranking. Take any $(\mu_i)_{i=1}^{|J|}$ such that $\sum_{i=1}^{|J|}\mu_i x_i=0$ and $\sum_{i=1}^{|J|}\mu_i=0$. Relabel alternatives so that $\mu_1\geq \mu_2\geq\cdots\geq \mu_{|J|}$, and define $\lambda_i=\sum_{k=1}^i\mu_k$ for all $i\in \{1,\ldots,|J|-1\}$.  Then $\lambda_i\geq 0$ for all $i$, and $\sum_{i=1}^{|J|-1}\lambda_i(x_i-x_{i+1})=0$. By Condition $(*)$, $\lambda_i=0$ for all $i$, and therefore $\mu_i=0$ for all $i$. This proves affine independence. Combining Steps 1 and 2 proves Lemma 3-(1).
\end{proof}

The proof of statement (2) is straightforward and is relegated to Section~\ref{proof:lemma3} of the online appendix.

\vspace{-0.2cm}

\section{Proof of Theorem 1-(ii) and Proposition 2}

\vspace{-0.2cm}

As mentioned in Section \ref{sec:menu-dependent}, the main results hold with menu-dependent fixed effects with a mild restriction. We prove Theorem 1-(ii) and Proposition 2 with general $\F$ defined by (\ref{def:Fixed}).

 To prove Theorem 1-(ii) and Proposition 2, we first prove the following two statements: (a) for any ranking $\pi$, any strict convex combination between $\rho^{\pi}$ and $\rho^{\pi^{-}}$ cannot be approximated arbitrarily well by any sequence of ARUMs with  fixed effects; and (b) moreover, if $\pi$ is not representable, then any strict convex combination between $\rho^{\pi}$ and $\rho^{\pi^{-}}$ cannot be approximated arbitrarily well by any sequence of random-coefficient ARUMs even with fixed effects. To prove statement (a), we prove a stronger result:

\begin{lemma}\label{lem:degenerate} Fix three alternatives $j,l,r$ such that for any $\eta \in \F$, $\eta_{(D,s)}=\eta_{(E,s)}$ for all $D, E \subseteq \{j,l,r\}$ and $s \in D \cap E$. Fix two rankings $\pi$ and $\pi'$ such that $\pi(j)>\pi(l)>\pi(r)$ and $\pi'(r)>\pi'(l)>\pi'(j)$.\footnote{Such three alternatives exist by assumption made in Section \ref{sec:menu-dependent}. The lemma could be stated without explicitly referring to them by letting \(\pi' = \pi^{-}\). However, we present the result in the current form because the lemma itself  is relevant to the literature, especially to \cite{norets2013surjectivity}. See Section \ref{section:menu-independent_fixed} in the online appendix for the details.
}  Let $\mu \in \M$ be a standard probability measure.  For any $\al \in (0,1)$, a random utility model $\rho^{\al}\equiv \al \rho^{\pi}+(1-\al)\rho^{\pi'}$ cannot be approximated by any sequence of ARUMs with fixed effects and the given $\mu$. Formally, for all $\alpha \in (0,1)$ and $\mu \in \M$,  $\rho^{\al} \not \in  \cl  \bigcup_{\eta \in \F} \P_{a}(\eta|\mu)$.
 \end{lemma}

This lemma obviously implies statement (a) by letting $\pi'= \pi^{-}$. This lemma shows the limitation of the representative power of ARUMs even with fixed effects.

To show statement (b),  we introduce the following concept:

\begin{definition}\label{def:adjacency}
\normalfont The two rankings $\pi$ and $\pi'$ are {\it adjacent} if there exist a real vector $t$ and a real number $a$ such that (i) $\rho^{\pi}\cdot t = \rho^{\pi'}\cdot t=a$, and (ii) for any $\hat{\pi}\in \Pi$, if $\hat{\pi}\not \in \{\pi, \pi'\}$, then $\rho^{\hat{\pi}}\cdot t > a$.\footnote{For any $\rho \in \P$, $\rho \cdot t= \sum_{(D,j) \in D \times J} \rho(D,j) t(D,j)$.}
\end{definition}

For example, in Figure \ref{fig:P_r} in the online appendix, $\rho^{\pi_1}$ and $\rho^{\pi_6}$ as well as $\rho^{\pi_i}$ and $\rho^{\pi_{i+1}}$ for each $i \le 5$ are adjacent and no other pairs are adjacent.\footnote{The figure is intended only as an illustration and does not represent the set of random utility models exactly.} Since $\pi$ and $\pi^{-}$ are reversed with each other, $\rho^{\pi}$ and $\rho^{\pi^{-}}$ seem very different.  It turns out, however, that they are adjacent:\footnote{Our discussions with Jean-Paul Doignon and Haruki Kono were very helpful for obtaining this result.}
\begin{lemma}\label{lem:adjacent}
For any ranking $\pi \in \Pi$, $\rho^{\pi}$ and $\rho^{\pi^-}$ are adjacent.
\end{lemma}

Lemma \ref{lem:adjacent} follows from Proposition 4.2 of \cite{doignion_adjacency}, who obtain the characterization of adjacency of vertices for the case $\D =2^J \setminus \{\emptyset\}$.  In the online appendix, we show that Lemma \ref{lem:adjacent} holds even for the case in which $\D \neq 2^J \setminus \{\emptyset\}$ as long as $\D$ contains all binary and ternary sets.\footnote{We are grateful to Haruki Kono for pointing out this fact.} 

By using Lemma \ref{lem:degenerate} and \ref{lem:adjacent}, we prove the following lemma, which implies  statement (b).

       \begin{lemma}\label{lem:unrep} Let $\mu$ be any element of $\M$. For any $\al \in (0,1)$ and any ranking $\pi$ that is not representable, there exists a neighborhood $U$ of $\al \rho^{\pi}+(1-\al)\rho^{\pi^-}$ such that any random utility model that belongs to $U$ cannot be approximated arbitrarily well by any sequence of random-coefficient ARUMs with fixed effects and the given  $\mu$.  Formally, $\forall \mu \in \M, \forall \alpha \in (0,1), \forall \pi: \text{unrepresentable}, \exists U \text{ s.t. }\al \rho^{\pi}+(1-\al)\rho^{\pi^-} \in U, \forall \rho \in \P_r \cap  U, \rho \not \in \cl \bigcup_{\eta \in \F} \P_{ra}( \eta|\mu)$.
\end{lemma}

%allows us to complete the proof of Lemma \ref{lem:unrep} as follows. If $\pi$ is not  representable, then $\pi^-$ is also not   representable. Although fixed effects are powerful enough to approximate each vertex $\rho^{\pi}$, we will prove that it is not powerful enough to approximate both $\rho^{\pi}$ and $\rho^{\pi^{-}}$ by using the same fixed effects, intuitively because $\rho^{\pi}$ and $\rho^{\pi^{-}}$ are reversed. Thus, no strict convex combination of $\rho^{\pi}$ and $\rho^{\pi^{-}}$ can be approximated arbitrarily well by the  random-coefficient ARUMs with standard probability measure $\mu$, no matter which fixed effects we use.  Notice that this conclusion does {\it not} follow if $\rho^{\pi}$ and $\rho^{\pi^{-}}$ are not adjacent since a strict convex combination of $\rho^{\pi}$ and $\rho^{\pi^{-}}$  may be represented in a different way. This proves statement (b) and thus, Lemma \ref{lem:unrep}.

%Lemmas \ref{lem:appxQ}, \ref{lem:real_x3}, \ref{lem:real_x}, and \ref{lem:unrep} prove statement (ii) of Theorem \ref{theo:1}, as the proof in the appendix formalizes.

Proposition 2 follows directly from Lemma 3-(1) and Lemma \ref{lem:unrep}. Theorem 1-(ii)  trivially follows from Proposition 2.  
 
\vspace{-0.2cm}

\subsection{Proof of Lemma \ref{lem:degenerate}}

\vspace{-0.2cm}

Fix any  $\mu \in \M$ and $\alpha \in (0,1)$. Fix the three alternatives $j,l,r \in J$  and $\pi$ and $\pi'$ as in the statement of the lemma. Suppose by contradiction that $\rho^{\al}\in \cl \bigcup_{\eta \in \F} \P_{a}(\eta|\mu)$. This implies that there exists a sequence $\{\rho_n\}_{n=1}^{\infty} \subset \bigcup_{\eta \in \F} \P_{a}(\eta|\mu)$ converging to $\rho^{\al}$.  

Let  $\beta_n$ and $\eta_n$ be the sequence of coefficients and fixed effects vector associated with $\rho_n$. Define $\gamma_{n,jl}\equiv \beta_n\cdot (x_j-x_l)+\eta_{n, (\{j,l\},j)}-\eta_{n, (\{j,l\},l)}$. (Remember that we allow menu dependent fixed effects in the proof.) Define $\gamma_{n,jr}$, $\gamma_{n,lr}$, $\gamma_{n,lj}$, $\gamma_{n,rl}$ and $\gamma_{n,rj}$ similarly. First consider the sequence $\{\gamma_{n,jl}\}_{n=1}^{\infty}$.

\noindent\textbf{Step 1}: The sequence $\{\gamma_{n,jl}\}_{n=1}^{\infty}$ is  bounded. 

\begin{myproof}  We prove this by contradiction. First, note that $\ep_l -\ep_j $ is a tight random variable: for each $\delta>0$, there exists a positive number $N_{\delta}$ such that $\mu\left(\{\epsilon|\ep_l -\ep_j \in (-N_{\delta},N_{\delta})^c\}\right)<\delta$. 

Fix a positive real number $\sigma$ such that $\sigma<1-\alpha$. 
Let $\delta=\min\{\frac{\al}{2},1-\al-\sigma\}$. Now, if $\{\gamma_{n,jl}\}_{n=1}^{\infty}$ is unbounded, then either there exists a subsequence $\{\gamma_{n_{k},jl}\}_{k=1}^{\infty}$ such that $\gamma_{n_{k},jl}>N_{\delta}$ for all $k$, or there exists a subsequence $\{\gamma_{n_{k},jl}\}_{k=1}^{\infty}$ such that $\gamma_{n_{k},jl}<-N_{\delta}$ for all $k$. In the first case,  we have $\rho_{n_k}(\{j,l\},j)=\mu\big(\{\ep| \gamma_{n_{k},jl}> \ep_l-\ep_j \}\big)\geq\mu\big(\{\ep| N_{\delta}> \ep_l-\ep_j \}\big)\geq 1-\delta\geq \alpha+\sigma>\al$; in the second case,  $\rho_{n_k}(\{j,l\},j)=\mu\big(\{\ep| \gamma_{n_{k},jl}> \ep_l-\ep_j \}\big)\leq\mu\big(\{\ep| -N_{\delta}> \ep_l-\ep_j \}\big)\leq \delta \leq \frac{\alpha}{2}<\al$. Either case  contradicts convergence to $\al= \rho^{\al}(\{j,l\},j)$. Thus $\{\gamma_{n,jl}\}_{n=1}^{\infty}$ must be bounded. 
\end{myproof}

In the same way, we can show that $\gamma_{n,lr}$, $\gamma_{n,lj}$, $\gamma_{n,rl}$, and $\gamma_{n,rj}$ are all bounded sequences. Given Step 1, we can select  convergent subsequences $\{ (\gamma_{n_k,jl},\gamma_{n_k,jr},\gamma_{n_k,lr},\gamma_{n_k,lj},\gamma_{n_k,rj},\gamma_{n_k,rl})\}_{k\in \N}.$ We denote the limits as $(\gamma^*_{jl},\gamma^*_{jr},\gamma^*_{lr},\gamma^*_{lj},\gamma^*_{rj},\gamma^*_{rl})$. We consider the corresponding stochastic choice functions $\rho_{n_k}$. Note that, by definition,  $\lim_{n_k}\rho_{n_k}= \rho^{\al}$.

For any $s,t\in  \{j,l,r\}$ and any  positive integer $n$, define a measurable set
$E_{n,st}=\{\ep|\gamma_{n,st}>\epsilon_t-\epsilon_s\}$. In words, $E_{n,st}$ is the event where $s$ is preferred over $t$. We further define $E_{st}=\{\ep|\gamma_{st}^*\ge\epsilon_t - \epsilon_s\}$, and $E'_{st}=\{\ep|\gamma_{st}^*>\epsilon_t-\epsilon_{s}\}$. Since $\mu\in \M$ is absolutely continuous with respect to the Lebesgue measure,  $\mu \{\ep|\gamma^*_{st}= \epsilon_t-\epsilon_s\}=0$. Thus $\mu(E_{st})=\mu(E'_{st})$. Note also by the definition of $\lim \inf E_{n,st}
$, we have $E'_{st} \subset \lim \inf E_{n,st} \subset E_{st}$.

\noindent\textbf{Step 2:}  (i) $E_{jl}=E_{jr}$ and $E_{rl}= E_{rj}$ up to a measure zero set; (ii) $\mu(E_{jl}\cap E_{jr})= \al$ and $\mu(E_{rl}\cap E_{rj})=1-\al$.
\begin{myproof}
First note that for each $n$, we have $\rho_n(\{j,l,r\},j)=\mu(E_{n,jl}\cap E_{n,jr})$ because of the assumption on the fixed effects: $\eta_{(\{j,l,r\},t)}=\eta_{(\{j,l\}, t)}= \eta_{(\{j,r\}, t)}$ for all $t \in \{j,l,r\}$. Thus, by Fatou's lemma, we have $\alpha =\rho^{\al}(\{j,l,r\},j) =\lim \sup \rho_n(\{j,l,r\},j)=\lim \sup \mu(E_{n,jl}\cap E_{n,jr}) 
   \leq   \mu(\lim \sup (E_{n,jl}\cap E_{n,jr})) \leq \mu(E_{jl}\cap E_{jr})$, and $\alpha =\rho^{\al}(\{j,l\},j) =\lim \inf \mu(E_{n,jl}) \geq  \mu(\lim \inf E_{n,jl})\geq \mu(E'_{jl})=\mu(E_{jl})$. By an identical argument applied to $\{j,r\}$, we have $\al \ge \mu(E_{jr})$.  Thus we have $\mu(E_{jl}\cap E_{jr}) \geq \alpha \geq \max\{\mu(E_{jl}),\mu(E_{jr})\} \ge \mu(E_{jl}\cap E_{jr})$. It follows that $\mu(E_{jl}\cap E_{jr})=\al= \mu(E_{jl})= \mu(E_{jr})$; thus $E_{jl}=E_{jr}$ up to a measure zero set. By applying the same argument to $\{r,l\}$ and $\{r,j\}$, we obtain $\mu(E_{rl}\cap E_{rj})=1-\al$ and $E_{rl}=E_{rj}$ up to a measure zero set. 
\end{myproof}

For each $\ep \in \Re^{J}$, define a projected column vector $\epsilon_{jlr}= (\ep_j ,\ep_l ,\ep_r)$. Further define $
A= \left\{ \epsilon \big| U \epsilon_{jlr} \ge c \right\}$ and $B=  \left\{\epsilon  \big| U \epsilon_{jlr}\le c
 \right\}$, where 
\begin{equation}\label{eq:c}
U =\begin{bmatrix}
     1 & -1 & 0 \\
     0 & 1 & -1 \\
    \end{bmatrix}, 
\quad c =    
    \begin{bmatrix}
    \gamma_{lj}^* \\
    \gamma_{rl}^*   \end{bmatrix}
.
\end{equation}
%}
In words, $A$ is the event that $j$ is preferred over $l$ and $l$ is preferred over $r$, and $B$ is the event with the reverse ordering. Note that both sets are closed and thus measurable. 

\noindent\textbf{Step 3}: $\mu(A)=\alpha$, $\mu(B)=1-\al$, and $\mu(A \cup B)=1$.

\begin{myproof}
By Step 2, $\mu(E_{jl}\cap E_{jr})=\alpha$ and  $\mu(E_{rl}\cap E_{rj})=1-\alpha$. Remember $E_{st}$ is the event that $s$ is chosen over $t$ in the binary set $\{s,t\}$. Notice $E_{jl}\cap E_{jr}$ and $E_{rl}\cap E_{rj}$ have measure zero intersections, so the two events partition the probability space (ignoring measure zero events). 

Notice that $\al=\rho^{\al}(\{r,l\},l)= \mu(E_{lr})$.  Since the event $E_{lr}$ is incompatible with the event $E_{rl}\cap E_{rj}$ up to a measure zero set, $E_{lr}$ must completely lie within the event $E_{jl}\cap E_{jr}$ (up to a measure zero set).  Moreover, since $\mu(E_{lr})=\al=\mu(E_{jl}\cap E_{jr})$, the event $E_{lr}$ coincides with the event $E_{jl}\cap E_{jr}$ (ignoring measure zero events). Finally, notice the event $A$ is the intersection of $E_{jl}\cap E_{jr}$ and $E_{lr}$. Thus, $\mu(A)=\al$.
%{\color{red} Thus $\textbf{1}\{\lim \beta_n \cdot (x_j-x_l)+(\eta_x-\eta_y) \ge \ep_l -\ep_j \}=  \textbf{1}\{\lim \beta_n \cdot (x-z)+(\eta_x-\eta_z) \ge \epsilon_z-\ep_j \}= \textbf{1}\{\lim \beta_n \cdot (z-y)+(\eta_z-\eta_y) <\ep_l -\epsilon_z\}$ almost surely.  This implies that the probability of the event (A) is $\alpha$.}
In a similar way, we can show that $\mu(B)=1-\al$. Thus we have $\mu(A\cup B)=1$.
%\footnote{First, we can show that notice that $1-\al=\rho_{\al}(\{j,l\},l)= \mu (E_{lj})$. Since the event $E_{lj}$ is not compatible with the event $E_{jl}\cap E_{jr}$, $E_{lj}$ must coincide with the event $E_{rl}\cap E_{rj}$ (ignoring measure zero events). Finally notice that the event $B$ is the intersection of $E_{rl}\cap E_{rj}$ and $E_{lj}$, up to a measure zero events. Thus $\mu_{\Omega}(B)=1-\al$.}  
\end{myproof}

Define $A'\equiv \left\{\epsilon\left| U \epsilon_{jlr}> c
\right. \right\}$  and $B'\equiv  \left\{\epsilon\left| U \epsilon_{jlr} < c\right. \right\}$.

%\footnote{Consider two hyperplanes $H_1\equiv\{(\ep_j ,\ep_l,\ep_r) | \ep_j -\ep_l= \gamma^*_{lj}\}$ and $H_2\equiv \{(\ep_j ,\ep_l,\ep_r) | \ep_l-\ep_r= \gamma^*_{rl}\}$. Notice that the set $A'$ is the intersection of two half spaces $H_1^+ \cap H_2^+$; similarly, $B'= H_1^- \cap H_2^-$, where $H_1^+ \equiv\{(\ep_j ,\ep_l,\ep_r) | \ep_j -\ep_l> \gamma^*_{lj}\}$ and $H_1^- \equiv\{(\ep_j ,\ep_l,\ep_r) | \ep_j -\ep_l< \gamma^*_{lj}\}$. ($H_2^+$ and $H_2^-$ can be defined in a similar way.)}

\noindent\textbf{Step 4}: There exist $\ep^a\in A' \cap \supp \mu$ and $\ep^b\in B'\cap \supp \mu$ such that $\ep^{\lambda}\equiv\lambda \ep^a+(1-\lambda)\ep^b  \not\in A \cup B $ for some $\lambda\in [0,1]$.

\begin{myproof}
Since $\mu$ is a standard probability measure, $\supp \mu$ contains an open set in $\R^J$. Moreover, since $U$ has full row rank, there exist $\epsilon^a\in A'\cap \supp \mu$ and $\epsilon^b\in B'\cap \supp \mu$ such that three points $U\epsilon^a_{jlr}$, $U\epsilon^b_{jlr}$ and $c$ in $\R^2$ are not  collinear, where $\ep^a_{jlr}=\left(\ep^a_j, \ep^a_l,\ep^a_r \right)$ and $\ep^b_{jlr}=\left(\ep^b_j, \ep^b_l,\ep^b_r \right)$ are projections of $\ep^a$ and $\ep^b$, respectively.\footnote{That $\supp \mu$ contains an open set follows from the definition of standard probability measures. By Definition 2, $\mu$ is absolutely continuous with respect to Lebesgue measure and has convex support. A convex set with positive Lebesgue measure must have nonempty interior (Theorem 6.2 of \cite{rockafellar2015}), and therefore contains an open ball.}
%\footnote{\textcolor{red}{Step 3 and absolute continuity imply $\mu(A')>0$ and $\mu(B')>0$, since the relevant boundaries are contained in finitely many hyperplanes. Because $U$ has full row rank, the preimage of any line in $\mathbb R^2$ has Lebesgue measure zero. Hence $\varepsilon^a$ and $\varepsilon^b$ can be chosen so that $U\varepsilon^a_{jlr}$, $U\varepsilon^b_{jlr}$, and $c$ are not collinear.}}  

For all $\lambda\in [0,1]$, define  $\ep^{\lambda} \equiv\lambda \ep^a+(1-\lambda)\ep^b$  and let $\epsilon^\lambda_{jlr}$ denote the corresponding projection. %Note that $U\epsilon^\lambda_{jlr}\not =c$ for all $\lambda\in (0,1)$ since otherwise we have, for some $\lambda\in (0,1)$, $\lambda U\epsilon^a_{jlr}+ (1-\lambda) U\epsilon^b_{jlr}=c \to U\epsilon^a_{jlr}-c=\frac{1-\lambda}{\lambda}\left(U\epsilon^b_{jlr}-c\right)$, violating the fact that $U\epsilon^a_{jlr}$, $U\epsilon^b_{jlr}$ and $c$ are not collinear.
It remains to show that there exists \(\lambda^*\in(0,1)\) such that $
\epsilon^{\lambda^*}\notin A\cup B$. For each \(\lambda\in[0,1]\), define $y^\lambda
\equiv
U\epsilon^\lambda_{jlr}-c
=
\lambda\bigl(U\epsilon^a_{jlr}-c\bigr)
+
(1-\lambda)\bigl(U\epsilon^b_{jlr}-c\bigr)
\in \R^2$.
Write \(y^\lambda=(y^\lambda_1,y^\lambda_2)\). Since \(\epsilon^a\in A'\) and \(\epsilon^b\in B'\), we have
$y^1_1>0$, $y^1_2>0$, and $y^0_1<0$, $y^0_2<0$.  Hence, for each coordinate \(i=1,2\), the affine function $\lambda\mapsto y_i^\lambda$ is negative at \(\lambda=0\),  positive at \(\lambda=1\), and  $y_i^\lambda$ is strictly increasing and continuous with respect to $\lambda$ for all $i$. Therefore, there exists a unique \(\lambda_i\in(0,1)\) such that $
y_i^{\lambda_i}=0$. We claim that \(\lambda_1\neq \lambda_2\). Indeed, if \(\lambda_1=\lambda_2\), then $
y^{\lambda_1}=0$, or equivalently,
$U\epsilon^{\lambda_1}_{jlr}=c$.
This would imply that \(c\) lies on the line segment joining \(U\epsilon^a_{jlr}\) and \(U\epsilon^b_{jlr}\), contradicting that three points $U\epsilon^a_{jlr}$, $U\epsilon^b_{jlr}$ and $c$ in $\R^2$ are not  collinear. Thus \(\lambda_1\neq \lambda_2\). Without loss of generality, suppose \(\lambda_1<\lambda_2\).  Since $y^1_i>0>y^0_i$, the affine function $\lambda\mapsto y_i^\lambda$ is strictly increasing for all $i$. Choose any $\lambda^*\in(\lambda_1,\lambda_2)$. Since $y_i^\lambda$ is strictly increasing,  $y_1^{\lambda_1}=0$, and  $y_2^{\lambda_2}=0$, we have  $
y_1^{\lambda^*}>0$ and $y_2^{\lambda^*}<0$. Thus \(y^{\lambda^*}\) has mixed signs. Hence \(U\epsilon^{\lambda^*}_{jlr}-c\) is neither weakly positive componentwise nor weakly negative componentwise. By the definitions of \(A\) and \(B\), this implies $\epsilon^{\lambda^*}\notin A\cup B$. This proves the desired claim.
\end{myproof}

By Step 4, it follows from the convexity of $\supp \mu$ that  $\ep^{\la} \in \supp \mu \cap (A\cup B)^c $ for some $\lambda\in [0,1]$.  Moreover, $(A\cup B)^c$ is open. It follows from the definition of the support that there exists $r>0$  such that the ball centered at $\ep^{\la}$ with radius $r$, $B_{r}(\ep^{\la})$,  satisfies $\mu\left(B_{r}(\ep^{\la})\right)>0$ and $B_{r}\left(\ep^{\la}\right) \subset \left(A \cup B\right)^c$. This contradicts  $\mu(A \cup B)=1$.

\vspace{-0.2cm}

\subsection{Proof of Lemma \ref{lem:unrep}}

\vspace{-0.2cm}

Let $\mu \in \M$.  Let $\pi$ be a ranking that is not representable.  Fix $\alpha \in (0,1)$. Suppose by way of contradiction that $\rho^{\pi}_{\al}$ is approximated  arbitrarily well by a sequence of random-coefficient ARUMs with fixed effects; formally, $\rho^{\pi}_{\al} \in \cl \bigcup_{\eta \in \F} \mathcal{P}_{ra}( \eta|\mu)$. 

We first show the following claim:

\noindent\textbf{Claim:} {\it 
There exist two sequences of ARUMs with the same fixed effects that converge to $\rho^{\pi}$ and $\rho^{\pi^-}$, respectively.
}

\begin{myproof}
Let $\rho_n$  be the elements of  $\bigcup_{\eta \in \F} \mathcal{P}_{ra}( \eta|\mu)$ converging to $\rho_{\al}^{\pi}$. Let  $\eta_n$ be the sequence of the fixed effects associated with $\rho_n$. By Proposition \ref{cor:red} of the online appendix,  for each $n$, we have $ \mathcal{P}_{ra}( \eta_n|\mu)=co\mathcal{P}_{a}( \eta_n|\mu)$; thus  $\rho_n\in co \mathcal{P}_{a}( \eta_n|\mu)$ for all $n$. Therefore, we can write $\rho_n$ as $\sum_{i=1}^{M}\sigma_{ni}\rho_{ni}$, where $\rho_{ni} \in \mathcal{P}_{a}( \eta_n|\mu)$ and $M=dim\mathcal{P}_r+1$.

Since the sets of weights and random utility models are compact, we can extract converging subsequences such that for all $i$, $\sigma_{n_ki} \to \sigma^*_i$ and $\rho_{n_ki} \to \rho^*_i$ as $n_k \to \infty$. Thus, $\sum_{i}\sigma_{n_ki}\rho_{n_ki}\to \sum_{i}\sigma^*_i\rho^*_i = \rho^{\pi}_\alpha$. Moreover, we have $\rho_{n_ki}\in \P_{a}( \eta_{n_k}|\mu)$ for each $i$; and thus $\rho^*_i\in \cl \bigcup_{\eta \in \F}\P_{a}( \eta|\mu)$ for each $i$.

In the following, we will show that there exist some $i,j$ such that $\rho^*_i= \rho^{\pi}$ and $\rho^*_j= \rho^{\pi^-}$. This completes the proof since it implies $\rho_{n_ki} \to \rho^{\pi}$ and $\rho_{n_k
j} \to \rho^{\pi^-}$ as $n_{k}\to \infty$; and moreover both sequences can be chosen to share the same sequence of fixed effects $\eta_{n_k}$ because $\rho_{n_ki}, \rho_{n_kj}\in \P_{a}( \eta_{n_k}|\mu)$ for each $n_k$.

Suppose, by contradiction and without loss of generality, that  $\rho^*_i \neq \rho^{\pi}$ for any $i$.\footnote{The proof for the other case is exactly the same after changing $\rho^{\pi^-}$ to $\rho^{\pi}$ and $\rho^{\pi}$ to $\rho^{\pi^-}$.}

By Lemma \ref{lem:adjacent}, $\rho^{\pi}$ and $\rho^{\pi^{-}}$ are adjacent.  Let $(t,a)$ be as in Definition \ref{def:adjacency} with  the pair $(\rho^{\pi}, \rho^{\pi^{-}})$ of adjacent rankings. We will consider two cases. 

\case 1: $\rho^*_i \neq \rho^{\pi^-}$ for any $i$.
Thus, we have $\rho^*_i \not \in \{\rho^{\pi}, \rho^{\pi^-}\}$ for all $i$. 
Since $\rho^*_i\in  \cl \bigcup_{\eta \in \F}\P_{a}(\eta|\mu)$ for all $i$, Lemma \ref{lem:degenerate} implies $\rho^*_i\not \in \{\rho^{\pi}_{\al}| \al \in (0,1)\}$ for all $i$. It follows that $\rho^*_i\not \in \{\rho^{\pi}_{\al}| \al \in [0,1]\}$  for all $i$. Since $\pi$ and $\pi^{-}$ are adjacent vertices, by Definition \ref{def:adjacency},  we have $\rho^*_i \cdot t>a$ for all $i$. Thus, $\Big(\sum_{i} \sigma^*_i \rho^*_i\Big) \cdot t = \sum_{i} \sigma^*_i \rho^*_i \cdot t >a$. On the other hand by Definition \ref{def:adjacency}, $\Big(\sum_{i} \sigma^*_i \rho^*_i\Big) \cdot t=\rho_{\al}^{\pi}\cdot t= a$. This is a contradiction.

%By Lemma\ref{coro:fixed_effect1} (ii), $\rho^*_i \cdot t>a$ for all $i$. Thus, $\Big(\sum_{i} \mu^*_i \rho^*_i\Big) \cdot t = \sum_{i} \mu^*_i \rho^*_i \cdot t >a$. On the other hand by Definition \ref{def:adjacency}, $\Big(\sum_{i} \mu^*_i \rho^*_i\Big) \cdot t=\rho_{\al}^{\pi}\cdot t= a$. This is a contradiction.

\case 2: $\rho^*_i = \rho^{\pi^-}$ for some $i$. Define $I=\{i \in \{1,\dots, M+1\} | \rho^*_i = \rho^{\pi^-}\}$. First notice that there exists $i\in \{1,\dots, M\} \setminus I$ such that $\sigma^*_i>0$. (If such $i$ does not exist, then $\rho^{\pi^-}=\sum_{i} \sigma^*_i \rho^*_i= \rho_{\al}^{\pi}$, which contradicts  $\al \not \in \{0,1\}$.) Then, $a= \rho_{\al}^{\pi}\cdot t= \sum_{i} \sigma^*_i \rho^*_i \cdot t=\sum_{i \in I}\sigma^*_i\rho^{\pi^-}\cdot t+ \sum_{i \not \in I}\sigma^*_i \rho^*_i\cdot t>a$, where the strict inequality holds because $\rho^*_i\cdot t>a$ for all $i \not \in I$ by Definition \ref{def:adjacency}. This is a contradiction.
\end{myproof}

Then by the claim above, there exist a sequence $\{\eta_n\}$ of fixed effects and
two sequences $\{\rho_n\}$ and $\{\rho'_n\}$ in $\bigcup_{\eta \in \F} \mathcal{P}_{a}(\eta|\mu)$
such that (i) $\rho_{n} \to \rho^{\pi}$; (ii) $\rho_{n}' \to \rho^{\pi^{-}}$; and
(iii) $\rho_{n}, \rho_{n}' \in \mathcal{P}_{a}(\eta_n|\mu)$ for each $n$. (We write $n$
instead of $n_k$ for simplicity.) Let $\beta_n$ be the coefficient vector associated
with $\rho_n$ and $\beta'_n$ the coefficient vector associated with $\rho'_n$; by (iii),
both share the fixed effects $\eta_n$.

\begin{comment}
\textcolor{red}{INCORRECT: Given  (i), by exactly the same argument as Step 2 of Lemma 2, we can prove that there exists a large positive integer $N_1$ such that for any $n \ge N_1$, we have $\beta_n\cdot x_j+\eta_{n,(\{j,l\},j)} > \beta_n\cdot x_l+\eta_{n,(\{j,l\},l)}$ for any $j,l \in J$ such that $\pi(j)>\pi(l)$.  Similarly by  (ii), there exists a large $N_2$ such that for any $n \ge N_2$, we have $\beta'_n \cdot x_j+\eta_{n,(\{j,l\},j)} > \beta'_n \cdot x_l+\eta_{n,(\{j,l\},l)}$ for any $j,l \in J$ such that $\pi^{-}(j)>\pi^{-}(l)$. Fix any $j,l \in J$ such that $\pi(j)>\pi(l)$. Fix any number  $n_{jl}\ge \max\{N_1,N_2\}$. Then for any $n \ge n_{jl}$, we have $\beta_n\cdot x_j+\eta_{n,(\{j,l\},j)} > \beta_n\cdot x_l+\eta_{n,(\{j,l\},l)}$. Since $\pi^{-}(l)>\pi^{-}(j)$, we have $-\beta'_n\cdot x_j-\eta_{n,(\{j,l\},j)} > -\beta'_n\cdot  x_l-\eta_{n,(\{j,l\},l)}$. Summing the two inequalities, we have  $(\beta_n-\beta'_n)\cdot x_j > (\beta_n-\beta'_n)\cdot x_l$. Because the number of binary choice sets is finite, we can find $n^*> n_{jl}$ for any $j, l \in J$ with  $\pi(j)>\pi(l)$,  such that $(\beta_{n^*}-\beta'_{n^*})\cdot x_j > (\beta_{n^*}-\beta'_{n^*})\cdot x_l$. This contradicts the fact that $\pi$ is not representable.}
\end{comment}

For each pair $j,l$ with $\pi(j)>\pi(l)$, define $
    Z_{lj}:=\ep_l-\ep_j$,
    $\gamma_{n,jl}
    :=
    \beta_n\cdot(x_j-x_l)+\eta_{n,\{j,l\},j}-\eta_{n,\{j,l\},l}$, and $\gamma'_{n,lj}
    :=
    \beta'_n\cdot(x_l-x_j)+\eta_{n,\{j,l\},l}-\eta_{n,\{j,l\},j}$. The convergence to $\rho^\pi$ and $\rho^{\pi^-}$ gives $\mu(\{\ep| Z_{lj}<\gamma_{n,jl}\})\to 1$ and $\mu(\{\ep| Z_{lj}>-\gamma'_{n,lj}\})\to 1$. If $\gamma_{n,jl}+\gamma'_{n,lj}\leq 0$ for infinitely many $n$, then along that infinite subsequence the two sets $\{\ep| Z_{lj}<\gamma_{n,jl}\}$ and $\{\ep| Z_{lj}>-\gamma'_{n,lj}\}$ are disjoint, so their probabilities have sum at most one. This contradicts the fact that both probabilities converge to one. Therefore, for this pair $j,l$, $\gamma_{n,jl}+\gamma'_{n,lj}>0$ for all sufficiently large $n$. Since there are finitely many pairs, there exists a single $N$ such that for all $n\geq N$ and all $j,l$ with $\pi(j)>\pi(l)$, $\gamma_{n,jl}+\gamma'_{n,lj}>0$. The fixed effects cancel, so  $\gamma_{n,jl}+\gamma'_{n,lj}
    =
    (\beta_n-\beta'_n)\cdot(x_j-x_l)>0$.
Thus $\beta_n-\beta'_n$ represents $\pi$ for all sufficiently large $n$, contradicting the unrepresentability of $\pi$.

%\begin{singlespace}
\begin{spacing}{1}
 
\bibliographystyle{ecma}
\bibliography{bib_arxiv_0803_2026}

%\end{singlespace}
\end{spacing}

\clearpage 
\pagenumbering{arabic}% resets `page` counter to 1
\renewcommand*{\thepage}{A-\arabic{page}}
\appendix
\setcounter{table}{0}
\renewcommand{\thetable}{A.\arabic{table}}
\setcounter{figure}{0}
\renewcommand{\thefigure}{A.\arabic{figure}}

\begin{center}
\begin{Large}
\title{\textbf{For Online Publication}\\
Online Appendix for ``Approximating Choice Data by Discrete Choice Models"}
%}
\end{Large}
\end{center}
\begin{center}
\begin{large}
%\author{Haoge Chang, Yusuke Narita, and Kota Saito}
\end{large}
\end{center}
\section{Generalization}

\subsection{Menu-dependent Fixed Effects}\label{section:menu-independent_fixed}

In Section \ref{sec:menu-dependent}, we stated that our key results continue to hold with menu-dependent fixed effects under a mild locality restriction: for some triple $\{j,l,r\}\subseteq J$, the fixed effects $\eta_{(D,i)}$ are constant across all menus $D\subseteq\{j,l,r\}$ and for each $i\in D$. Here, we drop this restriction and allow $\eta_{(D,i)}$ to vary arbitrarily across menus $D$ and items $i\in D$.

By a  result in \cite{norets2013surjectivity}, fully menu-dependent fixed effects are rich enough to approximate any random utility model over the choice sets in $D$, even without random coefficients. To see this, recall that our goal is to choose $m(\beta)$ and $(\eta_{(D,i)})_{i\in D}$ for each menu $D\in\D$  so that
\begin{align*}
\int\mu\Bigl(\{\varepsilon:\beta\cdot x_j+\eta_{(D,j)}+\varepsilon_j>\beta\cdot x_\ell+\eta_{(D,\ell)}+\varepsilon_\ell,\ \forall\,\ell\in D\setminus\{j\}\}\Bigr)\,dm(\beta)
\end{align*}
can approximate $\rho(D,j)$ arbitrarily well for every menu $D\in \D$ and $j\in D$.
Observe that we may absorb the systematic term \(\beta\cdot x_j\) into the fixed effect by defining \(\tilde\eta_{(D,j)}=\eta_{(D,j)}+\beta\cdot x_j\). Let the random coefficient distribution place all its mass at $\beta = 0$. The model then reduces to
\begin{equation}\label{eq:norets}
\mu\Bigl(\{\varepsilon:\tilde{\eta}_{(D,j)}+\varepsilon_j>\tilde{\eta}_{(D,\ell)}+\varepsilon_\ell,\ \forall\,\ell\in D\setminus\{j\}\}\Bigr).\footnote{Fixed effects are menu dependent but not random: \(m\) is not a distribution over \(\eta\). If one were to allow \emph{random} fixed effects (i.e., draw \(\eta\) from a distribution), then by matching that distribution to a distribution over rankings one can trivially replicate any RUM. Our focus here is the nonrandom \(\eta\) case.} 
\end{equation}
Mathematically, given the \(|D|\) target probabilities \(\{\rho(D,j)\}_{j\in D}\), we seek \(|D|\) location parameters \(\{\tilde{\eta}_{(D,j)}\}_{j\in D}\) solving \eqref{eq:norets}. This is precisely a surjectivity problem for the multinomial choice mapping from location shifts to choice probabilities. \citet{norets2013surjectivity} shows that under a weak regularity condition that $\epsilon$ is continuously distributed with respect to the Lebesgue measure, the mapping contains the interiors of all probability simplices associated with $\D$ and hence can approximate any random utility model. %Hence we obtain:

As the results above show, allowing menu-dependent fixed effects can substantially increase the model’s expressive power. With fully flexible menu-dependent fixed effects, the model can even approximate choice behavior that is inconsistent with any RUM, thereby achieving perfect in-sample fit. In counterfactual applications, however, researchers typically use the estimated random-coefficient distribution together with fixed effects that are either set to a priori constants or inherited from observed choice sets. In this case, our main theorem implies that, unless the affine-independence condition is satisfied, there exist random utility models on the relevant counterfactual choice sets that cannot be approximated arbitrarily well. For example, suppose that we observe data only for the full menu  $J$, but wish to make counterfactual predictions for all binary and ternary choice sets. Then affine independence is required for perfect approximation. We note, however, that this requirement depends on the collection of choice sets for which counterfactual predictions are of interest. For instance, if we observe choice frequencies for all binary and ternary choice sets and only wish to conduct counterfactual predictions for the full menu, then convex independence is sufficient.

%Estimation then becomes challenging—overparameterization can weaken identification and invite overfitting, with the fixed effects \(\eta_{(D,j)}\) absorbing most of the variation and leaving little signal to identify the systematic utility component. In turn, counterfactual credibility may suffer. These concerns motivate the local constancy restriction introduced earlier.

\subsection{Restricting Possible Rankings}\label{sec:res_ranking}

\vspace{-0.2cm}

As highlighted in the main text, there are situations where a researcher deems certain rankings as unreasonable, leading them to constrain the possible set of rankings to $\hat{\Pi} \subset \Pi$. To accommodate this case, 
define $\P_r(\hat{\Pi})\equiv \co\{\rho^{\pi}| \pi \in \hat{\Pi}\}$. We call an element of $\P_r(\hat{\Pi})$ {\it a random utility model on $\hat{\Pi}$}. 

In our main results (Theorem 1 and Proposition 1), we considered the case where $\hat{\Pi}=\Pi$ for simplicity. In this section, we provide a necessary and sufficient condition for the approximation of the random utility models on $\hat{\Pi}$.

Since the set $\P_r(\hat{\Pi})$ of random utility models on $\hat{\Pi}$ is also a polytope, we can generalize  Lemma \ref{lem:appxQ} for $\P_r(\hat{\Pi})$ by simply changing $\Pi$ to $\hat{\Pi}$. That is, we have the following result:

\begin{lemma}\label{lem:gen_appxQ} Let $\hat{\Pi} \subset \Pi$. Let $\mu \in \M$ be any standard probability measure. Then $\P_r(\hat{\Pi}) \subset \cl \P_{ra}(0| \mu)$ if and only if $\rho^{\pi} \in \cl \P_a(0| \mu)$ for any $\pi \in \hat{\Pi}$. 
\end{lemma}
The proof is essentially the same as the proof of Lemma \ref{lem:appxQ}. This result, together with Lemma 2 and Proposition 2, implies the following result.

\begin{corollary}\label{coro:gen_restricted} Let $\hat{\Pi} \subset \Pi$. 
\bit
\item[(i)] Let $\mu \in \M$ be any standard probability measure. If every ranking $\pi \in \hat{\Pi}$ is representable in $\D$, then any random utility model on $\hat{\Pi}$ can be approximated arbitrarily well by a sequence of  random-coefficient ARUMs with the given distribution $\mu$. Moreover, the approximation can be done  without fixed effects (i.e., $\eta=0$). 
\item[(ii)] If some ranking $\pi \in \hat{\Pi}$ is not representable and $\pi^-  \in \hat{\Pi}$, then there exists a random utility model $\rho$ on $\hat{\Pi}$ such that for any standard probability measure $\mu \in \M$, $\rho$ cannot be approximated arbitrarily well by any sequence of  random-coefficient ARUMs with fixed effects and with the given distribution $\mu \in {\cal M}$. 
\eit
\end{corollary}

\begin{proof}
Statement (i) follows from Lemma 2-(1) and Lemma  \ref{lem:gen_appxQ}. Statement (ii) follows from Proposition~2. Indeed, take $\alpha\in(0,1)$ and set $\rho^*=\alpha\rho^\pi+(1-\alpha)\rho^{\pi^-}$. Since $\pi,\pi^-\in\hat\Pi$, we have $\rho^*\in\P_r(\hat\Pi)$. By Proposition~2, for every $\mu\in\M$, $\rho^*\notin\cl\bigcup_{\eta\in F}\P_{ra}(\eta\mid\mu)$.
\end{proof}

Corollary \ref{coro:gen_restricted} offers a testable condition for determining if the random-coefficient ARUMs can adequately approximate any random utility model on $\hat{\Pi} \subset \Pi$. If the researcher wishes to omit certain rankings from the analysis, this corollary may be more useful than  Theorem 1 and Proposition 1.

\section{More Comments on the Related Literature}\label{sec:related}

\subsection{\cite{mcfadden2000mixed}}

In the main body of the paper, we mentioned \cite{mcfadden2000mixed} in several sections; here we provide a summary.   

 \citet{mcfadden2000mixed} show that a mixed-logit model can approximate an arbitrary (nonparametric) continuous random-utility model arbitrarily well.  We also explain that this flexibility is achieved by using higher-order polynomials of arbitrarily high degrees, which corresponds to letting $K \to \infty$.

The key differences concern both the setup and the sharpness of the results. We provide a \emph{necessary and sufficient} characterization of when random-coefficient ARUMs can approximate arbitrary random-utility behavior, whereas \citet{mcfadden2000mixed} establish a \emph{sufficient} condition for approximation. 
    We obtain this sharper characterization by focusing on approximation for a \textit{fixed, finite set of alternatives}. In addition, we analyze approximation of choice probabilities \textit{across choice sets}. This cross-menu perspective with fixed alternatives is central for applications in assortment optimization and revenue management \citep{farias2009non,jagabathula2022nonparametric}. Our affine-independence condition is necessary and sufficient, applies beyond mixed-logit, and is tailored to environments with fixed finite alternatives. By contrast, \citet{mcfadden2000mixed} provide asymptotic approximation guarantees, focusing primarily on settings with continuous covariates, and they offer no comparable guidance on how to construct random-coefficient ARUMs to achieve approximation in the finite-alternatives, cross-menu setting.

\subsection{\cite{norets2013surjectivity}}\label{sec:onlineapp_norets}

As explained in the main body of the paper, \citet{norets2013surjectivity} show that for a \emph{single} multinomial choice problem with menu \(D\), under a mild regularity condition, one can always find an ARUM that matches the given dataset. In particular, given any interior probability vector \(\{\rho(D,j)\}_{j\in D}\), there exists a vector of fixed effect parameters \(\{\eta_{(D,j)}\}_{j\in D}\) such that the ARUM with the fixed effect vector but without random coefficients generates those probabilities on that menu.

The essential differences are the choice-set configuration under which the approximation problem is studied; and the approximation class they use. \citet{norets2013surjectivity} study the approximation power of ARUMs \textit{on a single choice set}. On the other hand, we study the approximation power of random-coefficient ARUMs with fixed effects \emph{across multiple choice sets simultaneously} given a fixed set of alternatives.

In Lemma~\ref{lem:degenerate}, we examine the ARUM without random coefficients across \emph{multiple} menus. While Norets--Takahashi’s surjectivity ensures that, menu by menu, any interior vector \(\{\rho(D,j)\}_{j\in D}\) can be matched by some \(\{\eta_{(D,j)}\}_{j\in D}\), Lemma~\ref{lem:degenerate} shows that such per–menu constructions \emph{cannot, in general, be made mutually consistent across menus}. 

\subsection{\cite{lu2021pure}}

 \cite{lu2021pure} differ from our analysis in three key respects. First, they work in a continuous characteristics space, as in \cite{mcfadden2000mixed}. Second, they restrict the approximation target to \emph{pure-characteristics} models (i.e., continuous random-utility models). Third, they focus on parametric mixed-logit specifications whose systematic utility is a polynomial of degree at most a given bound. 

In particular, \citet{lu2021pure} ask,  within the class of pure-characteristics models, which subclasses can be approximated by a degree-\(d\) mixed-logit? Their analysis is conducted under continuity restrictions that are natural in a continuous characteristics space but are not well defined on our finite alternative domain. 

By contrast, we ask for necessary and sufficient conditions under which any RUM can be approximated by a popular parametric class—random-coefficient ARUMs—when only a finite collection of menus is observed. Consequently, there is no exact finite-domain analogue of the question studied in \citet{lu2021pure}.

In what follows, we explain the similarities and differences in more detail, using a formal mathematical comparison. Both \cite{lu2021pure}  and our paper establish a baseline approximation inclusion that holds for every \(d\):
\begin{equation}\label{eq:uni-bridge}
\cl\,\P_{ml}(d)\ \supseteq\ \P_r(d),
\end{equation}
where \(\P_{ml}(d)\)  is the class of mixed-logit models whose systematic utility is a polynomial of degree at most \(d\) in characteristics (random coefficients allowed), and \(\P_r(d)\) the subclass of target models whose systematic utility likewise has degree at most \(d\).\footnote{For the definition $\P_r(d)$, see Definition 1 in \cite{lu2021pure}.}
Intuitively, the result states that a degree-\(d\) mixed-logit can approximate any degree-\(d\) target model (in closure). This statement is the bridge connecting the two frameworks.

\paragraph{Continuous domain (Lu--Saito).}
In the continuous setting of pure-characteristic models, Lu--Saito show that intersecting with the continuous class yields an equality:
\begin{equation}\label{eq:cont}
\P_{cr}\ \cap\ \cl\,\P_{ml}(d)\ =\ \P_r(d),
\end{equation}
where $\P_{cr}$ is the set of pure-characteristics models (i.e., continuous random-utility models). The result means that within the continuous pure-characteristics domain, the closure of degree-\(d\) mixed-logits exactly recovers the degree-\(d\) target class. Moreover, their “degree matching’’ results clarify that approximation quality hinges on the polynomial degree of the systematic utility.

\paragraph{Discrete domain (our paper).}
On a finite alternative set, continuity imposes no restriction. The affine independence of the set of characteristics plays a key role. Formally, let $x_j \in \R^K$ be characteristics and let $x_j^{(d)}$ be the vector of monomials of $x_j$ of degrees at most $d$. Define 
\begin{equation}\label{eq:Xd}
X(d)=\{x_j^{(d)}|j \in J\}.
\end{equation}
Our central discrete-domain statement is:
\begin{equation}\label{eq:discrete}
\P_r(d)\ =\ \P_r
\quad\Longleftrightarrow\quad
X(d)\ \text{is affinely independent},
\end{equation}
where \(\P_r\) is the set of RUMs on the finite domain. Combining \eqref{eq:discrete} with the common inclusion \eqref{eq:uni-bridge} yields
\[
\cl\,\P_{ml}(d)\ =\ \P_r
\quad\Longleftrightarrow\quad
X(d)\ \text{is affinely independent},
\]
which is the discrete analogue of \eqref{eq:cont} and a special case of our main theorem. In words: degree-\(d\) mixed-logits (indeed, random-coefficient ARUMs with degree-\(d\) systematic utility) approximate \emph{every} finite-domain RUM in closure if and only if  \(X(d)\) is affinely independent. 

\paragraph{Similarity and difference.}
The similarity is the shared approximation set-inclusion \eqref{eq:uni-bridge}: in both papers, degree-\(d\) mixed-logits are as expressive as degree-\(d\) targets, up to closure. The difference lies in the domain-specific condition needed to sharpen inclusion to equality. In the continuous case, equality follows after intersecting with \(\P_{cr}\) (continuity). In the discrete case, equality holds when \(X(d)\) defined in \eqref{eq:Xd} is affinely independent.\footnote{To see this, note that under  affine independence of $X(d)$, we have 
$\P_r \subset \P_r(d) \subset \cl\,\P_{ml}(d) \subset \P_r$, which implies $\P_r = \P_r(d) = \cl\,\P_{ml}(d)$.
Thus, continuity in Lu--Saito and affine independence of \(X(d)\) in our paper play analogous roles: each pinpoints when the degree-\(d\) mixed-logit family attains the full approximation frontier relevant to its domain.}

\section{Proof of Proposition \ref{theo:2} (ii)}\label{proof:proposition3}

To prove statement (ii)(a), suppose that ${x_j:j\in J}$ is not convex-independent. Then there exists an alternative $j^*\in J$ such that $x_{j^*}\in \co\{x_l|l\in J\setminus\{j^*\}\}$. Thus there are nonnegative weights $(\lambda_l)_{l\neq j^*}$ satisfying $\sum_{l\neq j^*}\lambda_l=1$ and $x_{j^*}=\sum_{l\neq j^*}\lambda_l x_l$. Let $L:={l\neq j^*: \lambda_l>0}$. Define $\rho^*\in\P_r$ to be the deterministic random utility model defined by  $\rho^*(J,j^*)=1$ and $\rho^*(J,l)=0$ for all $l\neq j^*$. We show that $\rho^*\notin\cl\P_{ra}(0\mid\mu)$ for every $\mu\in\M$ such that $\textbf{0} \in \text{int} (\supp \mu)$.

Fix any $\mu\in\M$ such that $\textbf{0} \in \text{int} (\supp \mu)$. There exists $r>0$ such that $B_r(\textbf{0})\subseteq \supp\mu$, where $B_r(\textbf{0})$ denotes the open Euclidean ball of radius $r$ centered at the zero vector. For each $l\in L$, define $O_l:={\epsilon\in B_r(\textbf{0}):\epsilon_l>\epsilon_{j^*}}$. The set $O_l$ is nonempty and open. Since $O_l\subseteq\supp\mu$, the definition of support implies $\mu(O_l)>0$. Because $L$ is finite, we can define $\delta:=\min_{l\in L}\mu(O_l)>0$.

Now fix any coefficient vector $\beta\in\R^K$. By the convex-combination representation of $x_{j^*}$, $\beta\cdot x_{j^*}=\sum_{l\in L}\lambda_l \beta\cdot x_l\leq\max_{l\in L}\beta\cdot x_l$. Hence there exists $l(\beta)\in L$ such that $\beta\cdot x_{l(\beta)}\geq \beta\cdot x_{j^*}$. On the event $O_{l(\beta)}$, we also have $\epsilon_{l(\beta)} > \epsilon_{j^*}$. Therefore, on $O_{l(\beta)}$, $\beta\cdot x_{l(\beta)}+\epsilon_{l(\beta)}>\beta\cdot x_{j^*}+\epsilon_{j^*}$. Thus alternative $j^*$ cannot be chosen from $J$ on the event $O_{l(\beta)}$. It follows that, for the ARUM with coefficient vector $\beta$ and no fixed effects, the probability that $j^*$ is chosen from $J$ is bounded by $1-\mu(O_{l(\beta)}) \leq  1-\delta$. This upper bound is uniform in $\beta$.

Now consider any random-coefficient ARUM without fixed effects and with mixing distribution $m$ over coefficient vectors. Its probability of choosing $j^*$ from $J$ is 
\[
\int \mu\bigl(\{\epsilon| \beta\cdot x_{j^*}+\epsilon_{j^*}>\beta\cdot x_l+\epsilon_l,\ \forall l\neq j^*\}\bigr)dm(\beta).
\]
By the uniform bound above, this integral is at most $1-\delta$. Hence every element $\rho\in\P_{ra}(0\mid\mu)$ satisfies $\rho(J,j^*)\leq 1-\delta$. Since every element of $\cl\P_{ra}(0\mid\mu)$ also satisfies $\rho(J,j^*)\leq 1-\delta$, $\rho^*\notin \cl\P_{ra}(0\mid\mu)$.

    Statement (ii) (b) can be proved as follows. Consider any stochastic choice function $\rho$ on $\{J\}$. Then there exists a sequence of stochastic choice functions $\{\rho_n\}$ such that $\rho_n \to \rho$ and $\rho_n(J,j)>0$ for any $j \in J$.  Fix $\mu \in \M$. By Corollary 1 of \citet{norets2013surjectivity},  $\rho_n$ can be represented as  ARUMs for each $n$.

\section{Proof of Lemma  \ref{lem:real_x}}\label{proof:lemma3}

\subsection{Details of Proof of  Step 1 of Lemma  \ref{lem:real_x}-(1)}

We use the following lemma:
 
\begin{lemma} \label{lem:theo_alt_Q} Let $A$ be an $r\times n$ real matrix, $B$ be an $l\times n$ real matrix, and $E$ be a real $m\times n$ matrix.  Exactly one of the following alternatives is true: (1) There is $u\in \Re^n$ such that $A u = 0$, $B u \geq 0$, $E u \gg 0$; (2) There is $\theta \in \Re^r$, $\eta \in \Re^l$, and $\lambda \in \Re^m$ such  that $\theta  A+  \eta B+ \la E = 0$, $\la>0$ and  $\eta \geq 0$, where $\gg 0$ means all entries are positive, $>0$ means all entries are nonnegative and positive for some entry, and $\ge$ means all entries are nonnegative. 
\end{lemma}

\noindent See Theorem 1.6.1 of \cite{stoer2012convexity} for the proof.

For simplicity of notation, let $J=\{1,2, \dots, |J|\}$. For any ranking $\pi \in \Pi$, by relabeling $J$ if necessary, we assume  that $\pi(i)>\pi({i+1})$ for all $i \le |J|-1$ without loss of generality. We  label the following condition  as Condition ($*$):  if $\la_1 x_1+\sum_{i=2}^{|J|-1} (\la_i-\la_{i-1}) x_i -\la_{|J|-1}x_{|J|}=0$ and $\la_i \ge 0$ for all $i \in \{1,\dots, |J|-1\}$, then $\la_i=0$ for all $i \in \{1,\dots, |J|-1\}$. 

\step 1: For each $\pi \in \Pi$, Condition ($*$) holds if and only if $\pi$ is representable.

\begin{proof}
Since $\D$ contains all binary sets, $\pi \in \Pi$ is representable if and only if there exists $\beta$ such that for any $j, l \in J$, $\pi(j)>\pi(l) \Leftrightarrow \beta \cdot x_j > \beta \cdot x_l$. Fix $\pi \in \Pi$.
\begin{eqnarray*}
\begin{array}{lll}
&\!\!\!\!\!\!\!\exists \beta\  \big[x_1 \cdot \beta > x_2 \cdot \beta > \dots> x_{|J|-1}\cdot \beta>  x_{|J|} \cdot \beta\big]\\
&\!\!\!\!\!\!\!\lra \exists \beta\  \big[(x_1-x_2) \cdot \beta>0,\dots, (x_{|J|-1}-x_{|J|})\cdot \beta>0\big]\lra \exists \beta\  [E  \beta \gg 0]\\
&\!\!\!\!\!\!\!\lra \not \exists \la \in \Re^{|J|-1}\ [\la>0,\ \la  E=0]\lra \not \exists \la \in \Re^{|J|-1}\ \big[\la>0, \sum_{i=1}^{|J|-1} \la_i (x_i-x_{i+1})=0 \big]\\
&\!\!\!\!\!\!\!\lra \not \exists \la \in \Re^{|J|-1}\ \big[ \la> 0, \la_1 x_1+\sum_{i=2}^{|J|-1} (\la_i-\la_{i-1}) x_i-\la_{|J|-1}x_{|J|}=0 \big]\\
&\!\!\!\!\!\!\!\lra \text{Condition} (*),
\end{array}
\end{eqnarray*}
where $\lambda\equiv(\lambda_1, ..., \lambda_{|J|-1})$ and the third equivalence is obtained by using Lemma \ref{lem:theo_alt_Q} with $A,B=0$ and $E\equiv(x_1-x_2;\dots;x_{|J|-1}-x_{|J|})\in \R^{(|J|-1)\times K}$. 
\end{proof}

\vspace{-0.2cm}

\subsection{Proof of Lemma \ref{lem:real_x}-(2)}

\vspace{-0.2cm}

For any $j \in J$, $x_j \not \in \co(\{x_l| l \in J \setminus \{j\}\})$ $\Leftrightarrow$ $x_j$ is an extreme point of $\co(\{x_l| l \in J \})$ $\Leftrightarrow$ $x_j$ is an exposed point of $\co(\{x_l| l \in J \})$ $\Leftrightarrow  \exists \beta \ \forall l\in J \setminus \{j\}[\beta \cdot x_j  > \beta \cdot x_l]$$\Leftrightarrow$ all rankings $\pi$ on $\{J\}$, whose best alternative is $j$, are representable. The first and third equivalences are by the definitions of extreme points and exposed points, respectively, while the second equivalence is by the fact that $\co(\{x_l| l \in J \})$ is a polytope.

\section{Proof of Lemma  \ref{lem:adjacent}}

For each positive integer $n$, define $J_n=\{1,\dots, n\}$.  We prove the lemma by an induction on $n$.  Let $\Pi_n$ be the set of all rankings on $J_n$.  

Let $\pi_n$ be a ranking over $J_n$ such that $\pi_{n}(i)>\pi_{n}({i+1})$ for any $1 \le i \le n-1$. We will prove that $\rho^{\pi_n}$ and $\rho^{\pi^{-}_n}$ are adjacent. Since the labeling of the alternatives is  arbitrary, this proves the lemma.

\textbf{Induction Base:} Let us consider the case of $n = 3$. (Remember that all binary and ternary choice sets are in $\mathcal{D}_n$. The cases for $n=1$ and $n=2$ are trivial.) WLOG we consider $\pi(1)>\pi(2)>\pi(3)$ and its reverse. For $b > a > 0,$ let  $t_3(\{1, 2\}, 1) = a$, $t_3(\{2, 3\}, 2) = -b$, $t_3(\{1, 3\}, 1) = b - a$, and $t_3(\{1,2, 3\}, 2) = a + b$.  For all other $(D,j)\in \D \times J$, $t_3(D,j)=0$. A direct calculation shows that  $\rho^{\pi_3} \cdot t_3 = \rho^{\pi_3^{-}} \cdot t_3= 0,$ and $\rho^{\sigma_3} \cdot t_3 > 0$ for any $\sigma_3 \in \Pi_3\setminus \{\pi_3, \pi_3^{-}\}$.

Assume that $n\ge 4$. For each $i$ such that $3\le i\le n-1$, we define a set of sets $\D_i \subset 2^{J_i}\setminus \emptyset$ such  that (i) $\D_i \subset \D_{i+1}$ and $\D_n= \D$; (ii) for each $i$,  $\{j,l\} \in \D_i$ and $\{j,l,r\} \in \D_i$ for any $j, l, r \in J_i$.

\textbf{Induction Step:} WLOG, let $\pi_{n-1}$ be the ranking over $J_{n-1}$ such that $\pi_{n-1}(i)>\pi_{n-1}({i+1})$ for any $i \le n-2$. By the induction hypothesis there exists $t_{n-1} \in \mathbf{R}^{|\D_{n-1}| \times |J_{n-1}|}$ such that $\rho^{\pi_{n-1}} \cdot t_{n-1} = \rho^{\pi_{n-1}^{-}} \cdot t_{n-1} = 0,$ and $\rho^{\sigma_{n-1}} \cdot t_{n-1} > 0$ for $\sigma_{n-1} \in \Pi_{n-1} \setminus \{ \pi_{n-1}, \pi_{n-1}^{-}\}$.  Choose a positive number $\varepsilon_{n-1}$ such that  $0<\varepsilon_{n-1}<  \min_{\sigma_{n-1} \in \Pi_{n-1} \setminus\{ \pi_{n-1}, \pi_{n-1}^{-}\}} \rho^{\sigma_{n-1}} \cdot t_{n-1}$.  We define $t_{n} \in \mathbf{R}^{|\D_{n}| \times |J_{n}|}$ as follows: for each $(D, j) \in \D_{n} \times J_{n}$,
  \begin{eqnarray*}
    t_{n}(D, j)
    =
    \left\{
    \begin{array}{lll}
     t_{n-1}(D, j) &\text{ if } (D, j) \in (\D_{n-1} \times J_{n-1}) \setminus (\{1, 2\}, 1), \\
     t_{n-1}(D, j) + \varepsilon_{n-1} &\text{ if } (D,j) = (\{1, 2\},1),\\
     - \varepsilon_{n-1} / (n-1) &\text{ if } (D,j) = (\{i, n\},i) \text{ for some } i \in \{1, \cdots, n-1\}, \\
     2 \varepsilon_{n-1} &\text{ if } (D,j) = (\{{n - 2}, {n-1}, n \}, {n-1}),\\
     0  &\text{ otherwise.}
    \end{array}
    \right.
  \end{eqnarray*}

  It is clear that $\rho^{\pi_{n}} \cdot t_{n} = \rho^{\pi_{n}^{-}} \cdot t_{n} = 0$. Fix $\sigma_n \in \Pi_n \setminus \{\pi_n,\pi_n^-\}$. Let $j\in\{1,\dots,n\}$ be such that $n$ is the $j$th best alternative under $\sigma_n$, so there are exactly $j-1$ alternatives strictly preferred to $n$ in $\sigma_n$. Let $\sigma_{n-1} \in \Pi_{n-1}$ be the restriction of $\sigma_n$ to $J_{n-1}$.

First notice that by the definition of $t_n$ and $\rho^{\sigma_{n-1}}=\rho^{\sigma_{n}}$ on $\{1,2\}$, $\rho^{\sigma_n} \cdot t_{n}=\rho^{\sigma_{n-1}} \cdot t_{n-1}+\varepsilon_{n-1} \rho^{\sigma_{n-1}}(\{1, 2\}, 1)-\dfrac{\varepsilon_{n-1}}{n-1} (j - 1)+2 \varepsilon_{n-1} \rho^{\sigma_n}(\{{n - 2}, {n-1}, n \}, {n-1})$, where the second term of the right hand side follows since in $\sigma_n$, there are $j-1$ elements that are better than $n$. 

\case 1: $\sigma_{n-1} = \pi_{n-1}$. Note that $\rho^{\sigma_{n-1}}(\{1, 2\}, 1)=\rho^{\pi_{n-1}}(\{1, 2\}, 1)=1$. Note also that $\rho^{\sigma_n}(\{{n - 2}, {n-1}, n \}, {n-1})=0$. Thus, $\rho^{\sigma_n} \cdot   t_{n}= 0+ \varepsilon_{n-1} - \dfrac{\varepsilon_{n-1}}{n-1} (j - 1)+0 > 0$, where the last inequality holds because $j< n$. Note that if $j=n$, then $\sigma_{n-1} = \pi_{n-1}$ implies that $\sigma_{n}= \pi_{n}$.

\case 2: $\sigma_{n-1} = \pi^{-}_{n-1}$. Note that $\rho^{\sigma_{n-1}} \cdot   t_{n-1}=\rho^{\pi^{-}_{n-1}} \cdot   t_{n-1}=0$ and $\rho^{\sigma_{n-1}}(\{1, 2\}, 1)=\rho^{\pi^{-}_{n-1}}(\{1, 2\}, 1)=0$. Note also that $\rho^{\sigma_n}(\{{n - 2}, {n-1}, n \}, {n-1})=1$ because $n-1$ is the best element in $\sigma_n$ (except the case in which $n$ is the best element in $\sigma_n$ and $\sigma_{n-1}=\pi^{-}_{n-1}$, then the ranking in $\sigma_n$ coincides with $\pi^{-}_{n}$). Thus, $\rho^{\sigma_n} \cdot   t_{n}=0+0 - \dfrac{\varepsilon_{n-1} }{n-1}(j - 1) + 2 \varepsilon_{n-1} > 0$.

\case 3: $\sigma_{n-1} \not \in \{ \pi_{n-1}, \pi^{-}_{n-1}\}$. Thus, $\rho^{\sigma_n} \cdot   t_{n}> \varepsilon_{n-1} - \dfrac{\varepsilon_{n-1} }{n-1}(j - 1) \geq 0$, where the first inequality holds by $\rho^{\sigma_{n-1}} \cdot t_{n-1}>\varepsilon_{n-1}$ and the second inequality holds by $j\le n$. 
\section{Worst-case Approximation Error is Achieved at a Deterministic Preference}\label{section:worst_case}

We consider the worst-case approximation error
\begin{equation}
    \sup_{\hat{\rho}\in \mathcal{P}_r}\inf_{\rho \in \mathcal{P}_{ra}(\eta|\mu)} d(\hat{\rho},\rho),
\end{equation}
where $d(\cdot,\cdot)$ is the distance function defined in Section \ref{sec:measuring}. We show that the worst-case approximation error is attained at one of the deterministic preferences.

For each $\pi\in \Pi$, let $\rho^\pi$ denote the stochastic choice function associated with the deterministic preference ranking $\pi$.

\begin{proposition}
For any $\mu \in \M$ and $\eta \in \F$,
\begin{equation*}
  \sup_{\hat{\rho}\in \mathcal{P}_r}\inf_{\rho \in \mathcal{P}_{ra}(\eta|\mu)} d(\hat{\rho},\rho)
  \;=\;
  \sup_{\hat{\rho}\in \{\rho^\pi\}_{\pi\in\Pi}}\inf_{\rho \in \mathcal{P}_{ra}(\eta|\mu)} d(\hat{\rho},\rho).
\end{equation*}
\end{proposition}

\begin{proof}
Given an $\epsilon >0$, pick $\hat{\rho}_\epsilon \in \mathcal{P}_r$ such that
\begin{equation}\label{eqn:15}
    \inf_{\rho \in \mathcal{P}_{ra}(\eta|\mu)} d(\hat{\rho}_\epsilon,\rho)
    \;\ge\;
    \sup_{\hat{\rho}\in \mathcal{P}_r}\inf_{\rho \in \mathcal{P}_{ra}(\eta|\mu)} d(\hat{\rho},\rho)
    - \epsilon.
\end{equation}
Since $\mathcal{P}_r$ is the convex hull of $\{\rho^\pi\}_{\pi\in\Pi}$, we can write the vertex representation of $\hat{\rho}_\epsilon$ as
\[
  \hat{\rho}_\epsilon
  = \sum_{\pi \in \Pi} \lambda^\pi_\epsilon \rho^\pi,
\]
where the weights satisfy $\sum_{\pi \in \Pi} \lambda^\pi_\epsilon = 1$ and $\lambda^\pi_\epsilon \ge 0$ for all $\pi \in \Pi$.

For a sufficiently small $\delta>0$ and for each $\pi\in \Pi$, let $\rho^\pi_\delta \in \mathcal{P}_{ra}(\eta|\mu)$ be such that
\begin{equation}\label{eqn:16}
    d(\rho^\pi,\rho^\pi_\delta)
    \;\leq\;
    \inf_{\rho \in \mathcal{P}_{ra}(\eta|\mu)} d(\rho^\pi,\rho)
    +\delta.
\end{equation}

Combining this with \eqref{eqn:15} and \eqref{eqn:16} yields
\begin{align*}
 \sup_{\hat{\rho}\in \mathcal{P}_r}\inf_{\rho \in \mathcal{P}_{ra}(\eta|\mu)} d(\hat{\rho},\rho)- \epsilon
 &\leq \inf_{\rho \in \mathcal{P}_{ra}(\eta|\mu)} d(\hat{\rho}_\epsilon,\rho)  &(\because \eqref{eqn:15})\\
 &\leq
      d\Bigl(\hat{\rho}_\epsilon,\sum_{\pi\in \Pi}\lambda^\pi_\epsilon \rho^\pi_\delta\Bigr)\\
      &= d\Bigl(\sum_{\pi\in \Pi}\lambda^\pi_\epsilon \rho^\pi,\sum_{\pi\in \Pi}\lambda^\pi_\epsilon \rho^\pi_\delta\Bigr) \\
 &\leq \sum_{\pi\in \Pi}\lambda^\pi_\epsilon d\left(\rho^\pi,\rho^\pi_\delta\right)  &(\because \text{by convexity of }d )\\
 &\leq \sum_{\pi\in \Pi}\lambda^\pi_\epsilon
 \Bigl(\inf_{\rho \in \mathcal{P}_{ra}(\eta|\mu)} d(\rho^\pi,\rho)+\delta\Bigr)  &(\because \eqref{eqn:16}) \\
 &\leq \sup_{\hat{\rho}\in \{\rho^\pi\}_{\pi\in \Pi}}
 \inf_{\rho \in \mathcal{P}_{ra}(\eta|\mu)}d(\hat{\rho},\rho)+\delta.
\end{align*}
Since $\epsilon$ and $\delta$ are arbitrary, letting $\epsilon \to 0$ and $\delta \to 0$ yields the desired result.

\end{proof}

\section{Algorithms to Compute Approximation Errors}\label{sec:alg}

\subsection{EM Algorithm}\label{appendix:em}

To compute approximation errors in Table \ref{tbl:apx_error1}, we fit a finite-mixture logit model to each deterministic ranking by the method of maximum likelihood.
The data input is the stochastic choice function $\widehat{\rho}(D, j)$ and characteristics of each alternative $j$. We choose the number of mixtures, $M=18$, according to the theoretical upper bound suggested by Proposition \ref{cor:red} in Section  \ref{sec:simplify}. 
Given the number of mixtures, the model has two sets of parameters: (1) mixture weights $\{\lambda_i\}_{i=1}^M$ and (2) coefficients for each mixture $\{\beta_i\}_{i=1}^M$. 
The log-likelihood function of a finite mixture model with $M$ mixtures is
\[
\mathcal{L}(\{\lambda_i\}_{i=1}^M,\{\beta_i\}_{i=1}^M)\equiv \displaystyle \sum_{D \in \D} \sum_{j \in D}\widehat{\rho}(D, j)\log \sum^{M}_{i=1}\lambda_i\dfrac{\exp(\beta_i\cdot x_j)}{\sum_{l\in D}\exp(\beta_i\cdot x_l)}.
\]
%where $\mathcal{D}$ is the collection of choice sets. 
We estimate the parameters by the EM algorithm \citep{dempster1977maximum, train2009discrete}. We implement the algorithm according to Chapter 14 in \cite{train2009discrete}. 
We terminate the algorithm when the change of the implied $l_2$ distance between the estimated choice probability and the target choice probability becomes smaller than $\frac{1}{10^6}$ between two successive iterations.

Our use of the maximum likelihood method with the EM algorithm is motivated by the following observation: if the affine-independence condition is satisfied and the target choice probability is an interior random utility model $\hat{\rho}\in \rint \mathcal{P}_r$, then the model that maximizes the likelihood will yield a perfect fit to the target probability.  The maximum likelihood method  therefore  minimizes the approximation error metric in (\ref{eq:dist.1}).

To see this, notice that under the affine-independence condition, Proposition \ref{cor:red} and \ref{lem:dim_P_r} in Section \ref{sec:simplify} of the online appendix imply that any interior random utility model can be represented by a finite mixture of logit models with $M=\sum_{D \in {\mathcal{D}}} (|D|-1)+1$ mixtures. That is, if $M=\sum_{D \in {\mathcal{D}}} (|D|-1)+1$, there exists a set of parameters  $\{(\beta_i^*,\lambda_i^*)\}_{i=1}^M$ such that $\sum^{M}_{i=1}\lambda^*_i\dfrac{\exp(\beta_i^*\cdot x_j)}{\sum_{l\in D}\exp(\beta_i^*\cdot x_l)}=\hat{\rho}(D,j)$ for any $D\in\mathcal{D}$, $j\in D$.  The choice probability generated by this set of parameters yields a perfect fit of the target probability.  Hence this set of parameters maximizes the likelihood.\footnote{Recall that for any other stochastic choice function $\rho$, the likelihood is $    \sum_{D\in\mathcal{D}}\sum_{j\in D}\hat{\rho}(D,j)\log({\rho}(D,j))$. Note that $\hat{\rho}$ maximizes the likelihood since  
 \beq
 \begin{array}{lll}
        & \sum_{D\in\mathcal{D}}\sum_{j\in D}\hat{\rho}(D,j)(\log(\hat{\rho}(D,j))-\log({\rho}(D,j)))= \sum_{D\in\mathcal{D}}\sum_{j\in D}\hat{\rho}(D,j)\log\frac{\hat{\rho}(D,j)}{{\rho}(D,j)}\\
    =& - \sum_{D\in\mathcal{D}}\sum_{j\in D}\hat{\rho}(D,j)\log\frac{{\rho}(D,j)}{\hat{\rho}(D,j)}
    \geq  -\sum_{D\in\mathcal{D}}\sum_{j\in D}\hat{\rho}(D,j)(\frac{{\rho}(D,j)}{\hat{\rho}(D,j)}-1)\\
    =&  -\sum_{D\in\mathcal{D}}\sum_{j\in D}\rho(D,j) + \sum_{D\in\mathcal{D}}\sum_{j\in D}\hat{\rho}(D,j)=-|\mathcal{D}|+|\mathcal{D}|=0,
\end{array}
\eeq
where we use the fact $-\log(x)\geq -(x-1)$ when $x\geq 0$.} 

\subsection{Greedy Algorithm}\label{sec:greedy}
It is known that the EM algorithm may not converge to the global optimum. To alleviate this concern, we propose a second greedy algorithm inspired by \cite{barron2008approximation}. This algorithm solves a sequence of optimization problems and converges to the global optimal solution provided each inner problem is solved correctly. The structure of the random-coefficient RUMs is important for the proof. 

 The algorithm takes a stochastic choice function $\hat{\rho}$ and a fixed effects vector $\eta$ as input and returns a solution to (\ref{eq:dist.1}). The algorithm is iterative: each step seeks to optimize based on the results of previous steps:

\begin{itemize}
    \item \textbf{Step 1}: Given $\hat{\rho}$, choose $\rho^1$ such that $\rho^1=\arg\inf_{\rho\in \cl \mathcal{P}_{a}(\eta|\mu)} ||\hat{\rho}-\rho||^2$.
    \item \textbf{Step n}, $n\geq 2$: 
    \begin{itemize}
        \item Consider a set of grids $\alpha_n=\{\frac{2}{i+1}\}_{i=1}^n$.
        \item Find $(\alpha_n^*,\rho_n^*)=\arg
            \inf_{(\alpha,\rho)\in \alpha_n\times \cl \mathcal{P}_a(\eta|\mu)  }||\hat{\rho}-(1-\alpha)\rho^{n-1}-\alpha \rho||^2$.
        \item Define $\rho^n=(1-\alpha_n^*)\rho^{n-1}+\alpha_n^*\rho_n^*$ and let $\rho^{out}=\rho^n$.
    \end{itemize}
    \item Stop if a terminating criterion is reached. 
    %Note that the diameter (maximum distance between two points in the set) of the random utility polytope in our case is $\sqrt{22}$ (the square root of 11 choice sets $\times$ 2 per choice set), achieved by a pair of a degenerate ranking and its inverse ranking. 
    \item Return $\rho^{out}$ at the final step.
\end{itemize}

\noindent The next proposition shows that the algorithm will converge for our problems in Section \ref{sec:measuring}, provided each inner problem is solved correctly. Define $d(\rho,\hat{\rho})$ as in (\ref{eq:dist.1}). 
\begin{proposition}\label{prop:algorithm}
Let $\hat{\rho} \in \P$ be any stochastic choice function, $\eta \in \Re^{|J|}$, and $\mu \in \M$.  Define $d^*=\inf_{\rho \in \P_{ra}( \eta|\mu)} d(\rho, \hat{\rho})$. Let $n$ denote the number of steps and $\rho^n$ denote the output after the completion of the $n$-th step of the algorithm. Then, 
\begin{equation}\label{prop:greedy}
    d(\rho^n,\hat{\rho})-d^*\leq \sqrt{\frac{8}{n+1}}.
\end{equation}
\end{proposition}

For our implementation, the terminating criterion is when the number of steps taken reaches 1000. With 1000 steps, (\ref{prop:greedy}) implies the margin of error is within 0.09.\footnote{This is calculated by noting $\sqrt{8/1001}\approx 0.09$.}  When we approximate $\hat{\rho}$ without fixed effects, we let $\eta=0$. When we approximate  $\hat{\rho}$ with fixed effects, we couple the algorithm with a grid of fixed effects to search for the minimum.

\begin{myproof}
Since the set $\text{cl}\text{co} \P_a( \eta|\mu)$ is compact and convex, $\rho^*=\arg\inf_{\rho\in \text{cl}\, \text{co} \P_a( \eta|\mu)} d(\rho,\hat{\rho})$ $=\arg\inf_{\rho\in \text{cl}\, \text{co} \P_a( \eta|\mu)} ||\rho-\hat{\rho}||_2^2$ exists and it can be written as a convex combination of elements of $\text{cl} \P_a( \eta|\mu)$. By Caratheodory's theorem, it can be written as $\rho^* = \sum_{i=1}^M \lambda_i \rho_i$, where $\lambda_i\geq 0$, $\sum_i\lambda_i=1$ and $\rho_i\in \text{cl}\P_a( \eta|\mu)$, $M=\dim \P_r+1$.  For each step $n$, define $E_n=\|\hat{\rho}-\rho^{n}\|^2 - \|\hat{\rho}-\rho^{*}\|^2$, where $\|\cdot\|$ denotes the Euclidean $l_2$ norm.  For each step $n$, let $\al^*_n$ and $\rho_n^*$ be the minimizers over the grids $\{\al_n\}$ and $\cl \P_{a}(\eta|\mu)$, respectively.  Define $C=\sum_i \lambda_i \|\rho_i-\rho^*\|^2$ and $T =\max \{2E_1,4C\}$. $E_1$ and $C$ can be upper-bounded by the squared diameter of the random utility polytope. For example, $C\leq \sum_i \lambda_i||\rho_i-\rho^*||^2 \leq \sup_{\rho,\hat{\rho}\in\P_r}||\rho-\hat{\rho}||_2^2 = 2\times \text{the number of choice sets}$. The extremum is achieved by selecting $\rho$ to be a degenerate ranking and $\hat{\rho}$ its reverse ranking. Similarly $E_1$ can be upper bounded by $2\|\hat{\rho}\|^2 +2\|\rho^n\|^2\leq 4|\mathcal{D}|$. Hence we can choose $T=8|\mathcal{D}|$. 
\label{footnote:calculate T} Notice this implies $T=8\times (\text{the number of choice sets})$.

For each step $n$, let $\al_n=\frac{2}{n+1}$. Then, we have the following:
\beq
\begin{array}{llll}
%&=&\|\hat{\rho}-\rho^{n}\|^2 - \|\hat{\rho}-\rho^*\|^2\\
E_n&=& \|\hat{\rho}-(1-\alpha^*_n)\rho^{n-1}-\alpha^*_n\rho_n^*\|^2- \|\hat{\rho}-\rho^*\|^2\\
%&\le & \|\hat{\rho}-(1-\alpha_n)\rho^{n-1}-\alpha_n (\sum_i \la_i \rho_i)\|^2- \|\hat{\rho}-\rho^*\|^2\\
 &\leq& \sum_{i}\lambda_i \|\hat{\rho}-(1-\alpha_n)\rho^{n-1}-\alpha_n \rho_i\|^2- \|\hat{\rho}-\rho^*\|^2\\
 &=& \sum_i \lambda_i \{(1-\alpha_n)^2 \|\hat{\rho}-\rho^{n-1}\|^2 +2\alpha_n(1-\alpha_n)\big((\hat{\rho}-\rho^{n-1})\cdot(\hat{\rho}-\rho_i)\big) + \alpha_n^2 \|\hat{\rho}-\rho_i\|^2\}\\
 &&- \|\hat{\rho}-\rho^*\|^2 \\
% &=& (1-\alpha_n)^2 \|\hat{\rho}-\rho_{n-1}\|^2 + 2\alpha_n(1-\alpha_n)\sum_i \lambda_i  \big((\hat{\rho}-\rho_{n-1})\cdot(\hat{\rho}-\rho_i)\big) +  \alpha_n^2 \sum_i \lambda_i\|\hat{\rho}-\rho_i\|^2 - \|\hat{\rho}-\rho_*\|^2 \\
 &=& (1-\alpha_n)^2 \|\hat{\rho}-\rho^{n-1}\|^2 + 2\alpha_n(1-\alpha_n) \big((\hat{\rho}-\rho^{n-1})\cdot(\hat{\rho}-\sum_i \lambda_i  \rho_i)\big)\\
 &&+  \alpha_n^2 \sum_i \lambda_i\|\hat{\rho}-\rho_i\|^2 - \|\hat{\rho}-\rho_*\|^2 \\
 &\leq& (1-\alpha_n)^2\|\hat{\rho}-\rho^{n-1}\|^2 + \alpha_n(1-\alpha_n)(\|\hat{\rho}-\rho^{n-1}\|^2 + \|\hat{\rho}-\rho^*\|^2) - \|\hat{\rho}-\rho^*\|^2 \\
 &&+ \alpha^2_n \sum_i \lambda_i \|\hat{\rho}-\rho_i\|^2 \\
 &\leq& (1-\alpha_n)^2\|\hat{\rho}-\rho^{n-1}\|^2 + \alpha_n(1-\alpha_n)(\|\hat{\rho}-\rho^{n-1}\|^2 + \|\hat{\rho}-\rho^*\|^2) - \|\hat{\rho}-\rho^*\|^2 \\
 &&+ \alpha^2_n \sum_i \lambda_i \|\rho_i-\rho^*\|^2 + \alpha^2_n \|\rho^*-\hat{\rho}\|^2 \\
 &=& (1-\alpha_n) \|\hat{\rho}-\rho^{n-1}\|^2- (1-\alpha_n)\|\hat{\rho}-\rho^*\|^2 + \alpha^2_n \sum_i \lambda_i \|\rho_i-\rho^*\|^2 \\
 &=& (1-\alpha_n) E_{n-1} + \alpha^2_n \sum_i \lambda_i \|\rho_i-\rho^*\|^2. 
\end{array}
\eeq
Thus we have
\begin{equation}\label{eq:E_n}
    E_{n}\leq (1-\alpha_n)E_{n-1} + C \alpha^2_n.
\end{equation}
In the following, we will show $E_n\le \frac{T}{n+1}$ for each $n$. We prove this by induction.  The inequality holds with $n=1$. Fix $n$. Suppose $E_{n-1}\leq \frac{T}{n}$. By substituting $\al_n= \frac{2}{n+1}$  to (\ref{eq:E_n}), we have (i): $E_n \leq \frac{T}{n+1}$.\footnote{$E_n \leq \dfrac{n-1}{n+1}\dfrac{T}{n} + C\dfrac{4}{(n+1)^2}= \dfrac{(n^2-1)T+4Cn}{(n+1)^2n} \leq \dfrac{(n^2-1)T+Tn}{(n+1)^2n}\leq \dfrac{n^2T+Tn}{(n+1)^2n}= \dfrac{Tn(n+1)}{(n+1)^2n}= \dfrac{T}{n+1}$.} Let $d^n=d(\hat{\rho},\rho^n)$ and $d^*=d(\hat{\rho},\rho^*)$. Since $E_n= |\mathcal{D}|((d^n)^2-(d^*)^{2})$, we have (ii): $(d^n)^2-(d^*)^2\le \frac{T'}{n+1}$, where $T'=\frac{T}{|\mathcal{D}|}=8$. Then  we have $(d^n-d^*)^2  \leq (d^n-d^*)(d^n+d^*)= (d^n)^2-(d^*)^2 \leq \frac{8}{n+1}$, where we use the fact that $d^n\geq d^*$ and $d^*\geq 0$. This implies $d^n-d^*\leq \sqrt{\frac{8}{n+1}}$.
\end{myproof}

\subsection{Calculating the Maximal Substitution in (\ref{eq:dist.2})}\label{appendix:maximal_sub}

To calculate the maximal substitution (\ref{eq:dist.2}), we consider the problem 
\begin{equation}\label{eq:dist.3}    \inf_{\rho \in \mathcal{P}_{ra}(0|\mu)} \Bigg(\sum_{r\in J\setminus \{j,l\}}\rho(J\setminus\{j\},r)+\rho(J,l)\Bigg)^2,
\end{equation}
which can be readily solved by the greedy algorithm. Taking 1 minus the square root of the minimized value in (\ref{eq:dist.3}) gives the solution to the problem in  (\ref{eq:dist.2}). To see this, notice  
\[
\sup_{\rho \in \mathcal{P}_{ra}(0|\mu)} \left(\rho(J\setminus \{j\},l)-\rho(J,l)\right)=1-\inf_{\rho \in \mathcal{P}_{ra}(0|\mu)}\Bigg(\sum_{r\in J\setminus \{j,l\}}\rho(J\setminus\{j\},r)+\rho(J,l)\Bigg).
\]
Because $\sum_{r\in J\setminus \{j,l\}}\rho(J\setminus\{j\},r)+\rho(J,l)$ is nonnegative, minimizer(s) of $\sum_{r\in J\setminus \{j,l\}}\rho(J\setminus\{j\},r)+\rho(J,l)$ is the same as the minimizer(s) of the problem when the criterion is squared. 

We modify the greedy algorithm for maximal substitution as follows:

\begin{itemize}
\item \textbf{Step 1}: Choose $\rho^1$ as a solution of $\inf_{\rho\in \cl \mathcal{P}_{a}(0|\mu)}\Bigg(\displaystyle \sum_{r\in J\setminus \{j,l\}}\rho(J\setminus \{j\},r)+\rho(J,l)\Bigg)^2$.
\item \textbf{Step n}, $n\geq 2$: 
    \begin{itemize}
        \item Consider a set of grids $\alpha_n=\{\frac{2}{i+1}\}_{i=1}^n$.
        \item Find $(\alpha_n^*,\rho_n^*)$ as a solution of 
{\scriptsize
\[
\inf_{(\alpha,\rho)\in \alpha_n\times \cl \mathcal{P}_a(0|\mu) }\Bigg((1-\alpha)\Big(\sum_{r\in J\setminus \{j,l\}}\rho^{n-1}(J\setminus \{j\},r)+\rho^{n-1}(J,l)\Big) +\alpha\Big(\sum_{r\in J\setminus \{j,l\}}\rho(J\setminus\{j\},r)+\rho(J,l)\Big)\Bigg)^2.
\]
}
    \item Define $\rho^n=(1-\alpha_n^*)\rho^{n-1}+\alpha_n^*\rho_n^*$ and let $\rho^{out}=\rho^n$.
    \end{itemize}
    \item Stop if a terminating criterion is reached. 
    \item Return $\rho^{out}$ at the final step.
\end{itemize}
 
%Note that the diameter (maximum distance between two points in the set) of the random utility polytope in our case is $\sqrt{22}$ (the square root of 11 choice sets $\times$ 2 per choice set), achieved by a pair of a degenerate ranking and its inverse ranking. 

\section{Geometric Intuition for Theorem}

\subsection{Affine Independence Condition}\label{sec:intuition}

We first provide the intuition behind  Lemma \ref{lem:real_x}. To understand Lemma \ref{lem:real_x} (1) geometrically, see Figure \ref{fig:aff_ind}. In the figure, we assume that there are two original characteristic variables, say $(p_j,q_j)$ for each alternative $j \in J$. In Figure \ref{fig:aff_ind} (a) and (b), we consider the models with the original characteristics (i.e., $K=2$ and $x_j=(p_j,q_j)$ for each $j \in J$). In Figure \ref{fig:aff_ind} (a), the set $\{x_1,x_2,x_3\}$ is affinely independent. Thus, by Lemma 3 (1) (the ``if" part), any ranking is representable. For example, the ranking  $\pi(1) > \pi(2)> \pi(3)$ is representable by $\beta \in \Re^2$, which defines the parallel hyperplanes (indifference curves) in Figure \ref{fig:aff_ind} (a). On the other hand,  in Figure \ref{fig:aff_ind} (b), the set $\{x_1,x_2,x_3,x_4\}$ is not affinely independent.  The ranking $\pi(1)> \pi(4)> \pi(3)> \pi(2)$ is not representable. As the figure shows, no matter how one chooses $\beta \in \Re^2$ and draws  parallel hyperplanes as indifference curves, it does not hold that $\beta \cdot x_1 > \beta \cdot x_4 > \beta \cdot x_3> \beta \cdot x_2$. The existence of such an unrepresentable ranking is implied by the ``only if" part of Lemma 3 (1).

If we use ellipses as indifference curves, however, we can represent the ranking $\pi(1)> \pi(4)> \pi(3)> \pi(2)$ as in Figure \ref{fig:aff_ind} (c).\footnote{As the radius of the ellipses becomes larger, the represented utility of the alternatives   become larger.}  The existence of such  curves is again implied by the ``if" part of Lemma 3 (1) since ellipses can be defined with the quadratic polynomials  $\beta\cdot x_j$ with $x_j=(p_j, q_j, p_j^2, q_j^2, p_jq_j)$. Moreover, the generic  condition with quadratic polynomials is satisfied (i.e., $K=5\ge 3= |J|-1$ ) in  this example.
 Lemma \ref{lem:real_x} (2) is more straightforward.\footnote{To see this, notice that when $\D=\{J\}$, any $\pi \in \Pi$ is  representable in $\{J\}$ if and only if, for any $j\in J$, there exists $\beta$ such that $\beta \cdot x_j> \beta \cdot x_l$ for all $l \in J \setminus  \{j\}$, which means that $J$ is convex-independent.} By using Lemmas \ref{lem:appxQ}, \ref{lem:real_x3}, and \ref{lem:real_x}, we obtain statement (i) of Theorem \ref{theo:1} and Proposition \ref{theo:2}.

\begin{figure}[ht]
\centering
 \subfloat[]{
	\begin{minipage}[c][1\width]{
	   0.2\textwidth}
\begin{tikzpicture}[scale=.7]
\path (0,0) node[anchor=east] {$x_2$} (0,2) node[anchor=east] {$x_1$} (3,1) node[anchor=west] {$x_3$};

\fill (0,0) circle (1pt); \fill (0,2) circle (1pt);
\fill (3,1) circle (1pt); 

\draw[fill=white!20!white,opacity=1.0] (0,0) -- (0,2) --  (3,1) -- (0,0);

\draw[->, dotted, thick] (.7+.3, 15/8+1/20)-- (.2+.3, 15/8+1/20+.45) ;
\draw[red] (1.5,2.5) -- (-.5,.2); 
\draw[red] (3,2) -- (1,-.3);
\end{tikzpicture}
	\end{minipage}}
 \subfloat[]{
	\begin{minipage}[c][1\width]{
	   0.2\textwidth}
\begin{tikzpicture}[scale=0.45]

\path (0,0) node[anchor=east] {$x_3$} (5,0) node[anchor=west] {$x_4$} (3,4) node[anchor=west] {$x_2$ } (0,3.5) node[anchor=east] {$x_1$} (0,0);

\fill (0,0) circle (2pt); \fill (5,0) circle (2pt);
\fill (3,4) circle (2pt); \fill (0,3.5) circle (2pt);

\draw[fill=white!20!white,opacity=1.0] (0,3.5) -- (3,4) -- (5,0)-- (0,0)--(0,3.5);

\draw[red] (1,4.5) -- (-1,-.5);
\draw[red] (2.5,4.5) -- (.5,-.5);
\draw[red] (4.3,4.5) -- (2.3,-.5);

\draw[->, dotted, thick] (8/10,4)-- (8/10-1, 4+.48) ;

\end{tikzpicture}
	\end{minipage}}
 \subfloat[]{
	\begin{minipage}[c][1\width]{
	   0.2\textwidth}
    
\begin{tikzpicture}[scale=0.45]

\path (0,0) node[anchor=east] {$x_3$} (5,0) node[anchor=west] {$x_4$} (3,4) node[anchor=west] {$x_2$} (0,3.5) node[anchor=east] {$x_1$} (0,0);

\fill (0,0) circle (2pt); \fill (5,0) circle (2pt);
\fill (3,4) circle (2pt); \fill (0,3.5) circle (2pt);

\draw[fill=white!20!white,opacity=1.0] (0,3.5) -- (3,4) -- (5,0)-- (0,0)--(0,3.5);

%\draw[->, dotted, blue, thick] (1,3.8)--(-0.1, 4.8)  ;
%\draw[->, dotted, thick] (1.2,4)--(0.1, 5)  ;
%\draw[->, dotted, green, thick] (1.4,3.8)--(0.3, 4.8)  ;
\draw[->, dotted, thick] (1.9,3.3)--(0.8, 4.3)  ;

%\draw[color=red] (3,2) circle [x radius=5cm, y radius=25mm, rotate=40];
%\draw[color=red] (3.2,2) circle [x radius=2.5cm, y radius=12.5mm, rotate=40];
\draw[color=red] (3.2,2) circle [x radius=3.5cm, y radius=17.5mm, 
rotate=40];
\draw[color=red] (3.2,2) circle [x radius=3.9cm, y radius=20mm, rotate=40];
\draw[color=red] (3.2,2) circle [x radius=5.4cm, y radius=27mm, rotate=40];
\end{tikzpicture}

%\draw[color=blue] (2,2) circle [x radius=4cm, y radius=20mm, rotate=40];
%\draw[color=blue] (2,2) circle [x radius=1.3cm, y radius=7mm, rotate=40];
%\draw[color=blue] (2,2) circle [x radius=1.5cm, y radius=9mm, rotate=40];
	\end{minipage}}

\caption{Illustration of the  affine-independence condition. \label{fig:aff_ind}}
\end{figure}

\subsection{The Set of Random Utility Models}\label{sec:geometric}

Let \(\mathcal{P}_r := \operatorname{co}\{\rho^{\pi} : \pi \in \Pi\}\) denote the polytope of random utility models; its vertices are the deterministic choice rules \(\rho^{\pi}\) for \(\pi \in \Pi\) (left panel of Figure~\ref{fig:P_r}). Each point in \(\mathcal{P}_r\) corresponds to a random utility model. The figure is heuristic: each model is a high–dimensional vector indexed by pairs \((D,j)\) with \(j \in D \in \mathcal{D}\). It should not be confused with the characteristics space in Figure~\ref{fig:aff_ind}.

Our question is: under what conditions can all points in \(\mathcal{P}_r\) be represented or approximated by random–coefficient ARUMs with fixed effects? The analyst may choose a standard distribution \(\mu \in \mathcal{M}\) and a fixed–effect vector \(\eta \in \mathcal{F}\). Each pair \((\eta,\mu)\) induces a (convex) set of random–coefficient ARUMs with fixed effects, denoted \(\mathcal{P}_{ra}(\eta \mid \mu)\). Different choices of \(\eta\) and \(\mu\) generate different subsets of \(\mathcal{P}_r\) (right panel of Figure~\ref{fig:P_r}).

\begin{figure}[ht]
\begin{center}
\begin{tikzpicture}[mystyle/.style={draw,shape=circle,fill=black, inner sep=0pt, minimum size=4pt,  label={[anchor=center, label distance=3mm](90+360/\ngon*(#1-1)):#1}}]
\def\ngon{6}
\node[draw, regular polygon,regular polygon sides=\ngon,minimum size=3cm] (p) {};
\foreach\x in {1,...,\ngon}{
    \node[mystyle=] (p\x) at (p.corner \x){};
}

\node[anchor=west]  at (p.corner 1) {$\rho^{\pi_1}$};
\node[anchor=east]  at (p.corner 2) {$\rho^{\pi_2}$};
\node[anchor=east]  at (p.corner 3) {$\rho^{\pi_3}$};
\node[anchor=east]  at (p.corner 4) {$\rho^{\pi_4}$};
\node[anchor=west]  at (p.corner 5) {$\rho^{\pi_5}$};
\node[anchor=west]  at (p.corner 6) {$\rho^{\pi_6}$};

\draw[fill=yellow!20!white,opacity=1] (p.corner 1)-- (p.corner 2)--(p.corner 3)--(p.corner 4)--(p.corner 5)--(p.corner 6)--(p.corner 1);

\end{tikzpicture}
\begin{tikzpicture}[mystyle/.style={draw,shape=circle,fill=black, inner sep=0pt, minimum size=4pt,  label={[anchor=center, label distance=3mm](90+360/\ngon*(#1-1)):#1}}]
\def\ngon{6}
\node[draw, regular polygon,regular polygon sides=\ngon,minimum size=3cm] (p) {};
\foreach\x in {1,...,\ngon}{
    \node[mystyle=] (p\x) at (p.corner \x){};
}

\node[anchor=west]  at (p.corner 1) {$\rho^{\pi_1}$};
\node[anchor=east]  at (p.corner 2) {$\rho^{\pi_2}$};
\node[anchor=east]  at (p.corner 3) {$\rho^{\pi_3}$};
\node[anchor=east]  at (p.corner 4) {$\rho^{\pi_4}$};
\node[anchor=west]  at (p.corner 5) {$\rho^{\pi_5}$};
\node[anchor=west]  at (p.corner 6) {$\rho^{\pi_6}$};

\draw[fill=yellow!20!white,opacity=1] (p.corner 1)-- (p.corner 2)--(p.corner 3)--(p.corner 4)--(p.corner 5)--(p.corner 6)--(p.corner 1);

\draw[fill=blue!20!white,opacity=0.7] (p.corner 1)--(p.corner 2)--(p.corner 3)to[out=-55,in=175](p.corner 5)--(p.corner 1);

\draw[fill=red!20!white,opacity=0.7] (p.corner 2)--(p.corner 3)--(p.corner 4)to[out=20,in=230](p.corner 6)--(p.corner 1)--(p.corner 2);

\end{tikzpicture}

\caption{Illustration of $\P_r$ (yellow), $\P_{ra}(\eta_1|\mu)$ (red), and $\P_{ra}(\eta_2|\mu)$  (blue)}\label{fig:P_r}
\end{center}
\end{figure}

We consider the union \(\bigcup_{\eta \in\F} \mathcal{P}_{ra}(\eta \mid \mu)\) given arbitrarily chosen standard probability measure $\mu \in \M$ and ask whether it can approximate the entire polytope \(\mathcal{P}_r\). Lemma~\ref{lem:unrep} (Appendix) shows that if an unrepresentable ranking exists, then some random utility models cannot be approximated arbitrarily well—regardless of the choice of fixed effects or probability distributions.

\section{Additional Theoretical Results}\label{sec:simplify}

%We use the EM algorithm to estimate random-coefficient ARUM that maximizes the likelihood taking $\hat \rho$ as the observed choice probabilities. The Online Appendix shows that the resulting model is indeed a solution to (\ref{eq:dist.1}) when the affine-independence condition is satisfied. 

This section provides an additional theoretical result that upper-bounds the number of mixtures required for best possible approximation. The results in this section determine the number of mixtures to be used in the EM algorithm. 

%, which imply that the set of mixed-logit models coincides with the set of finite-mixture logit models with at most $\sum_{D \in \D}(|D|-1)+1$ mixtures. 

The first proposition (Proposition \ref{cor:red}) implies that any random-coefficient ARUM can be represented as a finite mixture of ARUMs.  The second proposition  (Proposition \ref{lem:dim_P_r}) gives us an upper bound on the number of mixtures required by calculating the dimension of the set of random utility models. To prove Proposition \ref{cor:red}, we need the following lemma:

\begin{lemma}\label{rem:logit_latent} Let $K$ be a fixed integer. For any bounded Borel set $C\subset \Re^K$, let $\Delta(C)$ denote the set of Borel probability measures over $C$.\footnote{In particular, $m(C)=1$ for any $m\in \Delta(C)$. } Then, 
\begin{equation*}
    \conv C = \Big\{\int x dm(x) \left| m\right. \in \Delta(C)\Big\},
\end{equation*}
where $\int x dm(x)$ denotes the $K$-dimensional vector whose $l$-th element is $\int x(l) dm(x)$ for any $l \in \{1,\dots,K\}$. 
\end{lemma}

The proof is in the next section.\footnote{Observe that the result is not true in an infinite dimensional space. To see this, let $\{e_i\}_{i=1}^{\infty}$ be the base of the infinite-dimensional real space. Define $C=\{e_i\}_{i=1}^{\infty}$. Define a measure $m$ on $C$ such that $m(e_i)= (1/2)^i$ for each $i$. Then, $\sum_{i=1}^{\infty} m(e_i)=1$, so that $m$ is a probability measure on $C$. $\int x dm $ cannot be represented as any finite mixture of elements of $C$. For any $y \in \co C$, there exists $i$ such that $y(e_i)=0$.} Recall the definition of $\P_{ra}(\eta|\mu)$ and $\P_{a}(\eta|\mu)$ from Definition \ref{def:mll2}. By using Lemma  \ref{rem:logit_latent}, we can show the following results that we use in the main body of the paper. Let $\dim \P_r$ denote the dimension of the set of the random utility models. 

\begin{proposition}\label{cor:red} 
For any $(\eta, \mu)$, 
\begin{align*}
    \mathcal{P}_{ra}(\eta|\mu)& =\co \P_a(\eta|\mu)\\
    & = \Big\{\sum_{m=1}^{\dim \P_r+1} \la_m \rho_m \Big|  \rho_m \in \P_a(\eta|\mu), \la_m \ge 0, \forall m=1,\dots, \dim \P_r+1,   \sum_{m =1}^{\dim \P_r+1} \la_m =1\Big\}.    
\end{align*}
\end{proposition}

\begin{proof}
The second equality follows from the Caratheodory's theorem. To show the first equality, fix $(\eta, \mu)$. The inclusion $\co \P_{a}(\eta|\mu)\subset \P_{ra}(\eta|\mu)$ holds because every discrete probability measure is a Borel probability measure. In the following, we will show $\P_{ra}(\eta|\mu) \subset \co \P_{a}(\eta|\mu)$.

Define a mapping $F:\mathbf{R}^K\to \P_a{(\eta|\mu)\subset \mathbf{R}^{|\D|\times J} }$ by
\begin{equation}
    F(\beta)(D,j)= \mu\left(\ep| \beta \cdot x_j+\eta_{(D,j)}+\ep_j >\beta \cdot x_l+\eta_{(D,l)}+\ep_l, \forall l \in D \setminus  \{j\}\right),
\end{equation}
and $0$ if $j\not \in D$. The mapping is continuous by the dominated convergence theorem given the fact that $\mu$ is absolutely continuous with respect to the Lebesgue measure. 

First note that $\P_{a}(\eta|\mu)$ is a Borel set because 
$\P_{a}(\eta|\mu)$ is the image of the continuous mapping $F$ from $\Re^K$ to $\Re^{|\D|\times J}$ defined in the proof. In particular, $\P_{a}(\eta|\mu)$ is the countable union of closed images of continuous mapping of compact cubes in $\Re^K$. 

To show $\P_{ra}(\eta|\mu) \subset \co \P_{a}(\eta|\mu)$,  let $\P_{a}(\eta|\mu)$ be the set $C$ in Lemma \ref{rem:logit_latent}, then we have  $\conv \P_{a}(\eta|\mu) = \big\{\int \rho dm(\rho) | m \in \Delta(\P_{a}(\eta|\mu))\big\}$. It suffices to show  $\P_{ra}(\eta|\mu) \subset \big\{\int \rho dm(\rho) | m \in \Delta(\P_a(\eta|\mu))\big\}$. 

Fix a given $m \in \Delta (\Re^K)$, we show $\int \rho_\beta dm(\beta) \in \big\{\int \rho dm(\rho) | m \in \Delta(\P_a(\eta|\mu))\big\}$, where $\rho_\beta \in \P_a(\eta|\mu)$. For any  Borel set $C\subset \P_a(\eta|\mu)$, define $\hat{m}(C)=m(F^{-1}(C))$, where $F^{-1}(C)$ is a Borel set because $F$ is continuous. Then, $\hat{m} \in \Delta(\P_a(\eta|\mu))$ and we have $\int \rho_\beta dm(\beta)=\int  \rho d\hat{m}(\rho)$, as desired.
\end{proof}

We now calculate the number $\dim \P_r$. We note that the set $\P_r$ defined in Definition \ref{def:ruf} is associated with a set of choice sets $\D$.

\begin{proposition}\label{lem:dim_P_r} $\dim \P_r = \sum_{D \in \D}(|D|-1)$.
\end{proposition}

The proof is in a later section.  Propositions \ref{cor:red} and  \ref{lem:dim_P_r} imply that in order to obtain the best approximating random-coefficient model to the observed choice probabilities, it is sufficient to consider finite mixture models with  at most $1+\sum_{D \in \D}(|D|-1)$ mixtures. For example, in Section \ref{sec:data}, we analyze a choice dataset with $|J|=4$ with $\D= 2^J \setminus \{\emptyset\}$. These results imply that it is enough to consider the finite mixture models with at most $18$ mixtures if one considers the whole choice sets.

\subsection{Proof of Lemma \ref{rem:logit_latent}}
By definition,  $\conv C$ is a subset of $\{\int x dm(x) | m \in \Delta(C)\}$. To show the reverse direction, we first establish a relaxed statement: $\Big\{\int x dm(x) | m \in \Delta(C)\Big\} \subset \cl \conv C$. Suppose by way of contradiction that $\int x dm(x) \not \in \cl \conv C$ for some $m \in \Delta(C)$. 
By the separating hyperplane theorem (Corollary 11.4.2 of \cite{rockafellar2015}), there exist a $t \in \Re^K\setminus\{0\}$ and $\al \in \Re$ such that $(\int x dm(x)) \cdot t = \al > x \cdot t$ for any $x \in \cl \conv C$. This is a contradiction because $\al=(\int x dm(x)) \cdot t= \int (x \cdot t) dm(x)<\int \al dm(x)=\al$.

We now prove $\Big\{\int x dm(x) | m \in \Delta(C)\Big\} \subset \conv C$ by an induction on the dimension of $\co C$. 

\textbf{Induction Base:} If $\dim \co C=1$, there exist $y=\inf \{x|x\in \co C\}$ and $z=\sup \{x|x\in \co C\}$ such that $\co C$ can be represented as the line segment between $y$ and $z$. We consider the case where the line segment contains  neither $y$ nor $z$. Proofs for the other cases are similar. For any $x \in  C$, there exists a weighting function $\al(x)=\frac{z-x}{z-y} \in (0,1)$ such that $x= \al(x) y+(1-\al (x))z$. The function $\al$ is bounded, nonnegative, and continuous in $x$. Hence it is measurable and integrable. 

Choose any $m \in \Delta(C)$. By continuity from below of measures, there exists $l$ and $u$ such that $y<l<u<z$ and $m(\{x|x\in (l,u)\})\geq 1-\epsilon$ for some $\epsilon<1$. Note here we use the fact that $m(C)=1$ and $m(\{y\})=m(\{z\})=0$ because we are assuming that the line segment does not contain both $y$ and $z$. Note that $\alpha(x)$ is uniformly bounded away from 0 and 1 for $x\in (l,u)$. Then it follows that $\int \al dm=\int \al(x) dm(x)$ exists and $0<\int \al(x) dm(x)<1$. Then, $\int x dm(x)= \int \al(x) dm(x) \times y+\int (1-\al (x)) dm(x) \times z\in \co C$, as desired. 

Let $l\geq 2$ be an integer. 

\textbf{Induction Step:} Suppose that $\Big\{\int x dm(x) | m \in \Delta(C)\Big\} \subset \conv C$ holds for any $C$ such that $\dim \co C \le l$. Then it holds for any $C$ such that $\dim \co C=l+1$.

To prove the step, choose any $m \in \Delta(C)$. We have $\int x dm(x)\in \cl \co C$. First consider the case where $\int x dm(x)\in \rint \cl \co C$. Since $\rint \cl \co C= \rint \co C$ (by Theorem 6.3 of \cite{rockafellar2015}), we have $\int x dm(x)\in \co C$, as desired.  

Next consider the case where $\int x dm(x)\not \in \rint \cl \co C$. Then, $\int x dm(x)\in \partial \cl \co C $ and by Theorem 11.6 of \cite{rockafellar2015}, there exists a hyperplane $H$ of $\cl \co C$ at $\int x dm(x)$ such that $\cl \co C \not \subset H$. There exist $t \in \Re^K\setminus \{0\}$ and $\al \in \Re$ such that $H= \{x |x \cdot t =\al \}$ and $\int x dm(x) \cdot t= \al >  x\cdot t$  for any $x \in \cl \co C \cap H^c$. This implies that $m(H)=1$ and hence $m(H \cap C)=1$. Since $H$ is a supporting hyperplane and $\cl \co C \not \subset H$, we obtain $\dim (H\cap \aff C)\le l$. Hence, $\dim \left( \co\left(H \cap C\right)\right) \le l$ because $\co \left(H\cap C\right) \subset \co H \cap \co C \subset H \cap \aff C$. Therefore, the induction hypothesis shows that $\int x dm(x) \in \co (H \cap  C) \subset \co C$, as desired.

\subsection{Proof of Proposition \ref{lem:dim_P_r}}
\label{sec:prop_4}

To prove Proposition \ref{lem:dim_P_r}, we prove two lemmas. Lemma \ref{rem:non_cons} is a technical lemma that facilitates the characterization of the affine hull of random utility polytopes in Lemma \ref{rem:affine}. The dimension of a set is defined as the dimension of its affine hull, and Proposition \ref{lem:dim_P_r} follows. Recall that we use $\rho^\pi\in \Re^{|\D| \times |J|}$ to denote the stochastic choice function associated with the strict preference ranking $\pi\in \Pi$ where $\Pi$ is defined in Section \ref{sec:math}.

\begin{lemma}\label{rem:non_cons}
For any $t \in \Re^{|\D| \times |J|}$, $\rho^{\pi} \cdot t= \rho^{\pi'} \cdot t$ for all $\pi,\pi' \in \Pi$ if and only if $t(D,j)= t(D, l)$ for all $D \in \D$ and  $j, l \in D$.\footnote{We are identifying each $\rho \in \P$ as an element of $\R^{\D\times J}$.}
\end{lemma}

\begin{myproof}

To prove the if part, assume $t(D,j)= t(D, l)$ for all $D \in \D$ and  $j, l \in D$. Define $t(D)=t(D,j)$ for any $j \in D$. Then for any $\pi\in \Pi$, $\rho^{\pi} \cdot t= \sum_{D \in \D} \sum_{j \in D}\rho^{\pi}(D,j) t(D,j)= \sum_{D \in \D} t(D) \sum_{j \in D}\rho^{\pi}(D,j)=\sum_{D \in \D} t(D)$, completing the proof of the if part. 

The only-if direction is trivially true when the set $D$ contains only one element. We prove the only-if direction for the remaining $D$, $|D|\ge 2$, by induction. Consider the sets in $\mathcal{D}$ with a size greater or equal to 2. Let $m$ be the smallest cardinality of the sets in $\D$, $m=\min\{|D|\big| D\in\D, |D|\geq 2\}$.\footnote{Note we do not induct from $|D|=1$ because $\D$ may not contain any set with one element.}

\textbf{Induction Base}: For any $D \in \D$ such that $|D|=m$ and any $j, l \in D$, $t(D,j)=t(D,l)$. 

\begin{proof}
To prove the claim, choose any $\pi, \pi' \in \Pi$ such that for $\pi(J\setminus D)>\pi(j)>\pi(l)>\pi(D\setminus \{j,l\})$, and $\pi'(J\setminus D)>\pi'(l)>\pi'(j)>\pi'(D\setminus \{j,l\})$.

Note that the only choice set that has different choice probabilities under $\pi$ and $\pi'$ is $D$. For all the remaining choice sets, either only one of $j$ or $l$ is in the set, or there exists an element that dominates $j$ and $l$ under both $\pi$ and $\pi'$. Notice here that we used the fact that $D$ has the smallest cardinality. Moreover, $\rho^{\pi}(D,j)=1$, $\rho^{\pi}(D,r)=0$ for any $r \in D \setminus\{j\}$ and  $\rho^{\pi'}(D,l)=1$, $\rho^{\pi'}(D,r)=0$ for any $r \in D \setminus\{l\}$. 

Hence, since $t \cdot \rho^{\pi} = t \cdot \rho^{\pi'}$,
\begin{align*}
  0 &= \sum_{(E,r) \in \D  \times J} t(E,r)(\rho^{\pi}(E,r)-\rho^{\pi'}(E,r))=\sum_{r\in D} t\left(D,r\right)(\rho^{\pi}(D,r)-\rho^{\pi'}\left(D,r)\right)\\
&= t(D,j)-t(D,l).   
\end{align*}

This completes the proof of the induction base.
\end{proof}

Let $k\geq m$.

\textbf{Induction Step:} Suppose that for any $D \in \D$ such that $|D|\leq k$ and any $j, l \in D$, $t(D,j)=t(D,l)$. Then the same claim holds for any $D\in\D$ such that $|D|=k+1$.
\begin{proof}
By the induction hypothesis, for any $E \in \D$, if  $|E|\le k$ then $t(E,j)=t(E,l)$ for any $j, l \in E$. Repeat the proof above by considering the choice sets $\mathcal{D}\setminus \{E\in\mathcal{D}||E|\leq k\}$.
\end{proof}

\end{myproof}

\begin{lemma}\label{rem:affine}
     The affine hull of $\P_r$ is 
\begin{equation*}
  \P_{\pm}\equiv \big\{q \in \Re^{ |\D| \times |J|}\big|\text{(i)}\sum_{j \in D}q(D,j)=1, \forall D \in \D;\text{(ii) }q(D,j)=0,  \forall j \not \in D \in \D \big\}.  
\end{equation*}     
\end{lemma}

\begin{myproof}
The set $\P_{\pm}$ is affine. So it suffices to show that for any affine set $A$, if $\P_r \subset A$, then $\P_{\pm}\subset A$.  For any affine set $A$, by Theorem 1.4 of \cite{rockafellar2015}, it has a representation $A= \{q \in \Re^{ |\D| \times |J|}| B q = b\}$, where $B$ is a $L \times (|\D|\times |J|)$ matrix, $b$ is a $L$-dimensional vector, and $L$ is an arbitrary positive integer.

For any $l\in \{1,\dots, L\}$, let $B_l(D,j)$ denote the $(l, (D,j))$ entry of $B$. Note that each column of $B$ is associated with a $(D,j) \in \D \times J$. So $B q  = b$ means that for any row index $l\in \{1,\dots, L\}$, 
\begin{equation}\label{eq:mat1}
\sum_{D \in \D} \sum_{j \in J} B_l(D,j) q(D,j)=b_l.
\end{equation}

By assuming $\P_r \subset A=\{q \in \Re^{\D \times J}| B q = b\}$, we will show that if $q$ satisfies (i) and (ii), then (\ref{eq:mat1}) holds for any $l\in \{1,\dots, L\}$.

\step 1: We show that  $B_l(D,j)=B_l(D,r)$ for any $l\in \{1,\dots, L\}$, $D \in \D$, and $j, r \in D$. For any $\pi\in \Pi$, $\rho^{\pi} \in \P_r \subset A=\{q \in \Re^{|\D| \times |J|}| B q = b\}$. Hence, (\ref{eq:mat1}) holds with $q=\rho^{\pi}$ for any $\pi \in \Pi$. Thus $\rho^{\pi}\cdot B_l=\rho^{\pi'}\cdot B_l$ for any $\pi,\pi' \in \Pi$ for any $l$.  By Lemma \ref{rem:non_cons}, this implies that $B_l(D,j)=B_l(D,r)$ for any $D \in \D$, and $j, r \in D$.

By Step 1, we can define $B_l(D)= B_l(D,j)$ for any $j \in D$.

\step 2: If $q$ satisfies property (i) and (ii), choose any $\pi\in \Pi$ and $l \in \{1,\dots, L\}$.  Since $\rho^{\pi} \in A$, then by (\ref{eq:mat1}),
\begin{equation}\label{eq:mat2}
b_l= \sum_{D \in \D} \sum_{r\in J} B_l(D,r)\rho^{\pi}(D,r)= \sum_{D \in \D} B_l(D),
\end{equation}
where the second equality holds by $\rho^{\pi}(D, r)=1$ if $r=\max_{D} \pi$ and $\rho^{\pi}(D, r)=0$ otherwise. Finally, by using these equalities, for each $l \in \{1,\dots, L\}$, we obtain the following equations: 
\beq
\begin{array}{llll}
\sum_{D \in \D} \sum_{r \in J} B_l(D,r)q(D,r)
&=& \sum_{D \in \D} \sum_{r \in D} B_l(D,r)q(D,r)&(\text{by property (ii)})\\
&=& \sum_{D \in \D} \sum_{r \in D} B_l(D)q(D,r) &(\text{by Step 1})\\
&=& \sum_{D \in \D} B_l(D) \sum_{r \in D}q(D,r) \\
&=& \sum_{D \in \D} B_l(D)=b_l. &(\text{by property (i)  and } (\ref{eq:mat2}))
\end{array}
\eeq
This establishes that $\aff \P_r= \{q \in \Re^{\D \times J} |\text{(i) and (ii)}\}$.

The  equalities in (i) and (ii) are independent. The dimension of $\{q \in \Re^{\D \times J} |\text{(ii)}\}$ is $\sum_{D \in \D}|D|$.  The number of equalities of (i) is $|\D|$. Hence, the dimension of $\P_r$ is $(\sum_{D \in \D}|D|)-|\D|=\sum_{D \in \D}(|D|-1)$.
\end{myproof}
Proposition \ref{lem:dim_P_r} follows from Lemma \ref{rem:non_cons} and Lemma \ref{rem:affine}.

%Based on Proposition \ref{prop:unrepresentable}, for the approximation error without fixed effects, we measure the distance to $\rho^{\pi}$, where $\pi$ is an unrepresentable ranking; for the approximation error with fixed effects, we measure the distance to $\frac{1}{2}\rho^{\pi}+\frac{1}{2} \rho^{\pi^{-}}$, where $\pi$ is an unrepresentable ranking $\pi$ and $\pi^{-}$ is the reverse ranking of $\pi$. 

\section{Further Empirical Analyses for Section \ref{sec:data}}

\subsection{Approximation Errors for $|J|=10$ and $K=3$}\label{sec:J10K2}

\begin{figure}[h]
    \centering
    \includegraphics[width=0.6\linewidth]{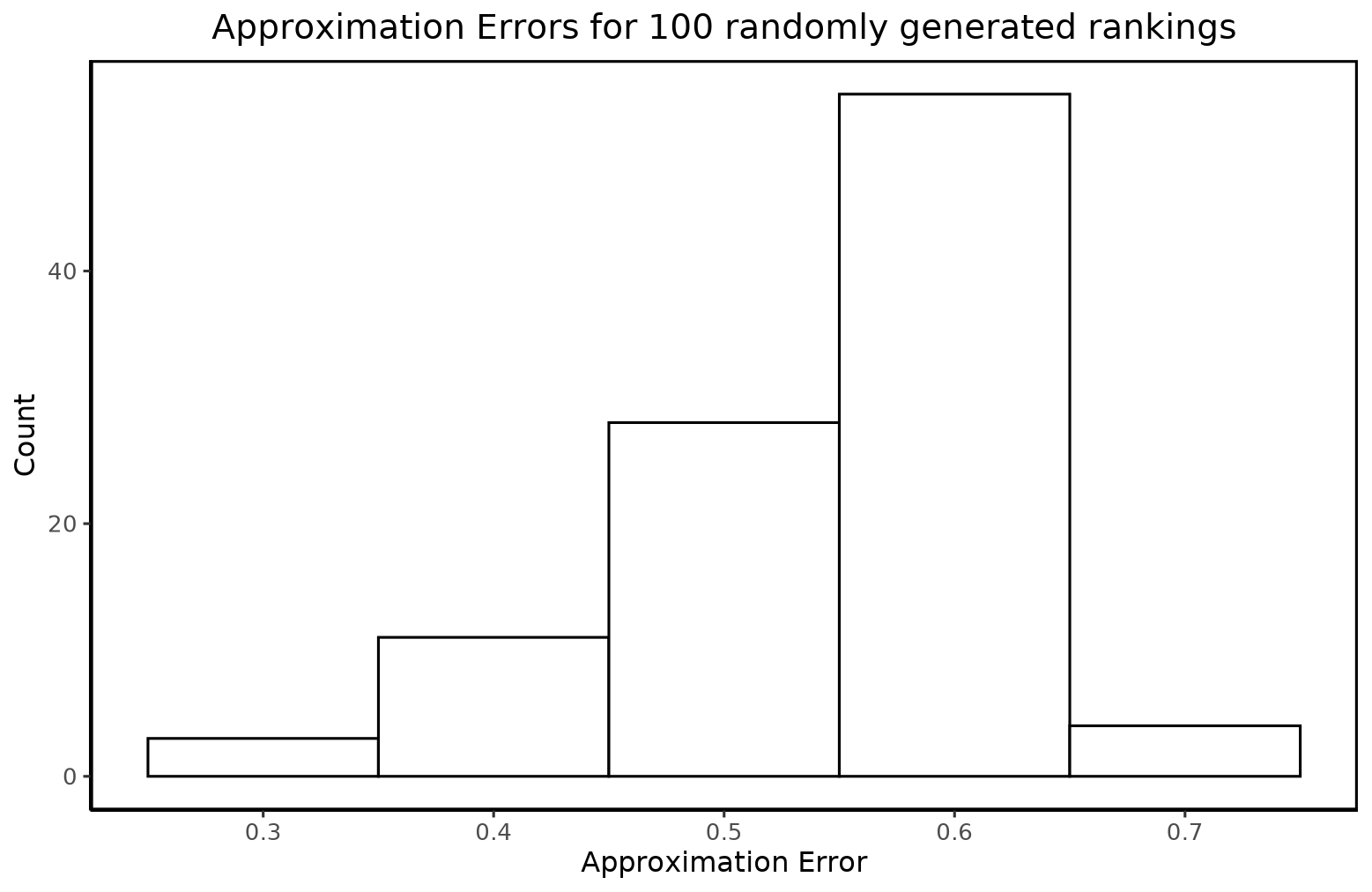}
    \caption{Approximation Errors of 100 randomly generated rankings with 10 alternatives ($|J|=10$) and 3 characteristics ($K=3$).}
    \label{fig:approx_error_J10}
\end{figure}

In Section \ref{sec:data}, we examined approximation errors in the setting with four alternatives and two characteristics. To assess how these errors behave with a larger number of alternatives, we now consider 10 alternatives and three characteristics. The 10 alternatives are selected from the first ten rows of Table 2 in \cite{farias2009non}, which correspond to the ten most expensive alternatives. For computational tractability, the analysis is restricted to binary and ternary choice sets. Figure \ref{fig:approx_error_J10} contains the histogram of approximation errors of linear mixed-logit models without fixed effects for 100 rankings  drawn uniformly at random from all possible rankings. The results indicate that all randomly generated rankings exhibit non-negligible approximation errors, ranging from $0.3$ to $0.7$ in the normalized Euclidean metric.

\subsection{Details of the DVD Dataset}\label{appendix_dvd_dataset}
We accessed the DVD dataset from Table 2 of \cite{farias2009non}, which we replicate as Table \ref{tab:amazon_dvd}. For our analysis in the main text, we selected the products from the first four rows. For our analysis in Section \ref{sec:J10K2}, we selected the products from the first ten rows. For the analysis with four products and two characteristics, we used \textit{Average Price per Disc} and \textit{Total Helpful Votes}. For the analysis with ten products and three characteristics, we additionally included \textit{Price}.
\begin{table}[h]
\centering
\caption{DVD dataset as in Table 2 of \cite{farias2009non}}
\label{tab:amazon_dvd}
\begin{tabular}{|c|c|c|c|}
\hline
\textbf{Product ID} & \textbf{Price } & \textbf{Avg. Price per Disc } & \textbf{Total Helpful Votes} \\
\hline
1  &  115.49 & 5.7747  & 462 \\
2  & 92.03  & 7.6694  & 20 \\
3  &  91.67  & 13.0955 & 496 \\
4  &  79.35  & 13.2256 & 8424 \\
5  &  77.94  & 6.4949  & 6924 \\
6  &  70.12  & 14.0242 & 98 \\
7  &  64.97  & 16.2423 & 1116 \\
8  &  49.95  & 12.4880 & 763 \\
9  &  48.97  & 6.9962  & 652 \\
10 &  46.12  & 7.6863  & 227 \\
11 &  45.53  & 6.5037  & 122 \\
12 &  45.45  & 11.3637 & 32541 \\
13 &  45.41  & 11.3523 & 69 \\
14 &  44.92  & 11.2292 & 1113 \\
15 &  42.94  & 10.7349 & 320 \\
\hline
\end{tabular}
\end{table}

\newpage

\section{Additional Empirical Application: Fishing-site Choice}\label{appendx:fish}

\vspace{-0.2cm}

In this section, we measure approximation errors with and without fixed effects by using a dataset on fishing-site choices from \cite{thomson1991results}.\footnote{The dataset is taken directly from the R package `mlogit' by \cite{croissant2020package}.} The dataset has been  used by \cite{herriges1999nonlinear} and \cite{cameron2005microeconometrics} (p.464). 

In the dataset, 1182 individuals choose among $4$ fishing modes, namely, $J=\{{\text{beach}}, {\text{boat}},$ ${\text{charter}}, {\text{pier}}\}$, which denote fishing from the beach, a private boat, a charter boat or a pier, respectively.  Each alternative $j \in J$ is  described by a vector of two characteristics $(p_j,q_j)$. The first characteristic $p_j$ is the fishing mode $j$'s price. The other characteristic $q_j$ is the {\it catch rate}, defined as a per-hour-fished basis for major species by fishing mode $j$.\footnote{In the original study, the values of $p_j$ and $q_j$ depend on each individual. For our analysis, we aggregate them by taking the average over individuals.} Our empirical analysis concentrates primarily on mixed-logit models. We assess the approximation errors using a linear mixed-logit model and a quadratic mixed-logit model, as in Section \ref{sec:data}.

\subsection{Approximation Errors without Fixed Effects}\label{sec:fish_without_fixed}

Table \ref{tbl:fish_apx_error1} shows the approximation errors of the linear or quadratic mixed-logit models. We calculated the errors by the greedy algorithm and the EM algorithm. In both algorithms, 
 the approximation errors under quadratic mixed-logit models are zero, as shown in columns (3) and (4) in the table. This finding is consistent with the theorem. For the linear mixed-logit models, the approximation errors for  representable rankings $\pi$ shown in the bottom row of the table are always zero, as the theorem predicts. 
 
On the other hand, the approximation errors for unrepresentable rankings $\pi$ are almost always larger than $0.4$, which means that even the best possible linear mixed-logit model remains far from the corresponding choice probabilities $\rho^{\pi}$. Some errors are much larger. For example,  the approximation errors of the two rankings $\pi(1)>\pi(2)>\pi(3)> \pi(4)$ and $\pi(1)>\pi(2)>\pi(4)> \pi(3)$ by the linear  models are greater than $0.67$. Notice that these two rankings are the only rankings in which alternative $1$ is the best and alternative $2$ is the second-best. This fact suggests that the substitution from alternative 1 to 2 would be difficult to capture. The next subsection further explores this issue. 

\begin{table}[ht]
\begin{center}
  \caption{Approximation errors to  $\rho^{\pi}$}\label{tbl:fish_apx_error1}
\scalebox{0.9}{
\begin{tabular}{|c|c|c|c|c|}
\hline
\multirow{2}{*}{Ranking $\pi$ } &\multicolumn{2}{c|}{Linear mixed-logit}&\multicolumn{2}{c|}{Quadratic mixed-logit} \\ \cline{2-5}
 & Greedy  & EM    & Greedy  & EM \\ 
   &  (1) &  (2) &(3) &(4) \\  
\hline
Linearly Unrepresentable Rankings &&&  &\\
$\pi(1)> \pi(2)> \pi(3)> \pi(4)$    &0.723&0.753& 0.000 &0.000   \\
$\pi(1)>\pi(2)>\pi(4)>\pi(3)$        &0.670&0.700& 0.000 &0.000\\
$\pi(1)> \pi(3) > \pi(2) > \pi(4)$   &0.425&0.381&0.000  &0.000\\
$\pi(1)> \pi(4)> \pi(2)> \pi(3)$   &0.418&0.547& 0.000 &0.000\\
$\pi(2)>\pi(1)> \pi(3)> \pi(4)$  &0.458&0.488& 0.000 &0.000\\
$\pi(2)> \pi(1) > \pi(4)> \pi(3)$  &0.391&0.408&0.000  &0.000\\
$\pi(3)>\pi(2)>\pi(4)>\pi(1)$ &0.302&0.318&0.000  &0.000\\
$\pi(3)>\pi(4) >\pi(1)> \pi(2)$  &0.401&0.425& 0.000 &0.000\\
$\pi(3)> \pi(4)> \pi(2) >\pi(1)$ &0.494&0.531&0.000  &0.000\\
$\pi(4)>\pi(2)> \pi(3)> \pi(1)$  &0.375&0.381&0.000  &0.000\\
$\pi(4)>\pi(3)> \pi(1)> \pi(2)$  &0.521&0.514&0.000  &0.000\\
$\pi(4)> \pi(3)> \pi(2)> \pi(1)$  &0.604&0.614&0.000  &0.000\\
\hline
Linearly Representable Rankings & 0.000 & 0.000 & 0.000& 0.000\\
\hline
\end{tabular}
}
\end{center}
\begin{scriptsize}
\textit{Note}: The numbers in the table show the approximation errors for each $\rho^{\pi}$, where each ranking $\pi$ is defined in the leftmost column. Alternative numbers 1, 2, 3, 4 denote beach, boat, charter, and pier, respectively. For each ranking, columns (1) and (2) show the approximation errors of the linear mixed-logit models computed by the greedy algorithm and the EM algorithm, respectively. Columns (3) and (4) show the approximation errors of the quadratic mixed-logit models calculated by each algorithm. All numbers are rounded to three decimal places. For the greedy algorithm we set the number of iterations to 1000. For the EM algorithm we set the number of random initial points to 10. The greedy algorithm sometimes produces larger approximation errors than the EM algorithm, which is possible with finitely many steps.
\end{scriptsize}
\end{table}

\subsubsection{Maximal Substitution}\label{sec:fish_substitution}

 Table \ref{tbl:fish_subs} shows the values of maximal substitution between the two alternatives $j$ and $l$. Some numbers in the table are close to one, which implies that the linear mixed-logit models are rich enough to capture flexible substitution from $j$ to $l$. Some other numbers are smaller. In particular, the maximal substitution between alternative $1$ (i.e., beach) and $2$ (i.e., private boat) as well as the substitution between alternative $3$  (i.e., charter) and $4$ (i.e., pier) are at most $0.3$. In fact, the maximal substitution from $1$ to $2$ is $0.12$. This means that no matter how the parameters of a linear mixed-logit model are chosen, the maximal substitution from alternative 1 to alternative 2 is very limited. This finding aligns with the result presented in Table \ref{tbl:fish_apx_error1}, where we observe substantial approximation errors for the two specific rankings: $\pi(1)>\pi(2)>\pi(3)> \pi(4)$ and $\pi(1)>\pi(2)>\pi(4)> \pi(3)$. In this way, identifying preferences that are hard to approximate with precision helps researchers in evaluating whether their models successfully capture relevant economic behaviors such as substitution patterns. 

\begin{table}[ht]
\begin{center}
  \caption{Maximal substitution  of the linear mixed-logit models}\label{tbl:fish_subs}
\scalebox{0.9}{
\begin{tabular}{|c|c|c|c|c|}
 \hline
\diagbox[width=10em, trim=l]{$j$ (drop)}{$l$ (choose)} & 1 & 2 & 3 & 4 \\
  \hline
 1 &-& 0.120 & 0.998 & 0.998 \\ 
     \hline 
2 & 0.317 &-& 1.000 & 0.997 \\ 
    \hline 
3 & 0.998 & 1.000 & -& 0.286 \\ 
    \hline 
4 & 0.994 & 0.998 & 0.137 & - \\ 
   \hline 
\end{tabular}
}
\end{center}
\begin{scriptsize}
\textit{Note}: The numbers in the table show the value of (\ref{eq:dist.2}) for each $j, l \in \{1,2,3,4\}$ s.t. $j\neq l$.  Alternative numbers 1, 2, 3, 4 denote beach, boat, charter, pier, respectively.  All numbers are rounded to three decimal places.   
\end{scriptsize}
\end{table}

\subsection{Approximation Errors with Fixed Effects}\label{sec:fish_with_fixed}

Table \ref{tbl:fish_apx_error2} shows the approximation error to $\frac{1}{2}\rho^{\pi}+ \frac{1}{2} \rho^{\pi^-}$ for each unrepresentable ranking. In both algorithms, the approximation errors to $\frac{1}{2}\rho^{\pi}+\frac{1}{2}\rho^{\pi^-}$ are always around $0.2$ if $\pi$ is not representable. This means that even the best possible linear mixed-logit model remains bounded away from  $\frac{1}{2}\rho^{\pi}+\frac{1}{2}\rho^{\pi^-}$ in the normalized Euclidean metric. On the other hand, the approximation errors to $\frac{1}{2}\rho^{\pi}+\frac{1}{2}\rho^{\pi^-}$  are almost zero if $\pi$ is  representable, as the theorem predicts. The approximation errors under quadratic mixed-logit models are also almost zero, as the theorem again predicts. 

\begin{table}[ht]
\caption{Approximation errors to random utility models $\frac{1}{2}\rho^{\pi}+\frac{1}{2}\rho^{\pi^-}$}\label{tbl:fish_apx_error2}
\begin{center}
\scalebox{0.9}{
\begin{tabular}{|c|c|c|c|c|}
\hline
\multirow{2}{*}{Ranking $\pi$ } &\multicolumn{2}{c|}{Linear mixed-logit}&\multicolumn{2}{c|}{Quadratic mixed-logit} \\ \cline{2-5}
 & Greedy  & EM    & Greedy  & EM \\ 
   &  (1) &  (2) &(3) &(4) \\  
\hline
Linearly unrepresentable rankings &&&&\\
$\pi(1)> \pi(2)> \pi(3)> \pi(4)$     &0.229&0.255& {0.000}& {0.000} \\
$\pi(1)>\pi(2)>\pi(4)>\pi(3)$        &0.229&0.250& {0.000}& {0.000}\\
$\pi(1)> \pi(3) > \pi(2) > \pi(4)$   &0.163&0.217& {0.000}& {0.000}\\
$\pi(1)> \pi(4)> \pi(2)> \pi(3)$     &0.163&0.173&{0.000} &{0.000}\\
$\pi(2)>\pi(1)> \pi(3)> \pi(4)$      &0.192&0.238&{0.000} &{0.000}\\
$\pi(2)> \pi(1) > \pi(4)> \pi(3)$    &0.192&0.198&{0.000} &{0.000}\\
\hline
Linearly representable rankings & 0.000 & 0.000 & {0.000}& {0.000}\\
\hline
\end{tabular}
}
\end{center}
\begin{scriptsize}
\textit{Note:} The numbers in the table show the approximation errors to $\frac{1}{2}\rho^{\pi}+ \frac{1}{2}\rho^{\pi^-}$, where $\pi$ is defined in the leftmost column.  All numbers are rounded to three decimal places. For the greedy algorithm we set the number of iterations to 1000. For the EM algorithm we set the number of random initial points to 10. We optimize over fixed effects from -10 to 10 with a step size of 1. 
\end{scriptsize}
\end{table}

\subsection{In-sample and Out-of-sample Fit}\label{sec:insamp}

%In this section, we evaluate in-sample and out-of-sample fit of our model to address the possible concern of over-fitting. 

 In this section, we evaluate the in-sample and out-of-sample fit of our model. We show that our method performs better or equally well compared to standard methods, not only in terms of in-sample fit but also in terms of out-of-sample fit. We use the same fishing choice dataset used above and predict choice probabilities using aggregated characteristics.

We estimate a random-coefficient logit model with arbitrary mixing distributions.  In the dataset, we have four alternatives and we consider only one choice set $\D=\{ J\}$. Thus by Proposition \ref{lem:dim_P_r} and Proposition \ref{cor:red}, it suffices to mix four logit models without fixed effects to represent any random utility model. We refer to a 4-mixture mixed-logit model as \textit{our method} hereafter. We also estimate several standard models for comparison. They include 1) a multinomial logit model, 2) nested logit model with two nests (charter and the rest), 3) a nested logit model with two nests (boat and the rest), 4) a random coefficient logit model with a log-normal mixing distribution for each variable, 5) a multinomial logit model with alternative fixed effects, 6) and a random coefficient logit model with log-normal mixing distributions and alternative fixed effects. We detail the definition of each specification in Section \ref{section:models}. 

To evaluate in-sample and out-of-sample fits, we adopt the following strategy. We randomly divide individuals in the sample into a training sample and a test sample of equal sizes. Separately for the training and testing samples, we average individual choices and characteristics to obtain aggregate data on choice probabilities and characteristics. We then estimate the models using the training sample. The models are estimated by maximizing the log-likelihoods. That is, for each model, we solve the problem $\max_{\theta\in\Theta}\sum_{j=1}^{|J|}\hat{\rho}_j\log\rho(j|\theta)$, where $j$ indexes fishing modes, $\theta$ is the parameter vector of the model, $\Theta$ denotes the set of possible parameter vectors, $\hat{\rho}_j$ is the observed market share for fishing mode $j$ in the training data, and $\rho(j|\theta)$ is the model-predicted choice probability for fishing mode $j$ with characteristic vector $x_j$. See Section \ref{section:models} for a likelihood expression for each model. 
For the standard models, we maximize the likelihoods with the nonlinear optimization package in R \citep{Rsolnp1, Rsolnp2}. 
For our model, we use the EM algorithm.\footnote{We prefer the EM algorithm over the greedy algorithm here because the EM algorithm is faster.} 

To evaluate the in-sample fit performance, we compute the predicted choice probabilities in the training sample $\hat{\rho}_{train}\in\mathbf{R}^{|J|}$ and compare them with the observed choice probabilities in the training sample $\rho_{train}\in\mathbf{R}^{|J|}$.\footnote{We only consider the single choice set case in this simulation. So the choice probability vector has length $|J|$.} For this comparison, we calculate the $l_2$ distance between the predicted choice probabilities and the aggregated observed choice probabilities $||\hat{\rho}_{train}-\rho_{train}||_2$. Similarly, to evaluate the out-of-sample performance, we compute the predicted choice probabilities using the testing sample $\hat{\rho}_{test}\in\mathbf{R}^{|J|}$ and compare it with the aggregated observed choice probabilities in the testing sample $\rho_{test}\in\mathbf{R}^{|J|}$. We use the $l_2$ metric  $||\hat{\rho}_{test}-\rho_{test}||_2$ for this comparison as well.

We repeat this exercise with 50 random splits. The results for in-sample fits are reported in Table \ref{tbl:in-sample_d1}. The results for out-of-sample fits are in Table \ref{tbl:out-sample}.

As expected, the in-sample fit of our model is perfect.  Several standard models, especially those without fixed effects, exhibit imperfect in-sample fit. For example, the random coefficient logit model with the log-normal distributions has the $l_2$ prediction error 0.038.

Table \ref{tbl:out-sample} shows that the out-of-sample prediction error of our model is positive but small. Standard models without alternative fixed effects have out-of-sample prediction errors substantially larger than our model. The two alternative models with fixed effects have out-of-sample prediction errors comparable to ours. This result suggests that even without using  fixed effects, our model performs as well as or better than standard models in this simulation, not only in terms of in-sample fit but also in terms of out-of-sample fit.
%\footnote{While we use the average characteristics for all exercises on algorithm 1, we use individual-level characteristics for estimation and validation for standard models. The reason for this is to compare out results with \citet{HK_1999_REStat}, who estimate their models using microdata.} 
\begin{table}[ht]
\caption{In-Sample Fit} 
\begin{center}
\scalebox{0.85}{
\begin{tabular}{|c|c|c|c|c|c|}
  \hline
  \multirow{2}{*}{Model } &\multicolumn{4}{c|}{Choice probabilities}&\multicolumn{1}{c|}{Prediction error} \\ \cline{2-6}
 & Beach  & Boat    & Charter  & Pier & \\ 
   &  (1) &  (2) &(3) & (4) & (5)\\  
  \hline
Our method & 0.114 & 0.353 & 0.383 & 0.151 & 0.000 \\ 
   & (0.009) & (0.015) & (0.017) & (0.010) & (0.000) \\ 
     \hline 
  Multinomial logit & 0.141 & 0.355 & 0.378 & 0.126 & 0.038 \\ 
   & (0.007) & (0.015) & (0.017) & (0.007) & (0.009) \\ 
    \hline 
  Nested logit & 0.114 & 0.353 & 0.383 & 0.150 & 0.001 \\ 
  (charter and others) & (0.010) & (0.015) & (0.017) & (0.011) & (0.002) \\ 
    \hline 
  Nested logit & 0.141 & 0.355 & 0.378 & 0.126 & 0.038 \\ 
  (boat and others) & (0.007) & (0.015) & (0.017) & (0.007) & (0.009) \\ 
   \hline 
  Mixed-logit with & 0.142 & 0.354 & 0.378 & 0.126 & 0.038 \\ 
  log normal distribution & (0.008) & (0.015) & (0.017) & (0.007) & (0.009) \\ 
    \hline 
  Multinomial logit & 0.113 & 0.353 & 0.383 & 0.151 & 0.000 \\ 
  with fixed effects & (0.009) & (0.015) & (0.017) & (0.010) & (0.000) \\ 
    \hline 
  Mixed-logit with log normal & 0.114 & 0.353 & 0.383 & 0.151 & 0.000 \\ 
   distribution and fixed effects & (0.009) & (0.015) & (0.017) & (0.010) & (0.000) \\ 
   \hline
\end{tabular}
}
\end{center}
\label{tbl:in-sample_d1}
\begin{tablenotes}
\item \begin{scriptsize} \textit{Note}: 
Table \ref{tbl:in-sample_d1} summarizes the in-sample fit of different models. The row ``our method'' presents choice probabilities predicted by the four-mixture mixed-logit model and the prediction error. The remaining rows present in-sample predicted choice probabilities and prediction errors obtained by standard models. In parentheses are standard deviations obtained by repeating the same analyses 50 times.
\end{scriptsize}
\end{tablenotes}
\end{table}

\begin{table}[ht]
\caption{Out-of-Sample Fit} 
\begin{center}
\scalebox{0.85}{
\begin{tabular}{|c|c|c|c|c|c|}
  \hline
  \multirow{2}{*}{Model } &\multicolumn{4}{c|}{Choice probabilities}&\multicolumn{1}{c|}{Prediction error} \\ \cline{2-6}
 & Beach  & Boat    & Charter  & Pier & \\ 
   &  (1) &  (2) & (3) & (4) & (5)\\  
  \hline
Our method & 0.114 & 0.353 & 0.383 & 0.151 & 0.049 \\ 
  & (0.009) & (0.015) & (0.017) & (0.010) & (0.017) \\ 
  \hline 
  Multinomial logit & 0.143 & 0.353 & 0.377 & 0.127 & 0.058 \\ 
  & (0.011) & (0.017) & (0.022) & (0.011) & (0.019) \\ 
    \hline 
  Nested logit  & 0.115 & 0.350 & 0.383 & 0.152 & 0.050 \\ 
  (charter and others)  & (0.012) & (0.022) & (0.018) & (0.014) & (0.020) \\ 
    \hline 
  Nested logit & 0.143 & 0.353 & 0.377 & 0.127 & 0.058 \\ 
  (boat and others) & (0.011) & (0.017) & (0.022) & (0.011) & (0.019) \\ 
    \hline 
  Mixed-logit with & 0.143 & 0.352 & 0.377 & 0.127 & 0.058 \\ 
  log normal distribution & (0.010) & (0.016) & (0.021) & (0.010) & (0.018) \\ 
    \hline 
  Multinomial logit & 0.116 & 0.350 & 0.380 & 0.154 & 0.048 \\ 
  with fixed effects & (0.014) & (0.019) & (0.022) & (0.019) & (0.022) \\ 
    \hline 
  Mixed-logit with log normal & 0.113 & 0.353 & 0.383 & 0.151 & 0.048 \\ 
  distribution and fixed effects & (0.008) & (0.015) & (0.018) & (0.010) & (0.017) \\ 
   \hline
\end{tabular}
}
\end{center}
\label{tbl:out-sample}
\begin{tablenotes}
\item \begin{scriptsize} \textit{Note}: 
Table \ref{tbl:out-sample} summarizes the out-of-sample fit of different models. The row ``our method'' presents choice probabilities predicted by the four-mixture mixed-logit model and the prediction error (\ref{eq:dist.1}). The remaining rows present out-of-sample predicted choice probabilities and prediction errors obtained by standard models. In parentheses are standard deviations obtained by repeating the same analyses 50 times.
\end{scriptsize}
\end{tablenotes}
\end{table}

\subsection{Definitions of Other Models}\label{section:models}
In each of the standard models used in our empirical section, the choice probability $\rho(J,j)\equiv \rho_j$ of alternative $j$ from $J$ is specified as follows:
\begin{itemize}
    \item Multinomial logit: $\rho_j = \dfrac{\exp{(\beta \cdot x_j)}}{\sum_{j^\prime \in J}\exp{(\beta \cdot {x_j}^{\prime}})}$
    \item Nested logit (charter and others): the choice probability of alternative $j$ that belongs to nest $J_g$ is specified as $$\rho_j= \frac{\exp{(\beta \cdot x_j/\lambda})}{\sum_{j^\prime \in J_g}\exp{(\beta \cdot {x_j}^{\prime}/\lambda})} \times \frac{\left[\sum_{j^\prime \in J_g} \exp{(\beta \cdot {x_j}^{\prime}/\lambda})\right]^{\lambda}}{\sum_{g^\prime \in G}\left[\sum_{j^\prime \in J_{g^\prime}} \exp{(\beta \cdot {x_j}^{\prime}/\lambda})\right]^{\lambda}}.$$
    The nest is defined by the partition $G=\left\{\left\{\textrm{charter}\right\}, \left\{\textrm{beach, boat, pier}\right\} \right\}.$
    \item Nested logit (boat and others): the nested logit model specified above, with the nest defined by $G=\left\{\left\{\textrm{boat}\right\}, \left\{\textrm{beach, charter, pier}\right\} \right\}.$
    \item Mixed-logit: $\rho_j= \int\dfrac{\exp{(\beta \cdot x_j)}}{\sum_{j^\prime \in J}\exp{(\beta \cdot {x_j}^{\prime}})}f(\beta)d\beta$ where $f$ is the density of the distribution of random coefficients. We use independent log-normal distributions for each coefficient. To evaluate the integral, we randomly draw 100 realizations from the random coefficient distribution.
    \item Multinomial logit with fixed effects: the above multinomial logit model with $x$ including dummies for each alternative (except for beach). 
    \item Mixed-logit with fixed effects: the random-coefficient logit model with log normal distributions. We also include fixed effects for each alternative (except for beach). To evaluate the integral, we randomly draw 100 realizations from the random coefficient distribution.
\end{itemize}

\clearpage

\section{Notation}

\begin{center}
\begin{table}[h]
    \centering
    \begin{tabular}{|c|c|}
\hline
    Notation & Meaning \\
    \hline
     $J$    & the set of alternatives \\
      \hline
     $D$    & a generic choice set \\
      \hline
     $\mathcal{D}$    & the set of generic choice sets \\
      \hline
     $\mathcal{M}$ & the set of standard probability measures \\
      \hline
     $K$   &the dimension of explanatory variables \\
      \hline
     $\mathcal{P}$ &the set of stochastic choice functions \\
      \hline
          $\mathcal{P}_r$ &the set of random utility models \\
           \hline
     $\rho$ & a generic stochastic choice function \\
      \hline
     $\Pi$ & the set of rankings\\
      \hline
     co\hspace{2pt}$C$ & the convex hull of a set $C$ \\
      \hline
     rint\hspace{2pt}$C$ & the relative interior of $C$ \\
      \hline
      cl\hspace{2pt}$C$ &  the closure of $C$ \\
      \hline
     $\mathcal{P}_{ra}(\eta|\mu)$ &
 \begin{tabular}{@{}c@{}} the set of random-coefficient ARUMs with fixed effects $\eta$ \\ and a probability measure $\mu$\end{tabular}
 \\
      \hline
     $\mathcal{P}_{a}(\eta|\mu)$ & the set of ARUMs with fixed effects $\eta$ and a probability measure $\mu$ \\
      \hline
     $\mathcal{P}_{ml}(\eta)$ & the set of mixed-logit models with fixed effects $\eta$ \\
    \hline
\end{tabular}
\end{table}
\end{center}

\end{document}